\documentclass[twocolumn,tighten]{aastex62}

\revised{\today}

\shorttitle{Interloper and cosmological parameter estimation}
\shortauthors{Grasshorn Gebhardt et al.}

\usepackage{graphicx}
\usepackage{amsmath}
\usepackage{amssymb}
\usepackage{verbatim}
\usepackage{rotate}
\usepackage{color}
\usepackage{bm}

\usepackage{savesym}
\savesymbol{tablenum}
\usepackage{siunitx}
\restoresymbol{SIX}{tablenum}
\usepackage{tikz}
\usetikzlibrary{calc,graphs,positioning}

\definecolor{RedWine}{rgb}{0.743,0,0}
\definecolor{GrassGreen}{rgb}{0.125,0.75,0.125}
\definecolor{RoyalBlue}{rgb}{0.25,0.41,0.88}

\newcommand{\be}{\begin{equation}}
\newcommand{\ee}{\end{equation}}
\newcommand{\bea}{\begin{eqnarray}}
\newcommand{\eea}{\end{eqnarray}}
\def\ba#1\ea{\begin{align}#1\end{align}}

\def\({\left(}
\def\){\right)}
\def\<{\left<}
\def\>{\right>}

\newcommand{\refeq}[1]{Eq.~(\ref{eq:#1})}          
\newcommand{\refeqs}[2]{Eqs.~(\ref{eq:#1})--(\ref{eq:#2})}          
\newcommand{\reffig}[1]{Fig.~\ref{fig:#1}}          
\newcommand{\refsec}[1]{Sec.~\ref{sec:#1}}
\newcommand{\refapp}[1]{App.~\ref{app:#1}}

\newcommand{\vs}{\nonumber\\} 
\newcommand{\plane}[2]{$#1$\nobreakdash-$#2$~plane}

\def\vr{\bm{r}}
\def\vx{\bm{x}}

\def\vvs{\bm{s}}
\def\vk{{\bm{k}}}

\def\bfq{\bm{q}}

\def\nhat{\hat{\bm{n}}}
\def\khat{\hat{\bm{k}}}

\def\fnl{f_\mathrm{NL}}

\def\orderof{\mathcal{O}}

\DeclareSIUnit \parsec {pc}
\DeclareSIUnit \h {\text{$h$}}
\DeclareSIUnit \year {yr}
\DeclareSIUnit \solarmass {M_\odot}
\DeclareSIUnit \Mpc {\mega\parsec}

\def\tot{\rm total}
\def\oii{\mathrm{OII}}
\def\oiii{\mathrm{OIII}}
\def\lae{\mathrm{LAE}}

\def\hae{\mathrm{HAE}}

\def\obs{\mathrm{obs}}
\def\proj{\mathrm{proj}}
\def\true{\mathrm{true}}

\def\max{\mathrm{max}}

\def\fill{\mathrm{fill}}

\def\xlae{x_\lae}
\def\xoii{x_\oii}
\def\Plin{P_{\rm Lin}}

\def\Like{\mathfrak{L}}
\def\f{\mathfrak{f}}
\def\g{\mathfrak{g}}
\def\ffid{\f_\mathrm{true}}
\def\gfid{\g_\mathrm{true}}

\def\dd{\mathrm{d}}

\newcommand{\incgraph}[2][0.49]{\includegraphics[width=#1\textwidth]{#2}}

\begin{document}

\title{
    Unbiased Cosmological Parameter Estimation from Emission-line Surveys
    with Interlopers
}

\author[0000-0002-8158-0523]{Henry S. Grasshorn Gebhardt} 
\email{hsg113@psu.edu}
\affil{Department of Astronomy and Astrophysics and 
    Institute for Gravitation and the Cosmos, \\
    The Pennsylvania State University, University Park, PA 16802,
    USA}

\author[0000-0002-8434-979X]{Donghui Jeong}
\email{djeong@psu.edu}
\affil{Department of Astronomy and Astrophysics and 
    Institute for Gravitation and the Cosmos, \\
    The Pennsylvania State University, University Park, PA 16802,
    USA}

\author{Humna Awan}
\affil{Department of Physics and Astronomy,
Rutgers, The State University of New Jersey, \\
136 Frelinghuysen Road,
Piscataway, NJ 08854-8019, USA}

\author{Joanna S. Bridge}
\affil{Department of Physics and Astronomy, 102 Natural Science Building, University of Louisville, Louisville, KY 40292, USA}
     
\author{Robin Ciardullo}
\affil{Department of Astronomy and Astrophysics and 
    Institute for Gravitation and the Cosmos, \\
    The Pennsylvania State University, University Park, PA 16802,
    USA}

\author{Daniel Farrow}
\affil{Max-Planck-Institut f\"ur extraterrestrische Physik, Postfach 1312 Giessenbachstrasse, D-85741 Garching, Germany}

\author{Karl Gebhardt}
\affil{Department of Astronomy, University of Texas at Austin, 2515 Speedway, Stop C1400, Austin, TX 78712, USA}

\author{Gary J. Hill}
\affil{Department of Astronomy, University of Texas at Austin, 2515 Speedway, Stop C1400, Austin, TX 78712, USA}
\affil{McDonald Observatory, University of Texas at Austin, 2515 Speedway, Stop 1402, Austin, TX 78712, USA}

\author{Eiichiro Komatsu}
\affil{Max-Planck-Institut f\"ur Astrophysik, Karl-Schwarzschild-Str.  1, D-85741 Garching, Germany}
\affil{Kavli Institute for the Physics and Mathematics of the
Universe, Todai Institutes for Advanced Study, the University of Tokyo, \\
Kashiwa 277-8583 (Kavli IPMU, WPI), Japan}

\author{Mallory Molina}
\affil{Department of Astronomy and Astrophysics and 
    Institute for Gravitation and the Cosmos, \\
    The Pennsylvania State University, University Park, PA 16802,
    USA}

\author{Ana Paulino-Afonso}
\affil{Max-Planck-Institut f\"ur extraterrestrische Physik, Postfach 1312 Giessenbachstrasse, D-85741 Garching, Germany}

\author{Shun Saito}
\affil{Max-Planck-Institut f\"ur Astrophysik, Karl-Schwarzschild-Str.  1, D-85741 Garching, Germany}

\author{Donald P. Schneider}
\affil{Department of Astronomy and Astrophysics and 
    Institute for Gravitation and the Cosmos, \\
    The Pennsylvania State University, University Park, PA 16802,
    USA}

\author{Greg Zeimann}
\affil{Hobby Eberly Telescope, University of Texas, Austin, TX 78712, USA}

\begin{abstract}
    The galaxy catalogs generated from low-resolution emission-line surveys
    often contain both foreground and background interlopers due to line
    misidentification, which can bias the cosmological parameter estimation. In
    this paper, we present a method for correcting the interloper bias by using
    the joint analysis of auto- and cross-power spectra of the main and the
    interloper samples. In particular, we can measure the interloper fractions
    from the cross-correlation between the interlopers and survey galaxies,
    because the true cross-correlation must be negligibly small. The
    estimated interloper fractions, in turn, remove the interloper bias in the
    cosmological parameter estimation. For example, in the Hobby-Eberly
    Telescope Dark Energy Experiment low-redshift ($z<0.5$)
    [O~II]~$\lambda3727$\AA{} emitters contaminate high-redshift ($1.9<z<3.5$)
    Lyman-$\alpha$ line emitters. We demonstrate that the joint-analysis method
    yields a high signal-to-noise ratio measurement of the interloper fractions
    while only marginally increasing the uncertainties in the cosmological
    parameters relative to the case without interlopers. We also show that
    the same is true for the high-latitude spectroscopic survey of the
    \emph{Wide-Field Infrared Survey Telescope} mission where contamination
    occurs between the Balmer-$\alpha$ line emitters at lower redshifts
    ($1.1<z<1.9$) and oxygen ([O~III]~$\lambda5007$\AA{}) line emitters at
    higher redshifts ($1.7<z<2.8$).
\end{abstract}

\section{Introduction}
Current and future spectroscopic surveys such as HETDEX \citep[Hobby-Eberly
Telescope Dark Energy Experiment;][]{HETDEX}, eBOSS \citep[Extended Baryon
Oscillation Spectroscopic Survey;][]{eBOSS}, DESI \citep[Dark Energy
Spectroscopic Instrument;][]{DESI}, PFS \citep[Prime Focus
Spectroscopy;][]{PFS}, WFIRST \citep[Wide-Field Infrared Survey
Telescope;][]{WFIRST}, SPHEREx \citep[Spectro-Photometer for the History of the
Universe, Epoch of Reionization, and Ices Explorer;][]{SPHEREx:2014}, and
Euclid~\citep{Euclid:2013} are designed to map the large-scale structure of the
universe by measuring the positions of millions of galaxies. The galaxy power
spectrum, which is the Fourier transform of the galaxy two-point correlation
function, is a leading statistical measure of the large-scale structure, which
can constrain a number of cosmological parameters. For example, several groups
have used the baryon acoustic oscillation \citep[BAO;][]{BAO/2dF,BAO/SDSS}
feature as a standard ruler to measure the Hubble expansion rate $H(z)$ and
angular diameter distance $d_A(z)$, while redshift-space distortion
\citep[RSD;][]{kaiser:1987} has been used to constrain the linear growth rate
parameter $f(z)$. These measurements provide, respectively, the geometrical and
dynamical test of dark energy \citep[for a review, see][]{DEreview:2013}. The
scale dependence of the galaxy power spectrum relative to the matter power
spectrum on large scales also provides constraints on the non-Gaussianities in
the initial density fluctuations \citep{dalal/etal:2008,PBSreview} and is a
unique approach to check the consistency for the general theory of relativity
on cosmological scales \citep{yoo/etal:2009,gaugePk,GRreview}.

Many of the spectroscopic surveys have a modest spectral resolution ($R\equiv
\lambda/\Delta\lambda<1000$) and limited bandwidth that often leave an
ambiguity in emission-line identifications at specific redshifts. As a result,
a fraction of the objects in galaxy catalogs constructed from these surveys
are foreground or background interlopers. Recently, \citet{pullen+2016}
investigated the effect of both foreground and background interlopers on the
galaxy power spectrum. They show that the interlopers would induce systematic
biases in the cosmological parameter estimation. \cite{lidz/taylor:2016} and
\cite{cheng/chang/+:2016} explored the possibility of cleaning the interloper
effect in the intensity power spectrum from the spurious anisotropies induced
by the interlopers.

In this paper, we demonstrate that we can eliminate such interloper bias
by considering the statistics of both interlopers and main survey galaxies.
By simultaneously analyzing the auto- and cross-power spectra of the main
survey galaxies and the interlopers, we can estimate the interloper fraction
and the cosmological parameters. The cosmological parameters measured in this
joint-analysis method are unbiased, albeit with slightly increased measurement
uncertainties.

Interlopers in the primary and secondary samples will cause a
non-negligible angular cross-correlation that would otherwise be vanishingly
small due to the two samples being widely separated in redshift. This is the
case for HETDEX where the interlopers ([O~II]~$\lambda$3727\AA{} emitters,
hereafter OIIEs) are at $z<0.5$, while the main survey galaxies
(Lyman-$\alpha$~$\lambda$1216\AA{} emitters, or LAEs) are at $1.9<z<3.5$. For a
program such as the high-latitude spectroscopic survey of the proposed \emph{WFIRST}
mission, the redshift ranges of H$\alpha$
($\lambda=\SI{6563}{\angstrom}$) and [O~III]~$\lambda5007$\AA{} may overlap,
but the corresponding galaxies have sufficient separation so that the
cross-correlation is negligible compared to the autocorrelations.

We focus here on galaxy surveys with a small footprint such as HETDEX and
\emph{WFIRST}, for which we can apply the Fourier analysis assuming the
flat-sky approximation. For simplicity, we ignore the redshift evolution of the
galaxy number density, interloper fraction, as well as the linear growth rate.
In our investigation, we mimic the angular cross-correlation by
projecting one population (OIIEs, for example) onto the redshift of the other
(LAEs). Throughout the paper, we use HETDEX as our main case study, but the
formalism we develop is applicable for any survey afflicted with interlopers.
As an example, we apply the same formalism to the \emph{WFIRST} mission.

We assume a flat $\Lambda$CDM model for our fiducial cosmology with parameters
in the \textsf{base\_plikHM\_TTTEEE\_low\linebreak[0]TEB\_lensing\_post\_BAO\_H080p6\_JLA}
column from \emph{Planck} 2015 \citep{planck:2015-overview,planck:2015-parameter}:
$\Omega_{\Lambda}=0.69179$, $\Omega_{b0}h^2=0.022307$,
$\Omega_{c0}h^2=0.11865$,
$\Omega_{\nu0}h^2=0.000638$,
$h=0.6778$, and $n_s=0.9672$. 
We calculate the linear power spectrum $\Plin(k)$ with
\texttt{CAMB}\footnote{\url{http://www.camb.info}} and normalize the linear
power spectrum by setting the root-mean-squared value of the smoothed
(spherical filter with radius \SI{8}{\per\h\mega\parsec}) linear density
contrast, $\sigma_8=0.8166$. 

We begin in \refsec{prelim} by discussing preliminaries, providing details for
the HETDEX survey, giving a precise definition of interloper fraction, and
discussing the projection effects of misidentification. In \refsec{pk}, we
present the effect of interlopers on the density contrast, the
configuration-space correlation functions, and the galaxy power spectrum
measurement, including galaxy bias and redshift-space distortion. We construct
the likelihood function and apply our method to HETDEX in \refsec{fitting}, and
to \emph{WFIRST} in \refsec{wfirst}. We conclude in \refsec{conclusion}.
\refapp{fgplane} discusses the transformation between misidentification and
interloper fractions. \refapp{shotnoise} provides a rigorous derivation of the
observed galaxy power spectra including the discrete nature of the galaxy
density field, and \refapp{pkerror} derives the measurement uncertainty on the
power spectrum. Finally, in \refapp{systematicbias}, we present a formula
estimating the systematic bias in cosmological parameters from a systematic
shift of the power spectrum.

\section{Preliminaries}\label{sec:prelim}
\subsection{HETDEX}\label{sec:HETDEX}

HETDEX is a blind, integral-field spectroscopic survey observing a
\SI{434}{deg\squared} footprint (\SI{294}{deg\squared} around
\SI{53}{\degree} decl.\ and \SI{140}{deg\squared} around \SI{0}{\degree}
decl.) with a filling factor of $1/4.5$ on sky, over the
wavelength range from \SI{3500}{\angstrom} to \SI{5500}{\angstrom}. The primary
target population for HETDEX are high-redshift ($1.9<z<3.5$) galaxies emitting
the Lyman-$\alpha$ line at rest-frame \SI{1216}{\angstrom}. With the fiducial
cosmological parameters, the total survey volume for LAEs is
$V_\text{survey}=\SI{2.95}{\per\h\cubed\giga\parsec\cubed}$, centered around
$z=2.7$, which corresponds to the fundamental frequency of
$k_F=\SI{0.00438}{\h\per\mega\parsec}$.

The same wavelength range also detects star-forming galaxies at low redshift
($0<z<0.5$) emitting [O~II]~$\lambda3727$\AA{}. If $z_\oii$ is the redshift of
an OIIE and $z_\lae$ is the redshift of a corresponding LAE, then the observed
wavelength of the line is
\ba
\label{eq:redshift-relation}
\lambda^\obs &= \lambda_\oii (1+z_\oii) = \lambda_\alpha (1+z_\lae),
\ea
where $\lambda_\oii$ and $\lambda_\alpha$ are the rest-frame wavelengths of
[O~II]~$\lambda 3727$\AA{} and the Lyman-$\alpha$ line at \SI{1216}{\angstrom},
respectively. For the same HETDEX footprint, OIIEs occupy the volume
$V_{\oii}=\SI{0.0688}{\per\h\cubed\giga\parsec\cubed}$.

Based on the observed luminosity function of high-redshift LAEs
\citep{ciardullo/etal:2012,sobral+:2018}, we expect HETDEX to observe
\num{\sim755000} LAEs and \num{\sim1500000} OIIEs \citep[also
see][]{comparat+:2015}, where we assume a flux limit of
\SI{5e-17}{erg\per\second\per\centi\meter\squared}. We set the linear galaxy
bias for LAEs to $b_\lae=2$, which is consistent with
\cite{guaita+:2010}. For OIIEs, we use the linear bias $b_\oii=1.5$, and we will
show that our method is robust to a change in this value.  For the Fourier
analysis, we include the Fourier modes below the maximum wavenumber $k_{\rm
\max}=\SI{0.4}{\h\per\mega\parsec}$ \citep{jeong/komatsu:2006}, but we also
check that the result stays robust for $k_{\rm
max}=\SI{0.3}{\h\per\mega\parsec}$, which is adopted for
the planning and design of HETDEX \citep{HETDEX}. For our fiducial
cosmological parameters, we find $\bar{n}_g P_g(k) > 1$ for
$k<\SI{0.1}{\h\per\mega\parsec}$.

Confusion arises when the line identification is ambiguous. For the majority of
objects detected by HETDEX, [O~III]~$\lambda$5007\AA{} and H$\beta$ fall
outside the spectral range. In addition, although [O~II]~$\lambda3727$\AA{} is
a doublet, the resolution ($R\sim700$) of the HETDEX spectrographs is too low
to resolve it \citep{hill+:2016}. \citet{leung+2017} investigated a Bayesian
approach to distinguish between OIIEs and LAEs, making use of a number of
factors, including the presence of other lines in the spectrum
and the rest-frame equivalent width of the candidate Ly$\alpha$
line, which tends to be greater than \SI{20}{\angstrom} for LAEs and less than
that for OIIEs \citep{gronwall+:2007,ciardullo+:2013}.
\citet{leung+2017} used this method to reduce the interloper fraction to
\SI{\sim0.5}{\percent} at the expense of missing \SI{\sim6}{\percent} of the
LAEs. For this interloper fraction, we predict that the joint-analysis method
introduced in this paper can measure the interloper fraction with high
significance (see, for example, \reffig{fg_bymodel}).

\subsection{Notation}\label{sec:notation}
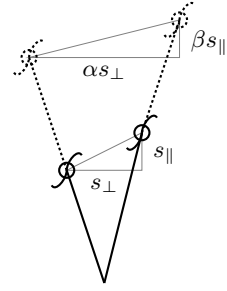
\begin{figure}
    \centering
    \newcommand{\gal}[2][]{%
        \draw[#1] (#2) circle[radius=0.1];
        \draw[#1] ($(#2) - (0.2,0.2)$) edge[out=300,in=270] (#2);
        \draw[#1] (#2) edge[out=90,in=120] ($(#2) + (0.2,0.2)$);
    }
    \begin{tikzpicture}[thick]
        \draw[help lines] (-0.5,1.5) -- (0.5,1.5) -- (0.5,2) -- cycle;
        \draw[help lines] (-1,3) -- (1,3) -- (1,3.5) -- cycle;
        \gal{-0.5,1.5};
        \gal[dash pattern=on 1pt off 1pt]{-1,3};
        \draw (0,0) -- (-0.5, 1.5);
        \draw[dash pattern=on 1pt off 1pt] (-0.5, 1.5) -- (-1,3);
        \gal{0.5,2};
        \gal[dash pattern=on 1pt off 1pt]{1,3.5};
        \draw (0,0) -- (0.5, 2);
        \draw[dash pattern=on 1pt off 1pt] (0.5, 2) -- (1,3.5);
        \node at (0,1.3) {$s_\perp$};
        \node at (0.8,1.7) {$s_{\parallel}$};
        \node at (0,2.8) {$\alpha s_\perp$};
        \node at (1.4,3.2) {$\beta s_{\parallel}$};
    \end{tikzpicture}
    \caption{\label{fig:contamination-projection}
        Illustration of the geometry of misidentification. The OIIEs (solid
        galaxy symbol) at lower redshifts are projected to higher redshifts
        (dashed galaxy symbol); they occupy a larger volume at a larger radius.
        In the figure, the observer is located at the bottom vertex. The true
        separations along the tangential direction ($s_\perp$) and radial
        direction ($s_{\parallel}$) are projected, respectively, to $\alpha
        s_\perp$ and $\beta s_{\parallel}$ when the OIIEs are misidentified as
        LAEs. The scaling factors $\alpha$ and $\beta$ are defined in terms of
        the geometrical quantities in \refeq{scaling_parameters}.
    }
\end{figure}

Throughout the paper, we shall use the following notation. First, we denote the
fraction of misidentified LAEs and OIIEs by, respectively, $\xlae$ and $\xoii$.
That is, if there are $N_\lae$ LAEs and $N_\oii$ OIIEs in the survey volume,
the observed number of LAEs ($N_\lae^\obs$) and OIIEs ($N_\oii^\obs$) are,
respectively, 
\ba
N_\lae^\obs =\,& (1-x_\lae)N_\lae + x_\oii N_\oii\,,
\label{eq:numlaeobs}
\\
N_\oii^\obs =\,& x_\lae N_\lae + (1-x_\oii)N_\oii\,.
\label{eq:numoiiobs}
\ea
Here, we use the superscript ``obs'' to denote the observed quantities in
contrast to their true value. We further define the overall interloper
fractions in the observed sample as 
\ba
\label{eq:f}
\f &\equiv \frac{\xoii N_\oii}{N_\lae^\obs} =\frac{x_\oii \, N_\oii}
{(1-x_\lae) \, N_\lae + x_\oii \, N_\oii}\,, \\
\label{eq:g}
\g &\equiv \frac{\xlae N_\lae}{N_\oii^\obs}=\frac{x_\lae \,  N_\lae}
{x_\lae \, N_\lae + (1-x_\oii) \, N_\oii}\,,
\ea
which will simplify the expressions for the observed density contrast. 

\begin{figure}
    \centering
    \incgraph[0.47]{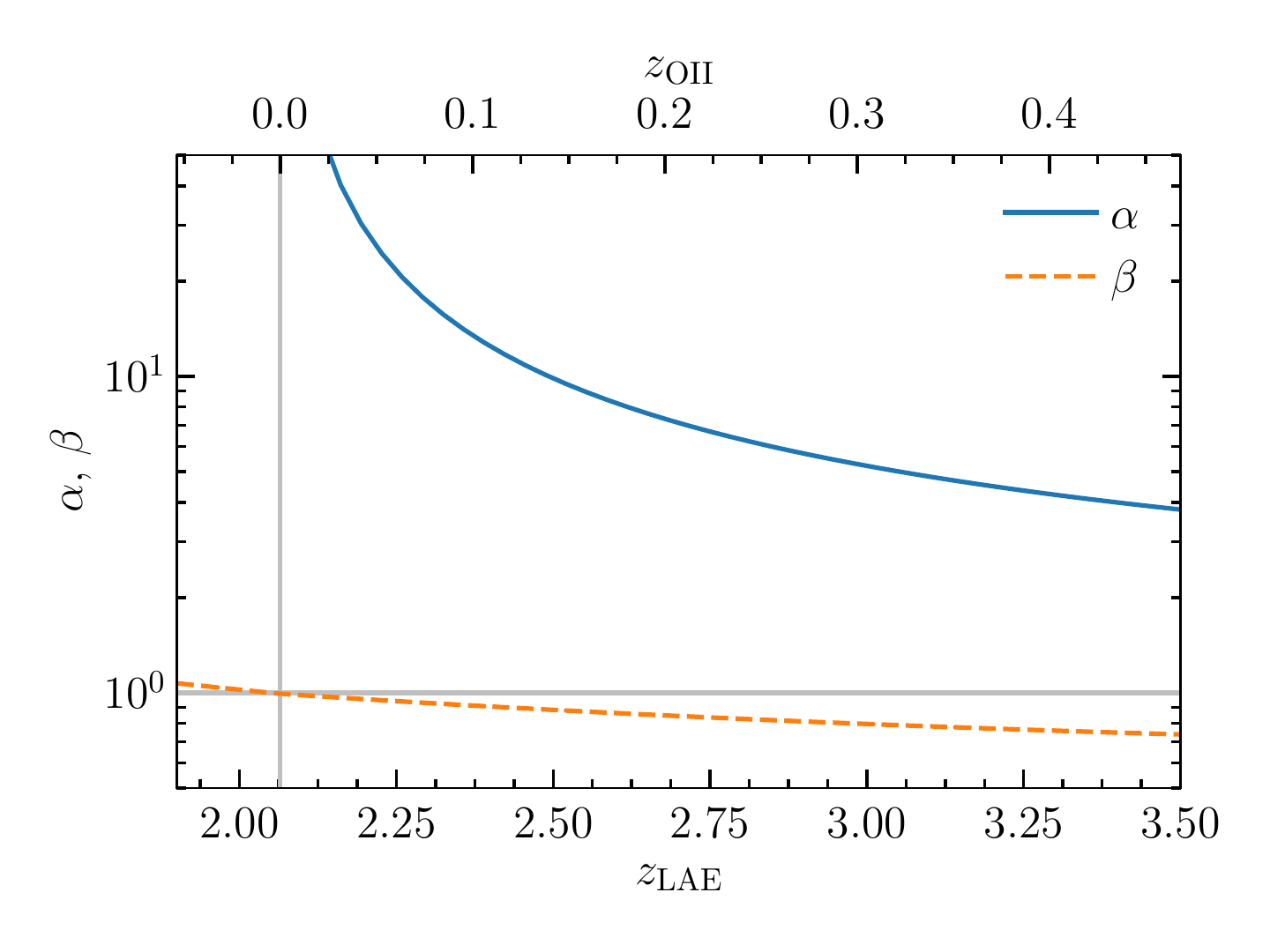}
    \caption{\label{fig:scaling_parameters}
        The scaling factors $\alpha$ (transverse) and $\beta$ (radial) as
        functions of redshifts ($z_\lae$ on the lower axis and $z_\oii$ on
        the upper axis) for the fiducial $\Lambda$CDM model. The gray
        horizontal line marks no rescaling ($\alpha=\beta=1$), while the gray
        vertical line is the limit below which no interloping OIIEs exist.
    }
\end{figure}

For sources in the galaxy-survey catalog, the most direct observables are the
angular coordinate and the redshift $z$. Misidentifying OIIEs and LAEs will
alter the estimated redshift and place the lower redshift objects (at $z_\oii$)
farther away, at the corresponding LAE redshift $z_\lae$ shown in
\refeq{redshift-relation}. As a result (see \reffig{contamination-projection}),
the misidentification stretches the tangential coordinate and the radial
coordinate of the lower redshift galaxies, respectively, by the factors
of\footnote{The angular separation $\Delta\theta$ and redshift difference
    $\Delta z$ are related to the comoving distances as
    \be
    \Delta s_\perp = d_A(z)\Delta \theta\,,\quad
    \Delta s_\parallel = \frac{\Delta z}{H(z)}\,.
    \label{eq:standard_ruler}
    \ee
    Therefore, when the redshifts of the galaxies are misidentified, the
    tangential and the parallel separations change with the scaling factors
    $\alpha$ and $\beta$.
}
\be
\alpha \equiv \frac{d_A(z_\lae)}{d_A(z_\oii)}
\,,\quad
\beta \equiv \frac{\lambda_\oii}{\lambda_\lae}\frac{H(z_\oii)}{H(z_\lae)}\,.
\label{eq:scaling_parameters}
\ee
We shall refer to these variables as scaling factors, where $d_A(z)$ is the
comoving angular diameter distance, and $H(z)$ is the Hubble expansion rate.
\reffig{scaling_parameters} displays the redshift dependence of the scaling
factors $\alpha$ and $\beta$ as a function of $z_\oii$ (upper abscissa) and
$z_\lae$ (lower abscissa) for the fiducial cosmology. At all redshifts of
interest, the change of coordinate is more significant in the tangential
direction ($\alpha$) than in the radial direction ($\beta$). For example, for
LAEs at redshift $z_\lae\sim2.7$, and OIIEs at redshift $z_\oii\sim0.2$,
$\alpha\sim7.1$ due to the change in angular diameter distance, whereas
$\beta\sim0.84$, because the change in $H(z)$ is largely compensated by
the ratio of the wavelengths in \refeq{scaling_parameters}. The
disparity in $\alpha$ and $\beta$ introduces yet another source of anisotropy
in the observed galaxy clustering.

\section{Observed Correlation functions with interlopers}\label{sec:pk}
The presence of low-redshift interlopers in the high-redshift galaxy sample
biases the clustering measurement. Similarly, the observed density contrast has
extra contributions produced from the density contrast of the interlopers. In
this section, we determine the effect interloping OIIEs have on the observed
density contrast and two-point correlation functions of LAEs. First, we derive
the effect on the observed density contrast and the two-point correlation
function in configuration space, then Fourier-transform these quantities to
derive the expression for the power spectrum.

Strictly speaking, one must fix the observed angular position and rescale only
the radial position for each interloper. In this paper, however, we focus on
the three-dimensional Fourier analysis by ignoring the opening-angle effect
and by applying the scaling factors at the median redshift ($z_\lae=2.7$ and
$z_\oii=0.2$) to all interloping galaxies. This approach projects the
interloping OIIEs from the true lower redshift cuboid volume to the
high-redshift cuboid volume of LAEs. This approximation provides a good
description for galaxy surveys with small sky coverage and a narrow range of
redshifts, with the correction only proportional to the square of the opening
angle, and the redshift bin size. It can be undoubtedly applied to galaxy
surveys such as HETDEX ($\simeq\SI{400}{deg\squared}$) and \emph{WFIRST}
($\simeq\SI{2000}{deg\squared}$). For galaxy surveys with broader sky
coverage, we must employ different statistics based on the Fourier-Bessel or
total angular momentum wave basis \citep{dai/kamionkowski/jeong:2012}.

\subsection{Observed density contrast with interlopers}
Let us denote by $\vvs$ the position vector for the galaxies in the observed
LAE sample and $\vvs'$ the true position vector of the OIIE interlopers;
i.e., $\vvs'$ refers to the same angular position on the sky as $\vvs$, but
with redshift $z_\oii$ instead of $z_\lae$. Using the scaling factors in
\refeq{scaling_parameters}, $\vvs$ and $\vvs'$ are related by
\ba
\label{eq:sprime}
\vvs=(\alpha\vvs'_\perp,\beta s'_{\parallel})\,,
\ea
where $\perp$ and $\parallel$ represent the components that are tangential and
radial to the line of sight, respectively.

Note that \refeq{sprime} also holds for
interlopers in redshift space, where the line-of-sight directional
peculiar velocity $v_\parallel$ shifts the observed redshift $z_{\rm obs}$
away from the true redshift $z$ by $\(1+z_{\rm
obs}\)=\(1+z\)\(1+v_\parallel/c\)$. As a result, the radial distance $r_s$ in 
redshift space is shifted relative to the real radial distance $r$ by $s
= r + v_\parallel/aH$. The same scaling factor $\beta$, therefore, applies to
both $r$ and $v_\parallel/aH$, when the observed redshift is misidentified.

With these position vectors and the variables defined in
\refeqs{numlaeobs}{g}, we can write the observed density contrast
$\delta_\lae^\obs(\vvs)$ as a function of the true density contrast of
LAEs $\delta_\lae(\vvs)$ and that of the OIIEs $\delta_\oii(\vvs')$ as:
\ba
\delta_\lae^\obs(\vvs)
&\equiv \frac{n^\obs_\lae(\vvs)}{\bar{n}^\obs_\lae}-1
\vs
&= (1-\f)\,\delta_\lae(\vvs) + \f\,\delta_\oii(\vvs')\,,
\label{eq:deltaobs}
\ea
where mean number densities are defined in the LAE volume as
$\bar{n}_\lae\equiv N_\lae/V_\mathrm{survey}$ and $\bar{n}_\oii\equiv
N_\oii/V_\mathrm{survey}$, where $V_\mathrm{survey}$ is the LAE volume.
Analogously, the observed OIIE density contrast is given by
\ba
\delta_\oii^\obs(\vvs')
&= \g\,\delta_\lae(\vvs) + (1-\g)\,\delta_\oii(\vvs')\,.
\label{eq:deltaobsoii}
\ea
The observed density contrast is a superposition of the true LAE density
contrast and the true OIIE density contrast, each contributing proportionally
by number of galaxies in the sample.

Our analysis assumes that the true mean number densities ($\bar{n}_\lae$,
$\bar{n}_\oii$), and the misidentification fractions $\xlae$ and $\xoii$,
remain constant over the survey volume $V_\mathrm{survey}$. For realistic
galaxy surveys, both the mean densities and the overall misidentification
fractions may vary across the survey volume, which results in a nontrivial
window function. We do not study the ramifications here, because the window
function effect can be, in principle, modeled very accurately up to our
knowledge of the survey conditions. For more discussion, see, for example,
chapter 7 of \citet{jeong:2010} and \citet{chiang/etal:2013}.

One subtle but important point is that we define the overall misidentification
fractions $\xlae$ and $\xoii$, as well as $\f$ and $\g$, in terms of the
underlying, continuous galaxy number density fields $N_\lae$ and
$N_\oii$. In reality, the observed galaxy density fields are the distribution
of discrete points (galaxies) that reflect these underlying continuous fields;
therefore, locally measured values of $\xlae$ and $\xoii$ are not necessarily
the same across the survey volume. For instance, in an infinitesimal volume
element we only expect a single LAE. Then, the misidentification fraction can
be either unity (if the galaxy is an OIIE) or zero (if the galaxy is a genuine
LAE). For the local quantities, therefore, the misidentification fractions
$\xlae$ and $\xoii$ are only statistical measures of the
probability of misidentification.

Of course, the density contrasts given in
\refeqs{deltaobs}{deltaobsoii} must lead to the correct result for the galaxy
two-point correlation function and power spectrum. The subtle difference
appears in the treatment of shot noise. For example, for galaxies that are
drawn randomly from a given continuous density field, the shot noise is
proportional to the reciprocal of the total number density of galaxies,
including the interlopers (shown in \refapp{shotnoise}).

\subsection{Observed two-point correlation function with interlopers}
The derivation in this section extends the Appendix A of
\cite{leung+2017}, including the two-point auto-correlation functions of both
the main sample and the interloper sample. 
We calculate the observed two-point correlation function
$\xi(\vvs)=\big<\delta(\vr)\delta(\vr+\vvs)\big>$ in the configuration space from \refeqs{deltaobs}{deltaobsoii} as
\ba
\xi_\lae^\obs(\vvs)
&= \left< \delta_\lae^\obs(\vr) \, \delta_\lae^\obs(\vr+\vvs) \right>
\vs &=
(1-\f)^2 \, \xi_\lae(\vvs)
+ \f^2 \, \xi_\oii^\proj(\vvs)
\vs&\quad
+ 2(1-\f)\f\,\xi_{\lae\times{}\oii}(\vvs)\,,
\label{eq:xiobs_all}
\ea
and
\ba
\xi_\oii^\obs(\vvs')
&= \left< \delta_\oii^\obs(\vr) \, \delta_\oii^\obs(\vr+\vvs') \right>
\vs &=
(1-\g)^2 \, \xi_\oii(\vvs')
+ \g^2 \, \xi_\lae^\proj(\vvs')
\vs&\quad
+ 2(1-\g)\g\,\xi_{\lae\times{}\oii}(\vvs')\,.
\label{eq:xiobsoii_all}
\ea
Here, 
\ba
\xi^\text{proj}_\oii(\vvs) &= \xi_\oii(\vvs') = \xi_\oii(\alpha^{-1}\vvs_\perp,\beta^{-1}s_{\parallel})\,,
\label{eq:xicontam}
\ea
and
\ba
\xi^\text{proj}_\lae(\vvs') &= \xi_\lae(\vvs) = \xi_\lae(\alpha\vvs_\perp',\beta s_{\parallel}')\,
\label{eq:xicontam_oii}
\ea
are the OIIE and LAE two-point correlation functions projected to the wrongly
assigned redshifts; therefore, they contaminate the two-point correlation
function of the respective sample. Note that the projection merely relabels
the coordinates (thus shifting the separation vector) while keeping intact the
amplitude of the two-point correlation function.

The other terms in \refeq{xiobs_all} and \refeq{xiobsoii_all},
$\xi_{\lae\times\oii}$, denote the cross-correlation between the LAEs and projected
OIIEs (evaluated at $\vvs$) and between the OIIEs and projected LAEs (evaluated at
$\vvs'$). They are much smaller than the respective auto-correlation functions
because the wide radial separation between LAEs and OIIEs suppresses the
true cross-correlation, and the cross-correlation from lensing is
small. We can therefore ignore this contribution.

Our final expressions for the observed galaxy two-point correlation functions 
of LAEs and OIIEs are
\ba
\xi_\lae^\obs(\vvs)
&=
(1-\f)^2 \, \xi_\lae(\vvs)
+ \f^2 \, \xi_\oii(\alpha^{-1}\vvs_\perp,\beta^{-1}s_{\parallel})\,,
\label{eq:xiobs}
\\
\xi_\oii^\obs(\vvs')
&=
(1-\g)^2 \, \xi_\oii(\vvs')
+ \g^2 \, \xi_\lae(\alpha\vvs_\perp',\beta s_{\parallel}')\,.
\label{eq:xiobsoii}
\ea
\refeqs{xiobs}{xiobsoii} show that the different scaling factors
($\alpha\neq\beta$) introduce anisotropies into the two-point correlation
functions. This is true even when $\xi(\vvs)$ only depends on $s$ --- for example,
without the RSD.

\subsection{Observed power spectrum with interlopers}
We initially calculate the observed galaxy power spectrum by the Fourier
transform of the corresponding two-point correlation function. 
The results of this section are consistent with those of \citet{pullen+2016} 
and \citet{leung+2017}.

The Fourier transform integrates over the respective observed coordinates---$\vvs$
for LAEs and $\vvs'$ for OIIEs---whose volume forms are related by the scaling
parameters [\refeq{sprime}]
\ba
\dd^3\!s &= \dd^2\!s_\perp \, \dd s_{\parallel} = \alpha^2 \beta \, \dd^3\!s'\,.
\ea
Using \refeq{xiobs}, we compute the observed LAE power spectrum as
\ba
P_{\lae}^\obs(\vk)
=
(1-\f)^2 P_{\lae}(\vk)
+
\f^2
P_\oii^\proj(\vk)\,,
\label{eq:pkobs_lae}
\ea
where $P_\oii^\proj(\vk)$ is the power spectrum of OIIEs 
(at $\vvs'$) projected onto the LAE coordinates ($\vvs$), or the Fourier
transform of \refeq{xicontam}:
\ba
P_\oii^\proj(\vk)
&\equiv
\int \dd^3\!s\,
e^{i\vk\cdot\vvs}
\xi_\oii(\vvs_\perp',s_\parallel')
\vs
&=
\alpha^2\beta
\int \dd^3\!s'
e^{i(\alpha\vk_\perp\cdot\vvs'_\perp+\beta k_\parallel s'_\parallel)}
\xi_\oii(\vvs_\perp',s_\parallel')
\vs
&=
\alpha^2\beta P_\oii(\alpha \vk_\perp,\beta k_\parallel)\,.
\label{eq:pkproj_oii}
\ea
Similarly, the power spectrum of the observed OIIEs is
\be
P_{\oii}^\obs(\vk)
=
(1-\g)^2 P_{\oii}(\vk)
+
\frac{\g^2}{\alpha^2\beta}
P_\lae\(\frac{\vk_\perp}{\alpha},\frac{k_\parallel}{\beta}\)\,.
\label{eq:pkobs_oii}
\ee
\refeq{pkproj_oii} and \refeq{pkobs_oii} demonstrate how misinterpreting the
emission-line redshifts leads to two effects. First, just as in the case for the
two-point correlation functions, the misinterpretation shifts the scales,
projecting small (large) scales onto larger (smaller) scales when
OIIEs (LAEs) are misinterpreted as LAEs (OIIEs). This effect is illustrated in
\reffig{contamination-projection}. Again, $\alpha\neq\beta$ introduces an
additional anisotropy into the observed power spectrum beyond RSD.
Second, the amplitude of the power spectrum is changed proportionally
to the ratio between the true volume and the projected volume. For
example, when OIIEs in a small, low-$z$ volume are projected into the larger,
high-$z$ volume, the projected power spectrum amplitude is boosted by a factor of
$\alpha^2\beta$.

\subsection{The Observed Cross-correlation Functions}\label{sec:crossPk}
As discussed earlier, we ignore the true cross-correlation between the
LAEs and OIIEs. Misidentification can, however, induce a cross-correlation
between the OIIEs and LAEs because both observed samples contain high-$z$ and
low-$z$ objects. Strictly speaking, such a cross-correlation must be measured
in the angular cross-correlation function or in the angular cross-power
spectrum. As we are adopting the flat-sky approximation throughout this paper,
we mimic the angular cross-correlation by the three-dimensional
cross-correlation between $\delta_\lae^\obs(\vvs)$ and
$\delta_\oii^\obs(\alpha\vvs_\perp,\beta\vvs_\parallel)$ artificially placed at
the corresponding LAE redshifts through \refeq{redshift-relation}. This
procedure resembles the angular cross-correlation because for a given LAE
redshift $z_\lae$, the OIIE redshift $z_\oii$ is uniquely
determined. Of course, one can also choose to correlate $\delta_\oii^\obs$ with
$\delta_\lae^\obs$ projected to the OIIE redshifts.

The cross-correlation function defined here is
\ba
\xi_{\lae\times\oii}^\obs(\vvs)
&= \left< \delta_\lae^\obs(\vr) \, \delta_\oii^\obs(\vr+\vvs) \right>
\vs &=
(1-\f)\g\xi_{\lae}(\vvs)
+
\f(1-\g)\xi_{\oii}^\proj(\vvs)
\label{eq:xiobs_LO}\,,
\ea
with $\xi_\oii^\proj(\vvs)$ given in \refeq{xicontam}, and the corresponding
cross-power spectrum
\ba
&P_{\lae\times\oii}^\obs(\vk)
\vs
&=
(1-\f)\g P_{\lae}(\vk)
+
\f(1-\g)\alpha^2\beta P_\oii(\alpha\vk_\perp,\beta k_\parallel)
\label{eq:pkobs_LO}\,.
\ea
The nonzero cross-correlation in \refeq{pkobs_LO} is the key for measuring the
interloper fractions $\f$ and $\g$. This property is an effective
indicator because we expect vanishingly small (contributions from the true
clustering and the lensing magnification) cross-correlation for perfect
($\f=\g=0$) LAE and OIIE samples.

Similarly, we define the observed cross-correlation coefficient using the OIIE
power spectrum projected into the LAE volume as 
\ba
r &\equiv \frac{P_{\lae\times\oii}^\obs(\vk)}{\sqrt{P_\lae^\obs(\vk) P_\oii^{\obs,\proj}(\vk)}}\,.
\ea
The value of $r$ varies continuously between $0$ and $1$ as a function of the
interloper fractions $\f$ and $\g$. The cross-correlation coefficient $r$
reaches maximum ($r=1$) when $\f+\g=1$, and minimum ($r=0$) for the
completely uncontaminated case ($\f=\g=0$) and the completely confused case
($\f=\g=1$)\,.

\subsection{Modeling the Redshift-space Galaxy Power Spectrum}
\label{sec:pkrsd}
\begin{figure*}[th!]
    \centering
    \includegraphics[width=0.85\textwidth]{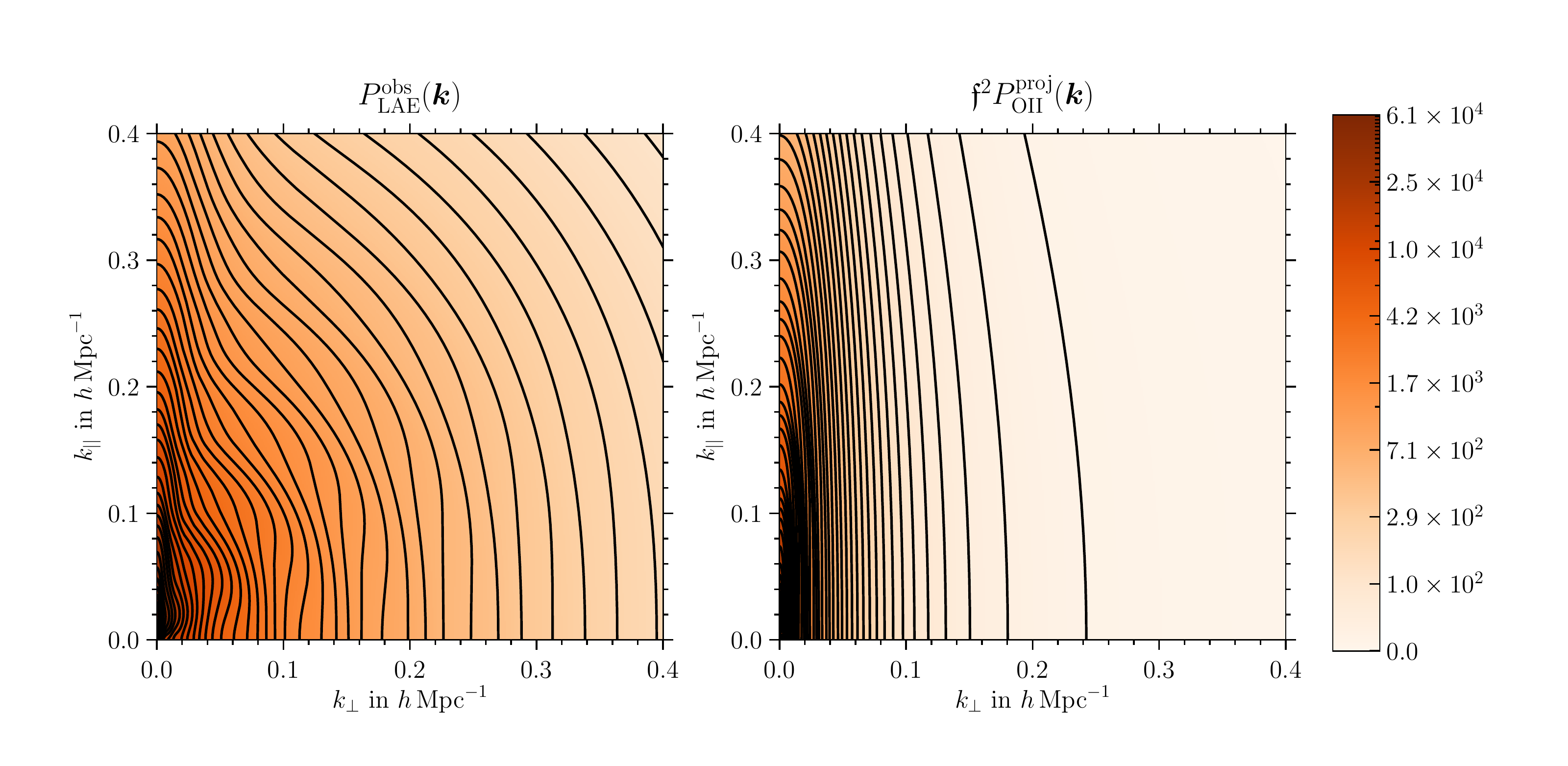}
    \\
    \vspace{-0.5cm}
    \centering
    \includegraphics[width=0.85\textwidth]{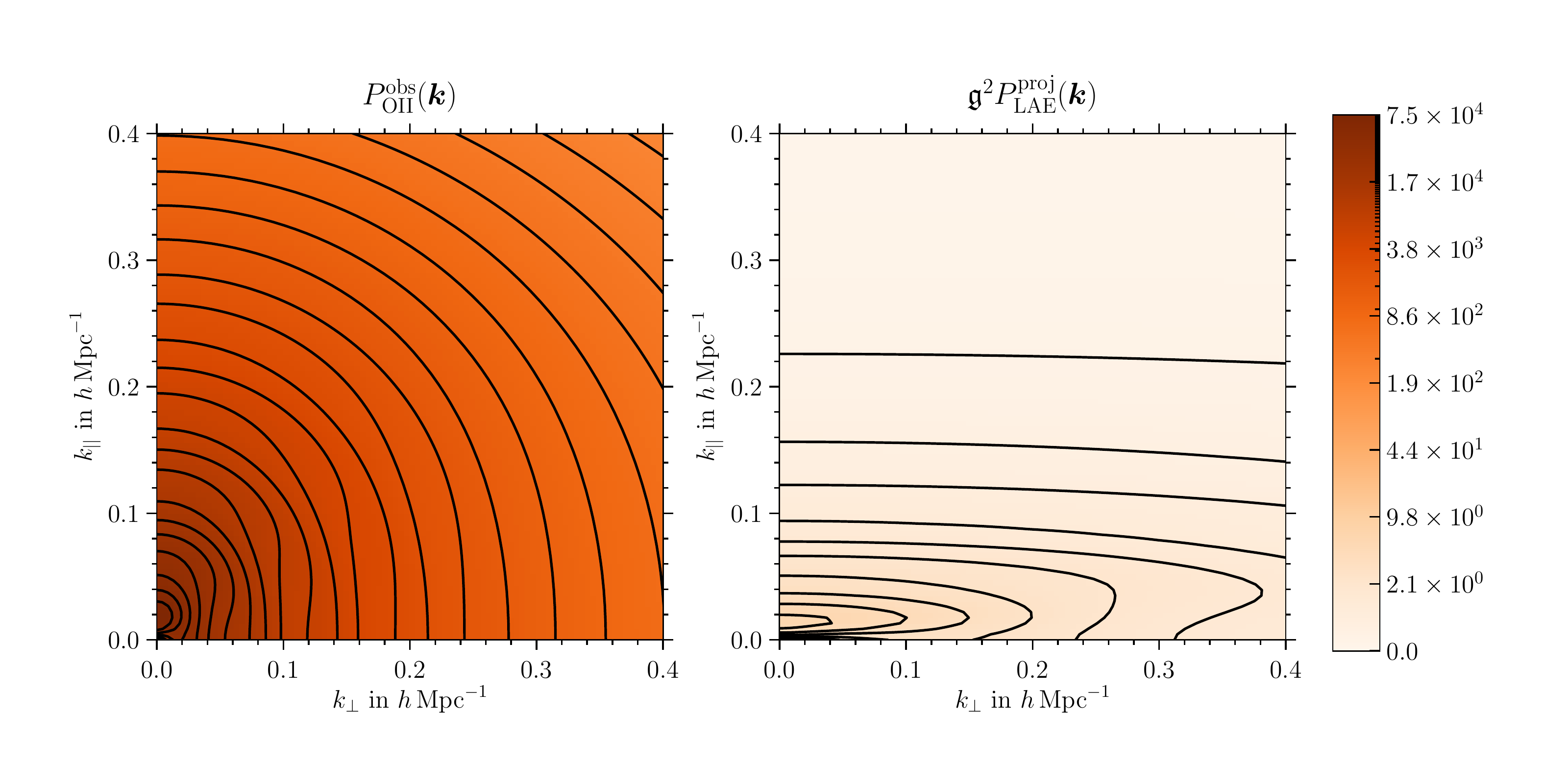}
    \caption{
        \emph{Left:} the observed two-dimensional power spectra of LAEs
        (\emph{top}, assuming $b_\lae=2$ at $z_\lae=2.7$) and OIIEs
        (\emph{bottom}, assuming $b_\oii=1.5$ at $z_\oii=0.207$) including
        RSD, FoG, and the contributions from interlopers. We assume
        the interloper fraction of $\f=\g=0.1$.
        \emph{Right:} contaminant power spectra from projected OIIEs
        (\emph{top}) and LAEs (\emph{bottom}) as they contribute to the
        observed power spectra in the left panels, i.e., the right panels
        contain an additional anisotropic feature, due to interlopers. The
        scaling factors are $\alpha=7.1$ and $\beta=0.84$.
    }
    \label{fig:pk2d-with-contam}
\end{figure*}
\begin{figure*}
    \centering
    \includegraphics[width=0.49\textwidth]{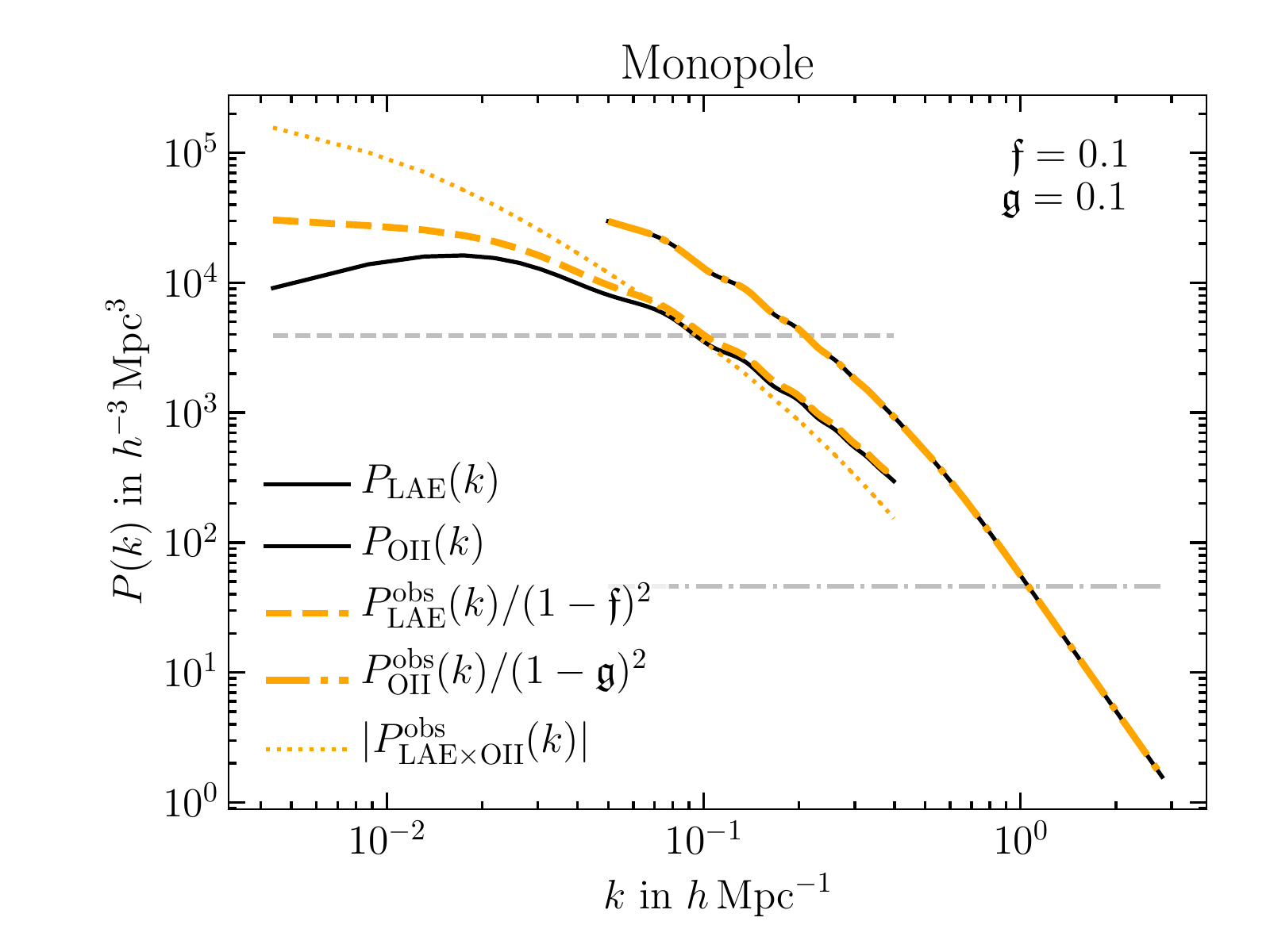}
    \includegraphics[width=0.49\textwidth]{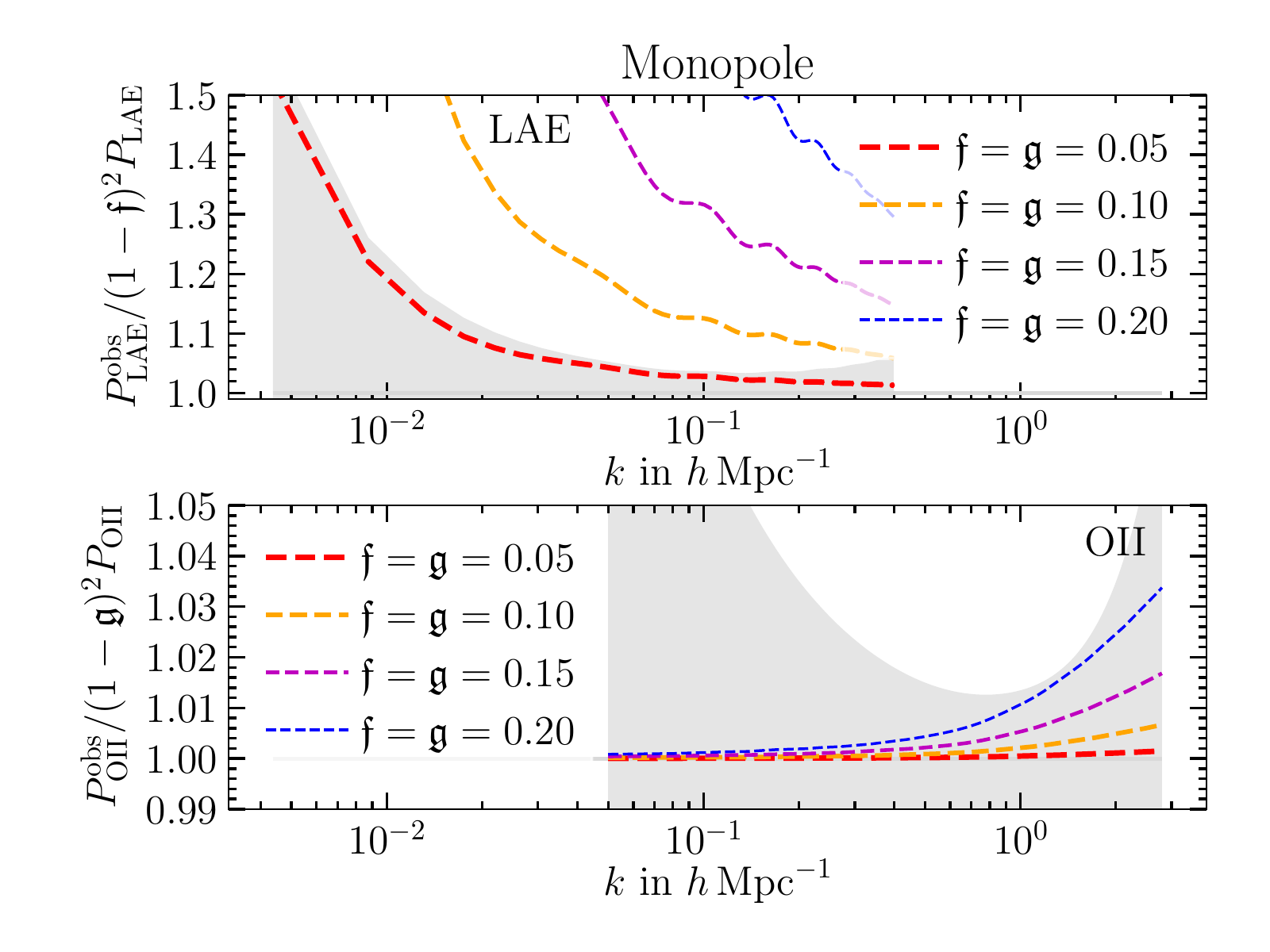}
    \includegraphics[width=0.49\textwidth]{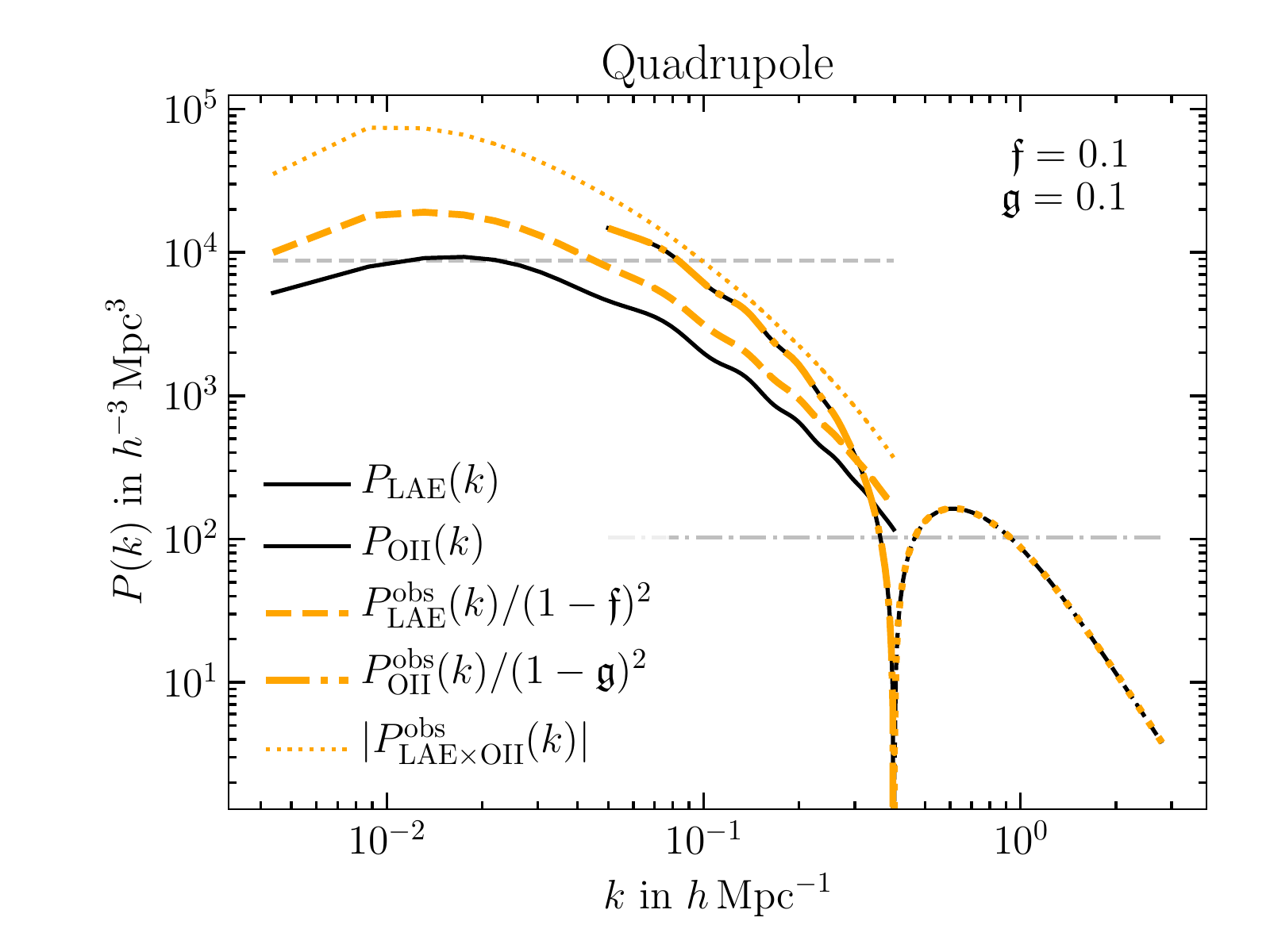}
    \includegraphics[width=0.49\textwidth]{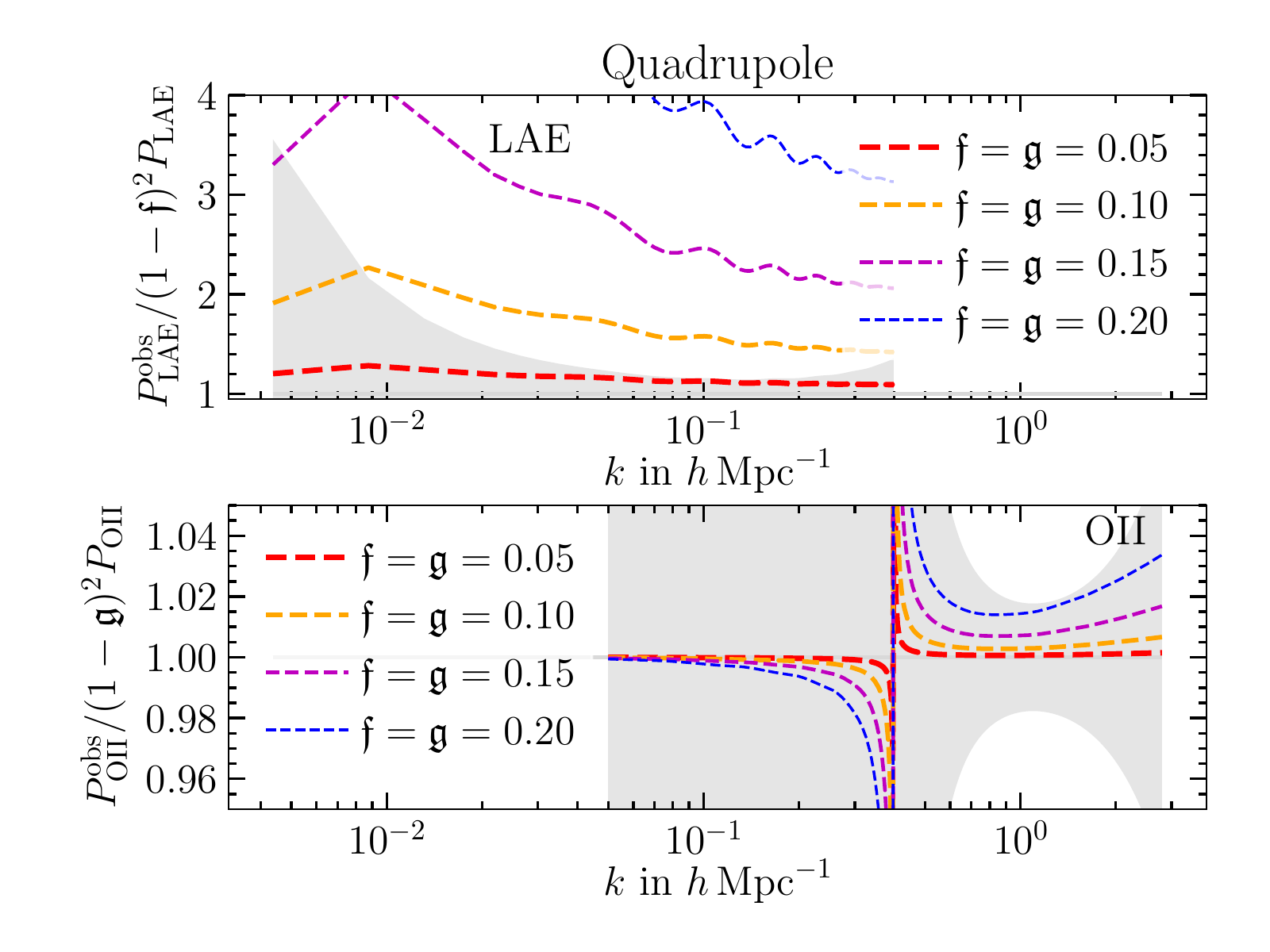}
    \caption{
        \emph{Top left:} The solid black curves represent the fiducial LAE and
        OIIE monopole power spectra for the HETDEX survey. The orange dashed
        line indicates the observed LAE power spectrum with interloper fraction
        $\f=0.1$; the orange dash-dotted line is the OIIE power spectrum with
        interloper fraction $\g=0.1$. We rescale the amplitude of the observed
        power spectra so that they match the corresponding fiducial (black
        solid) power spectra when the contaminant contribution is negligible.
        The dotted orange lines show the monopole and quadrupole of the
        cross-power spectrum. The grey horizontal lines are the shot noise
        contribution $P_{\rm shot}^{\rm monopole} = 1/\bar{n}_g$ and $P_{\rm
        shot}^{\rm quadrupole}=\sqrt{5}/{\bar{n}_g}$ with number density
        $\bar{n}_g$ (see \refapp{pkerror_multipole} for the general formula).
        For these values of the interloper fractions ($\f=\g=0.1$), the
        observed cross-correlation (orange dotted line) has a similar magnitude
        as the LAE power spectrum.
        \emph{Top right:} The two panels in the top right panel compare the
        residual deviation of the observed LAE and OIIE power spectra shapes to
        the true power spectra for $\f,\g=0.05,0.1,0.15,0.2$. The grey areas
        are the expected $1\sigma$ (68\% C.L.) uncertainty ranges.
        \emph{Bottom right:} The bottom panels show the same as the top panels,
        but for the quadrupole of the power spectrum. Here, the LAE redshift is
        $z_\lae=2.7$ and the corresponding OIIE redshift is $z_\oii=0.207$.
    }
    \label{fig:pk_with_contam}
\end{figure*}
\begin{figure*}
    \centering
    \includegraphics[width=0.49\textwidth]{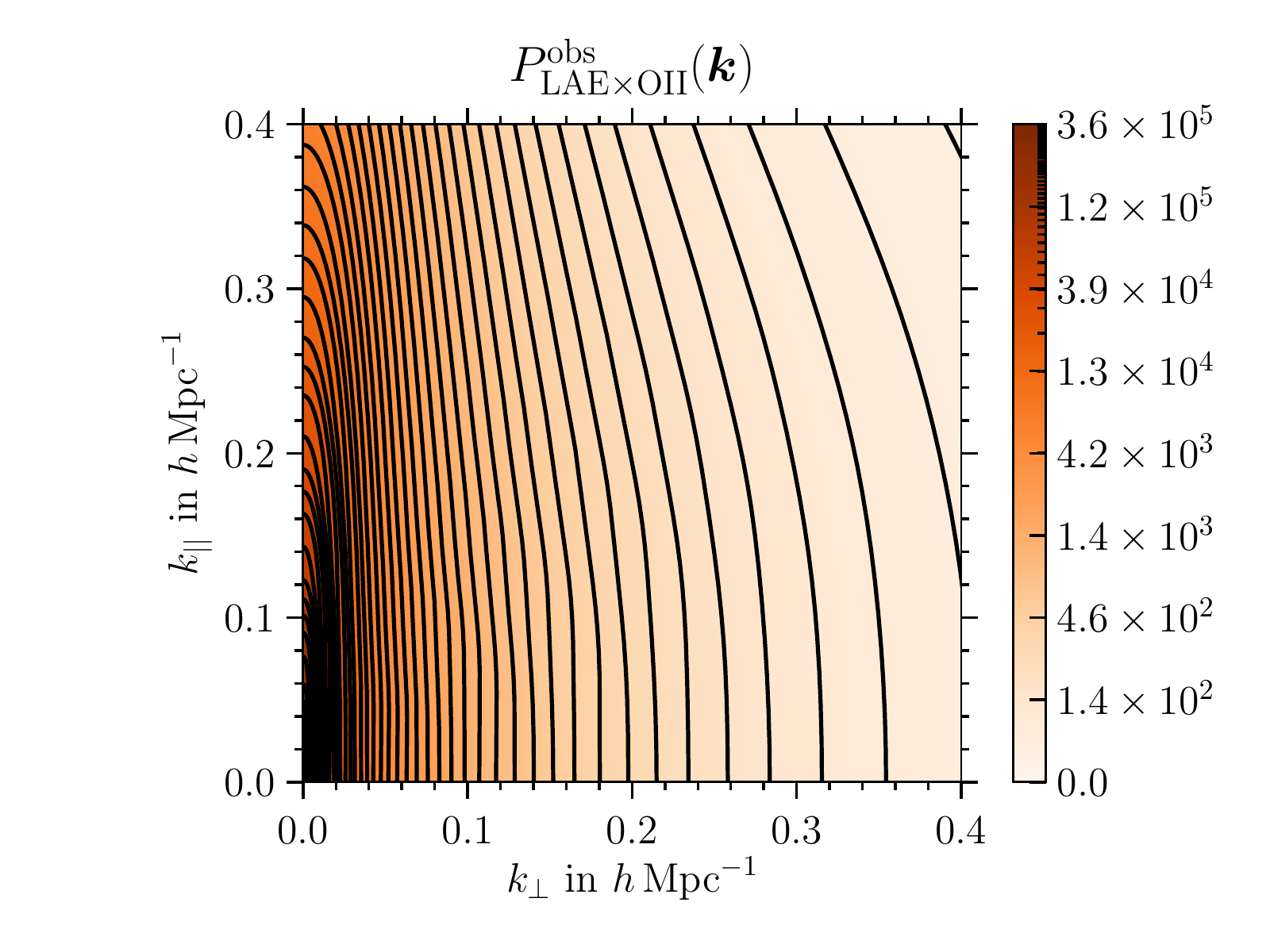}
    \includegraphics[width=0.49\textwidth]{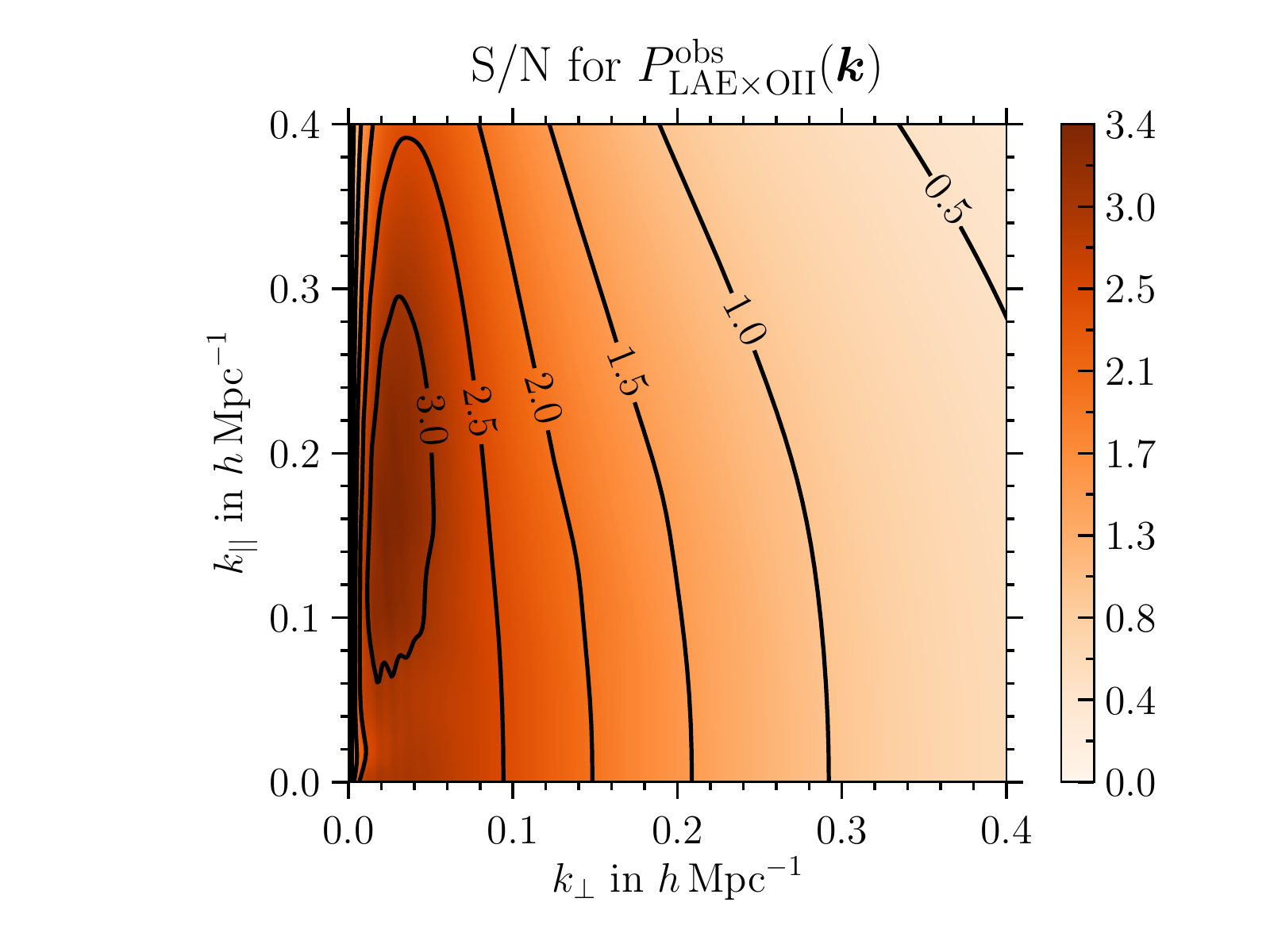}
    \caption{
        \emph{Left:} The observed two-dimensional cross-power spectrum
        between LAEs and OIIEs (projected into LAE volume) with the
        interloper fractions $\f=\g=0.1$. In the case for HETDEX, the
        observed cross-correlation is dominated by the OIIE power spectrum,
        due to the large volume factor.
        \emph{Right:} The bin-to-bin [$\Delta S = \(\partial^2S/\partial
        k_\perp\partial k_\parallel\) \Delta k_\perp \Delta k_\parallel$ with
        $\Delta k_\perp\Delta k_\parallel = k_F^2$] signal-to-noise (S/N)
        ratio for HETDEX with the interloper fractions $\f=\g=0.1$.
    }
    \label{fig:pkcross_2d}
\end{figure*}
The expressions for the observed auto- and cross-power spectra of LAEs
and OIIEs in terms of their true redshift-space power
spectra are given in \refeq{pkobs_lae}, \refeq{pkobs_oii}, and
\refeq{pkobs_LO}. Thus, we can now complete the calculation with expressions
for the true redshift-space power spectra of LAEs and
OIIEs.

As a baseline model for the true redshift-space galaxy
power spectrum, we adopt the linear bias model ($\delta_g(\vx)=b_g\delta_m(\vx)$,
where $\delta_g(\vx)$ is the galaxy number density contrast, $b_g$ the
linear galaxy bias parameter, and $\delta_m(\vx)$ the matter density contrast) with linear
RSD \citep{kaiser:1987} augmented by the Lorentzian
Finger-of-God (FoG) damping \citep{Jackson:1972}:
\ba
P_x(\vk) &= \frac{(1 + \beta_x \mu^2)^2}{1 + f^2(z_x) k^2 \mu^2 \sigma_{v,x}^2} \,
b_x^2 D^2(z_x) \Plin(k)\,.
\label{eq:pk-kaiser}
\ea
Here, $D(z)$ is the linear growth factor, $f(z)$ is the linear growth rate
($f(z)\equiv \dd\ln D/\dd\ln a$), and $\beta_x\equiv f(z_x)/b_x$, where we
use the subscript $x=L$ for LAEs and $x=O$ for OIIEs%
\footnote{We do not
distinguish between subscripts ``LAE'' and ``L'', and subscripts ``OII'' and
``O''. We prefer the former over the latter purely based on convenience.}. 
We define
$\mu=\hat{\vk}\cdot\hat{\bm{n}}=k_\parallel/k$ as the cosine of the angle
between the wave vector and the line-of-sight direction. We model the
FoG effect \citep{Jackson:1972} with a Lorentzian damping term via
the one-dimensional velocity dispersion
\ba
\sigma_{v,x}^2
&= \frac{p}{3}\,D^2(z_x)\int\frac{\dd^3k}{(2\pi)^3}\,\frac{\Plin(k)}{k^2},
\ea
with a fudge parameter $p$. We adopt $p=0.4$, which \citet{jeong:2010} measured
from the two-dimensional redshift-space power spectrum of a suite of $N$-body
simulations.

The left panels in \reffig{pk2d-with-contam} show the two-dimensional power
spectra of LAEs (top) and OIIEs (bottom) in the
\plane{k_\perp}{k_\parallel}, with interloper fractions of $\f=\g=0.1$. Also
displayed are the corresponding contamination from projected OIIEs
($P_\oii^\proj(\vk)$) and LAEs ($P_\lae^\proj(\vk)$) in the right panels. Here,
we use the linear bias of $b_\lae=2$ and $b_\oii=1.5$ and assume that all LAEs
are at $z=2.7$ while all OIIEs are at $z=0.2$, which yields the scaling factors
$\alpha=7.1$ and $\beta=0.84$. These anisotropic scaling factors squeeze or
stretch the power spectrum shapes. The effect is much larger along the
perpendicular direction due to $\alpha\gg\beta\simeq1$. To facilitate the
comparison between the anisotropies from contamination and the total
anisotropies in the observed power spectrum, we present the
contribution from the interlopers in the right panels. For both LAEs
and OIIEs, the small interloper fractions ($\f$ and $\g$) suppress the
contribution from interlopers. The overall contribution, however, is much
larger for the projected OIIEs (contaminating LAEs) because of the volume
factor $\alpha^2\beta \simeq 40$; when projecting the high-redshift objects
(such as LAEs) onto the lower redshift (to $z_\oii$), the power spectrum
amplitude is suppressed by the factor $\alpha^{-2}\beta^{-1}$.

The angle-averaged, monopole power spectra are displayed in the top-left panel
of \reffig{pk_with_contam}. These spectra are shown without (black) and with
(dashed orange) $\f=\g=\SI{10}{\percent}$ interloper fractions. For the
contaminated spectra, we divide by a factor of $(1-\f)^2$ for LAEs and
$(1-\g)^2$ for OIIEs to better compare the shape of the power spectra. The
dotted line represents the observed cross-correlation, and the dashed
(dash-dotted) grey horizontal lines indicate the shot noise for the LAEs
(OIIEs). Note that the observed cross-correlation can exceed the LAE power
spectrum.

To better demonstrate the effect of interlopers, the relative change of the
monopole power spectrum is presented in the two top-right panels of
\reffig{pk_with_contam}; the very top panel for LAEs, the bottom for OIIEs. The
grey areas indicate the fiducial error bars on the power spectrum that we
discuss in \refsec{Deltapk}. The colored dashed lines are the change for
several contamination fractions as indicated in the legends. The most notable
feature is that the LAE power spectrum is affected more strongly by interlopers
than the OIIE power spectrum.

To investigate the angular dependence of the effect of interlopers, we present
the quadrupole power spectrum in the bottom left panel of
\reffig{pk_with_contam}: the true quadrupoles (black lines), with
\SI{10}{\percent} interloper fractions (orange dashed for LAE, orange
dash-dotted for OIIE), and the cross-correlation quadrupole (orange dotted).
The graph reveals that, as for the monopole, the LAE power spectrum is more
strongly affected by interlopers than the OIIE power spectrum.

The two bottom-right panels of \reffig{pk_with_contam} display the relative
change of the power spectra for the quadrupole in the same manner as for the
monopole. Again, the LAE quadrupole power spectrum is more strongly affected by
contamination than the OIIE quadrupole.

Finally, the left panel of \reffig{pkcross_2d} shows the expected
cross-correlation for $\f=\g=0.1$. The figure demonstrates that for a survey
such as HETDEX, where the LAEs are at $z\sim2.7$ and the OIIEs at $z\sim0.2$,
the contribution from the projected OIIE power spectrum is expected to dominate
the observed cross-correlation.

\subsubsection{On using the linear model of \refeq{pk-kaiser}}
Strictly speaking, the linear theory prescription for the galaxy power spectrum
[\refeq{pk-kaiser}] breaks down on small scales and we must include the
nonlinear terms in our analysis. Indeed, for the HETDEX survey under
consideration here, high-redshift ($1.9<z<3.5$) LAEs probe the quasi-linear
scales $k\lesssim k_{\rm max} = \SI{0.4}{\h\per\mega\parsec}$, and
corresponding lower-redshift OIIEs probe the fully-nonlinear scales to $k_{\rm
max}\simeq \SI{2}{\h\per\mega\parsec}$ (the volume is smaller by a factor of
$\(\alpha^2\beta\)^{-1}$, see \reffig{scaling_parameters}). For the former, the
complete next-to-leading order expression is given in
\citet{desjacques/etal:pkgs} which includes nonlinearities in matter
clustering, galaxy bias, and redshift-space distortion. For the fully-nonlinear
regime, however, no such expression is known, and we may have to rely on
cosmological simulations \citep{springel/etal:2018}.

We therefore focus on analyzing the clustering of high-redshift LAEs rather
than low-redshift OIIEs. We further assume that the interlopers do not
dominate the high-redshift samples; this situation may be achieved by
applying astrophysically-motivated classifications of emission lines
\citep{pullen+2016,leung+2017}.

By focusing on the cosmological analysis from the LAE sample only,
the model in \refeq{pk-kaiser} suffices to study the effect of interlopers on
cosmological parameter estimation. As shown in \reffig{pk2d-with-contam} and
\reffig{pk_with_contam}, the misidentified low-$z$ interlopers (a) induce an
anisotropy into the power spectrum and (b) increase large-scale power. The
interlopers, therefore, affect the cosmological parameters measured from large
scales $k<k_{\rm eq}$, where $k_{\rm eq}$ is the wavenumber corresponding to
the matter-radiation equality (e.g., the local primordial non-Gaussianity
parameter $\fnl$), and from the anisotropies due to redshift-space distortion
(e.g., the angular diameter distance $d_A(z)$, the Hubble expansion rate
$H(z)$, and the linear growth rate $f(z)$). For the former, the linear theory
applies because $k<k_{\rm eq}$ corresponds to sufficiently large scales. For
the latter, the parameter estimation is relatively insensitive to the
non-linearities such as the constraint on geometrical quantities from the BAO
feature and Alcock-Paczynski (AP)-test \citep{AP:1979,shoji+2009}. We
explicitly check the effect of nonlinear redshift-space distortion in
\refsec{morersd} by marginalizing the 2D power spectrum over higher powers in
$\mu$.

\subsection{Variance of the power spectrum measurement}
\label{sec:Deltapk}
Ignoring the connected trispectrum, we use only the Gaussian part of the
covariance matrix where the power spectra at different wavevectors are
statistically independent. We then calculate the variance (diagonal part of
the covariance matrix) of the power spectrum as
\ba
\(\Delta P(\vk)\)^2
&= \frac{2}{N_{\vk}}
\left(P(\vk) + \frac{1}{{\bar n}_g}\right)^2
\label{eq:Deltapk}
\ea
(see \refapp{pkerror} for the derivation.) Here, $N_{\vk}$ is the number of
Fourier modes
\ba
\label{eq:Nk}
N_{\vk} &= 
\frac{V_\text{survey}}{(2\pi)^3} \Delta V_\vk\,,
\ea
with $\Delta V_\vk$ being the volume in Fourier space contributing to the
estimation of the power spectrum. For example, when computing the monopole
power spectrum with the Fourier bin of $\Delta k =
2\pi/V_\mathrm{survey}^{1/3}$, \be \Delta V_{\vk} = \Delta V(k,\Delta k)=4\pi
k^2\Delta k\,.
\ee 
When estimating the two-dimensional power spectrum 
$P(k_\perp,k_\parallel)$ with the Fourier bin of 
$(\Delta k_\perp,\Delta k_\parallel)$, 
\be
\label{eq:DeltaVk_2d}
\Delta V_{\vk} =
\Delta V(
k_\perp,k_\parallel,
\Delta k_\perp,\Delta k_\parallel
)=4\pi k_\perp \Delta k_\perp \Delta k_\parallel\,.
\ee
For the monopole matter power spectrum, \refeq{Deltapk} provides a good 
approximation to the measured variance from N-body 
simulations \citep{jeong/komatsu:2009}. 

The $1/{\bar n}_g$ term in \refeq{Deltapk} denotes the shot noise, assuming
that the galaxy distribution follows the Poisson statistics of the underlying
galaxy density field. When dealing with a galaxy sample containing the
interlopers, the shot noise contribution to the observed auto-power spectrum is
related to the number density $\bar{n}_g^{\tot}$ of the \textit{total} sample
including the interlopers. Also, there is no shot noise contribution to the
observed cross-power spectrum, even though both samples contain a mixture of
the two populations. We present the rigorous derivation in \refapp{shotnoise}.

In \refapp{pkerror}, we calculate the variance of the observed cross power
spectrum as
\ba
&\( \Delta P_{LO}(\vk)\)^2
\vs
&= \frac{1}{N_\vk} 
\biggl[
    \(P_\lae^\obs(\vk) + \frac{1}{\bar{n}_\lae^{\tot}}\) 
    \(P_\oii^{\obs,\proj}(\vk) + \frac{\alpha^2\beta}{\bar{n}_\oii^{\tot}}\)
\vs
&\quad+ \big(P_{\lae\times\oii}^\obs(\vk)\big)^2
\biggl]\,.
\ea
Here, $P_\oii^{\obs,\proj}(\vk)\equiv \alpha^2\beta
P_\oii^\obs(\alpha\vk_\perp,\beta k_\parallel)$ arises because we implement the
cross-correlation by projecting the OIIEs onto the LAEs redshift (see
\refsec{crossPk}). The same projection adds the factor $\alpha^2\beta$ to the
shot noise of OIIEs. The right panel of \reffig{pkcross_2d} presents the
signal-to-noise ratio for each $k_\perp$\nobreakdash-$k_\parallel$~mode of the observed
cross-power spectrum for the case $\f=\g=0.1$. This figure indicates that a
high S/N ratio measurement for the interloper fraction is possible from the
cross-correlation. We shall quantify this conclusion using a Fisher information
matrix formalism in the next section.

\section{Statistical Analysis for HETDEX}
\label{sec:fitting}
In this section, we present the statistical analysis for the cosmological
parameter estimation from all three power spectra: the auto-power spectra of
the main survey galaxies and the interlopers, and the cross-power spectrum
between the main galaxies and interlopers. 

In \refsec{likelihood} we first construct the likelihood function for the
dataset consisting of the main sample and the interlopers. We focus on the two
geometrical observables measured from cosmological distortion: the Hubble
expansion rate $H(z)$ and the angular diameter distance $d_A(z)$. These are
primary targets for current and future galaxy surveys. We describe the
cosmological distortion in the presence of the interloper population in
\refsec{cosmologymismatch}. In \refsec{model_pkcrossonly}, we present
(\textit{Case~A}) the proof of concept for measuring the interloper fractions
$\f$ and $\g$ from the cross-correlation between the main survey galaxies and
interlopers. We measure the interloper fractions by assuming only that the true
cross-correlation between the two populations vanishes.

In the following sub-sections, we present the projected uncertainties on $H(z)$
and $d_A(z)$ for several different treatments of nonlinearities in the
redshift-space galaxy power spectrum. In
Sec.~(\ref{sec:fullshape})--(\ref{sec:noknowledge}), we assume that the
redshift-space power spectrum is given by the linear Kaiser formula
\citep{kaiser:1987} with the Finger-of-God effect \citep{Jackson:1972} as in
\refeq{pk-kaiser}. In \refsec{fullshape}, we assume that we know the full shape
of the nonlinear power spectrum $P(k)$ for both LAEs and OIIEs
(\textit{Case~B}). Because OIIEs are lower redshift objects, their
population probes much smaller scales than LAEs, and their nonlinearities are
much stronger. We therefore investigate $H(z)$ and $d_A(z)$ after marginalizing
over the nonlinear OIIE power spectrum in \refsec{onlylaeshape}
(\textit{Case~C}), and we assume this case as a baseline for
HETDEX. In \refsec{noknowledge}, we study the pessimistic case of
marginalizing over both LAE and OIIE power spectra (\textit{Case~D}).
In this case, we can still measure the combination $d_A(z)H(z)$ given the
Alcock-Paczynski test. In \refsec{morersd} we test for the effect of non-linear
redshift-space distortion by including the higher-order dependence on the
angular cosine $\mu\equiv k_\parallel/k$.

Finally, we analyze the effect of interlopers on measuring other cosmological
parameters such as the linear growth rate $f=d\ln D/d\ln a$
(\refsec{fullshape}) and the primordial non-Gaussianity parameter $\fnl$
(\refsec{fnl}).

Throughout, we denote the true interloper fractions that we have assumed for the analysis by $\ffid$ and $\gfid$.

\subsection{Likelihood function}\label{sec:likelihood}
We construct the likelihood function by assuming that the galaxy power spectrum
completely specifies the statistics of the galaxy density contrast; i.e., we
ignore the effects from higher-order correlation functions. Since we are
focused on the galaxy power spectrum, this assumption suffices for the purpose
of this paper. 

To facilitate the calculation and to incorporate the auto- and cross-power
spectra in the same setting, we project all OIIEs into the LAE volume. We could
just as well have chosen to project all LAEs into the OIIE volume, or,
similarly, project both LAEs and OIIEs into any appropriate volume of our
choice. As seen from \refeq{nlnL} below, the choice of projection merely adds a
constant to the log-likelihood function.

The likelihood function is constructed from the observed density contrasts
$\delta^\obs_{x}(\vk_i)$, with $x=L$ for LAEs and $x=O$ for OIIEs, both of
which are contaminated by interlopers. For each observed wavemode $\vk$, we
define the observed data vector at the LAE redshift (with OIIEs projected to
that redshift) as 
\ba
\label{eq:Delta_DD}
\Delta(\vk) &= \begin{pmatrix} \Delta_L(\vk) \\ \Delta_O^\proj(\vk) \end{pmatrix}
= \begin{pmatrix}
	\delta^\obs_{L}(\vk_1) \\
	\vdots \\
	\delta^\obs_{L}(\vk_{N_\vk}) \\
        \delta^{\obs,\proj}_{O}(\vk_1) \\
	\vdots \\
        \delta^{\obs,\proj}_{O}(\vk_{N_\vk}) \\
\end{pmatrix}\,,
\ea
where $N_\vk$ is the number of Fourier-modes for a given bin centered on $\vk$.
We provide the explicit expression for $N_\vk$ in \refeq{Nk} and
\refeq{DeltaVk_2d}.

As shown in \refsec{Deltapk}, each element of the covariance matrix consists of
the observed power spectra plus Poisson shot noise
$P_{xy}^\mathrm{obs+shot}(\vk)=P^{\obs}_{xy}(\vk)+\delta^K_{xy}/\bar
n_x^{\mathrm{total}}$.  Here, $\delta^K$ is the Kronecker delta symbol. In
block-matrix form, the covariance matrix is 
\ba
&C(\vk)
= \langle\Delta(\vk) \Delta^\dagger(\vk)\rangle \vs
&= \frac{(2\pi)^3}{V_\vk}
\begin{pmatrix}
    P_L^\mathrm{obs+shot}(\vk) I_{N_\vk}    & P_{LO}^\mathrm{obs}(\vk) I_{N_\vk} \\
    P_{LO}^\mathrm{obs}(\vk) I_{N_\vk} & P_O^{\mathrm{obs+shot},\proj}(\vk) I_{N_\vk}
\end{pmatrix}\,,
\label{eq:covariancematrix}
\ea
where $I_{N_\vk}$ is the $N_\vk\times N_\vk$ unit matrix, and
$P_O^{\mathrm{obs+shot},\proj}(\vk)$ is the OIIE auto-power spectrum
projected into the LAE volume. The factor $V_\vk/(2\pi)^3$ appears from
averaging $\langle\delta^*(\bfq)\delta(\bfq')\rangle =
(2\pi)^3\delta^D(\bfq-\bfq')P(\bfq)$ over a cell of volume $V_\vk$ in
$\vk$-space. From the covariance matrix, we compute the log-likelihood
function
\ba
-\ln\mathfrak{L}
&= \frac12\ln\det C + \frac12\Delta^\dagger C^{-1} \Delta,
\ea
as 
\begin{widetext}
\ba
\label{eq:nlnL}
-\ln\mathfrak{L}
&=
\sum_\vk \frac{N_\vk}{2}\bigg[
    \ln\big[P_L^\mathrm{obs+shot} P_O^{\mathrm{obs+shot},\proj} - (P_{LO}^\obs)^2\big]
    \vs&\quad
    + \frac{
        P_O^{\mathrm{obs+shot},\proj} \hat P_L^\text{data+shot}
        + P_L^\mathrm{obs+shot} \hat P_O^{\mathrm{data+shot},\proj}
        - 2 P_{LO}^\obs \hat P_{LO}^\text{data}
    }{P_L^\mathrm{obs+shot} P_O^{\mathrm{obs+shot},\proj} - (P_{LO}^\obs)^2}
\bigg]\,.
\ea
\end{widetext}
Here, we replace the square of the data vector with the maximum-likelihood
estimators for the observed auto-power spectra:
\ba
\label{eq:pkLLdata}
\hat P_L^\text{data+shot}(\vk) &= \frac{V_\vk}{(2\pi)^3N_\vk}\sum_i|\delta_L^\obs(\vk_i)|^2
\,,\displaybreak[0]\\
\hat P_O^\mathrm{data+shot,\proj}(\vk) &= \frac{V_\vk}{(2\pi)^3N_\vk}\sum_i|\delta_O^{\obs,\proj}(\vk_i)|^2 \,,
\ea
and the cross-power spectrum
\ba
\hat P_{LO}^\text{data}(\vk) &= \frac{V_\vk}{(2\pi)^3N_\vk}\sum_i\Re\big(\delta_L^{\obs,*}(\vk_i)\delta_O^{\obs,\proj}(\vk_i)\big)\,,
\label{eq:pkLOdata}
\ea
where $\Re(z)$ is the real part of a complex number $z$.

\subsection{Cosmological distortion}
\label{sec:cosmologymismatch}
Cosmological distortion refers to the systematic change in the statistical
observables, such as the galaxy power spectrum, induced by adopting an
incorrect reference cosmology to convert the observed galaxy coordinate (RA,
Dec, $z$) into physical coordinates. Cosmological distortion allows us to
measure the Hubble expansion rate $H(z)$ and the angular diameter distance
$d_A(z)$ from features such as those produced by BAO
\citep{seo/eisenstein:2003,blake/glazebrook:2003,hu/haiman:2003} and the
AP-test \citep{AP:1979,shoji+2009} in the redshift-space galaxy power spectrum.
In this paper, we do not include the uncertainties in the cosmological
parameters such as $\Omega_{\rm m}h^2$ and $\Omega_{\rm b}h^2$ that determine
the shape of the linear matter power spectrum. In the real analysis, one must
include appropriate priors on these parameters from, e.g., the CMB analysis in,
e.g., \citep{planck:2015-parameter,planck:2018-parameter}.

We model the cosmological distortion in the galaxy power spectrum as follows.
Given a reference cosmology, we calculate the angular diameter distance
$d_{A,\text{ref}}(z)$ as well as the Hubble expansion rate $H_\text{ref}(z)$ at
redshift $z$. In general, the wavenumbers $k_\perp^\text{ref}$ and
$k_{\parallel}^\text{ref}$ measured from the reference cosmology differ from
the true wavenumbers $k_\perp$ and $k_{\parallel}$ by some factors $v(z)$ and
$w(z)$.
We determine the factors $v(z)$ and $w(z)$ and their effect on the power
spectrum in a manner similar to the projection effect discussed in \refsec{pk}
using $\alpha$ and $\beta$. For LAEs (subscript `L') and OIIEs (subscript `O'),
we define
\ba
\label{eq:vwL}
v_L &= \frac{d_{A,\text{ref}}(z_L)}{d_A(z_L)}\,, &
w_L &= \frac{H(z_L)}{H_\text{ref}(z_L)}\,,
\\
\label{eq:vwO}
v_O &= \frac{d_{A,\text{ref}}(z_O)}{d_A(z_O)}\,, &
w_O &= \frac{H(z_O)}{H_\text{ref}(z_O)}\,.
\ea
Using these variables, the Fourier space vector $(k_\perp^\text{ref},
k_{\parallel}^\text{ref})$ inferred from the reference cosmology is related
to the true Fourier vector as $(k_\perp, k_{\parallel}) = (v_L
k_\perp^\text{ref}, w_L k_{\parallel}^\text{ref})$, and the power spectrum
$P_L^\text{ref}(k^\text{ref}_\perp,k_\parallel^{\rm ref})$, measured by using
the reference cosmology, is
\ba
\label{eq:pkLLref}
P_L^\text{ref}(k_\perp^\text{ref}, k_{\parallel}^\text{ref})
&= v_L^2 w_L P_L(v_Lk_\perp^\text{ref}, w_Lk_{\parallel}^\text{ref})\,.
\ea

Similarly, we calculate the contribution from the projected interloper power
spectrum by defining $\alpha_\text{ref}$ and $\beta_\text{ref}$ as the
scaling factors $\alpha$ and $\beta$ [\refeq{scaling_parameters}] in the
reference cosmology. The projected OIIE power spectrum takes the following
form, where we include both the projection due to the misidentification [see
\refeq{pkproj_oii}], and the projection due to a cosmological distortion
[\refeq{pkOOref}]:
\ba
\label{eq:pkOOref}
&P_O^\text{ref,proj}(k_\perp^\text{ref}, k_{\parallel}^\text{ref})
\vs
&= \alpha_\text{ref}^2 \beta_\text{ref} P_O^\text{ref}(\alpha_\text{ref} k_\perp^\text{ref}, \beta_\text{ref} k_{\parallel}^\text{ref}) \vs
&= \alpha_\text{ref}^2 \beta_\text{ref} v_O^2 w_O
P_O(v_O\alpha_\text{ref} k_\perp^\text{ref}, w_O\beta_\text{ref} k_{\parallel}^\text{ref})\,.
\ea
The parameters $\alpha_{\rm ref}$ and $\beta_{\rm ref}$ are completely
degenerate with $v_L$, $v_O$ and $w_L$, $w_O$; therefore, we only include the
latter cosmological distortion  parameters in the analysis.

\subsection{Case~A: No prior knowledge on the shape of the galaxy power spectra}
\label{sec:model_pkcrossonly}
\begin{figure}
    \centering
    \incgraph[0.49]{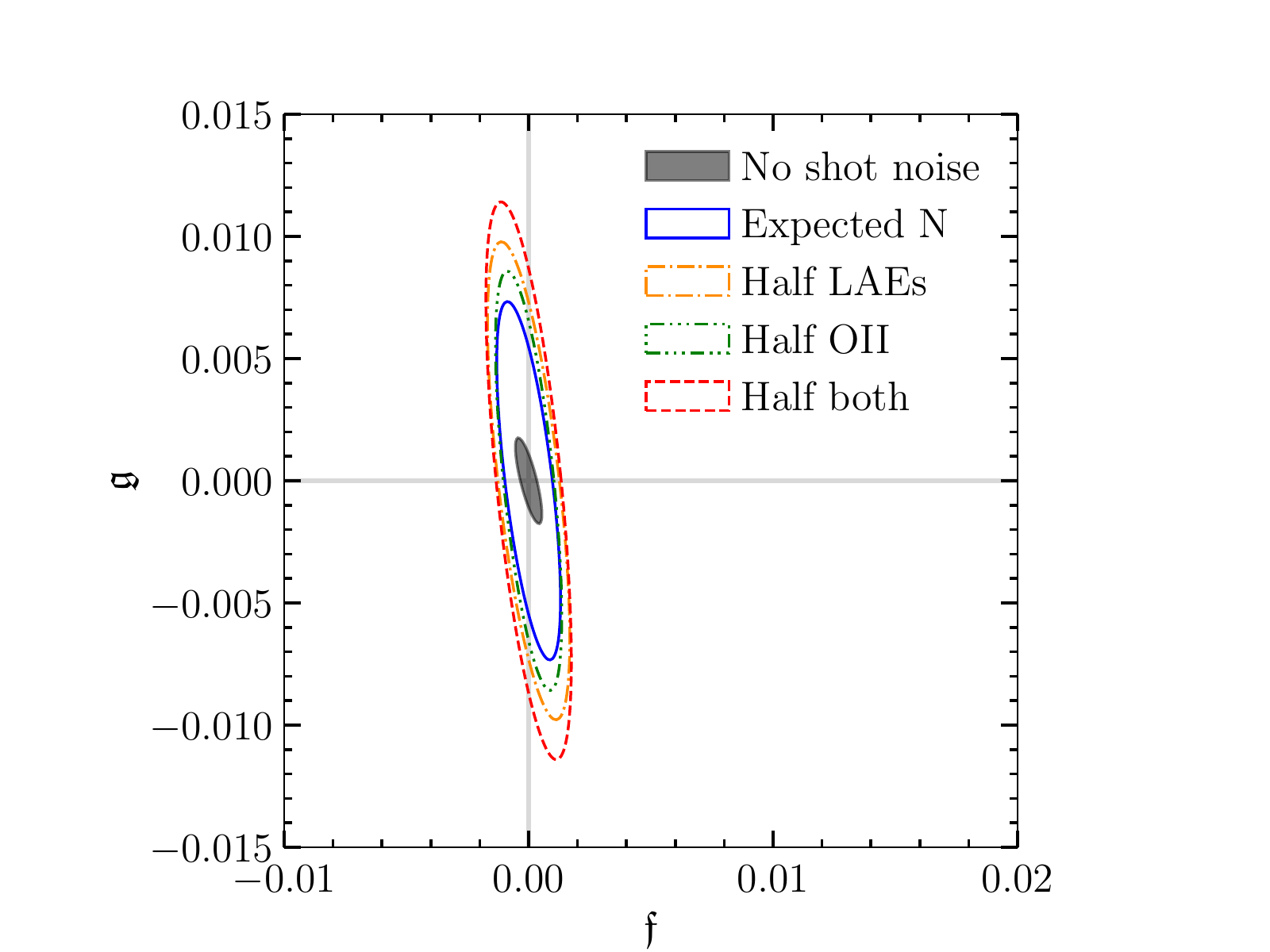}
    \caption{
        Constraints on the interloper fractions $\f$ and $\g$ for HETDEX. The
        shaded ellipse represents the $1\sigma$ (\SI{68}{\percent} C.L.)
        interval when there is no shot noise. The blue solid ellipse is the
        $1\sigma$ (68\% C.L.) interval for the expected numbers of LAEs and
        OIIEs, the orange dash-dotted is the same $1\sigma$ interval for when
        the number of LAEs is halved, the green line with several dots between
        dashes is the result when the OIIE number is halved, and the red dashed
        line indicates what occurs when both numbers are halved. Here,
        $\ffid=\gfid=0$.
    }
    \label{fig:pLO0_N}
\end{figure}
\begin{figure}
    \centering
    \incgraph[0.49]{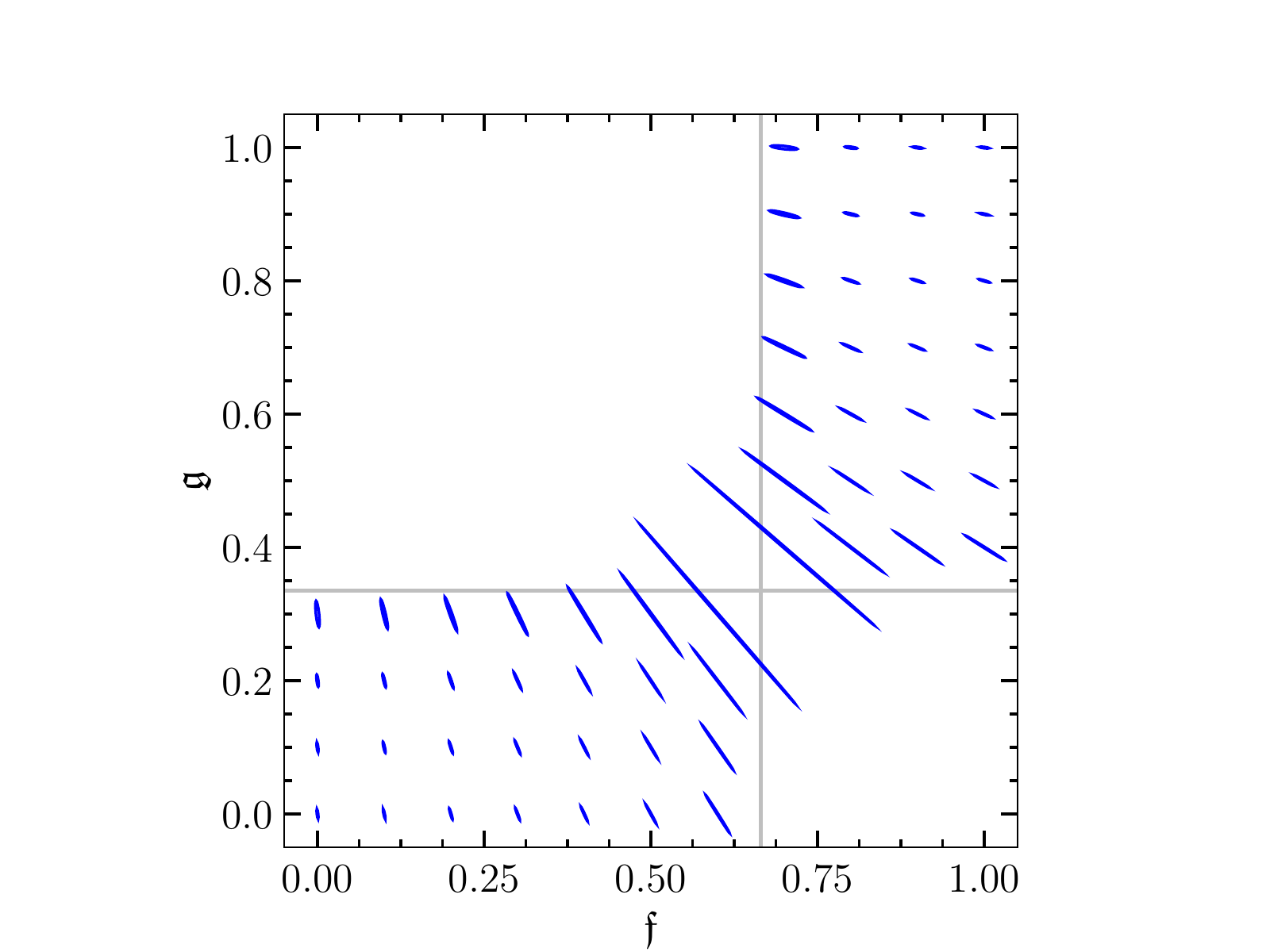}
    \caption{
        Constraints on the interloper fractions $\f$ and $\g$ for several
        interloper fractions spaced \SI{10}{\percent} apart. Since
        $N_\oii^\true/N_\lae^\true\approx2$ for HETDEX, only the
        interloper fractions bounded by the grey lines are physically possible
        (see \refapp{fgplane}). However, since the true number densities will
        not be known, and thus cannot be used in the fit, estimates of $\f$ and
        $\g$ may well fall outside those boundaries. $\f$ and $\g$ can be
        measured best near $(0,0)$ and $(1,1)$. On the diagonal $\f+\g=1$, the
        individual measurements become degenerate.
    }
    \label{fig:pLO0_fgplane}
\end{figure}
\begin{figure}
    \centering
    \incgraph[0.49]{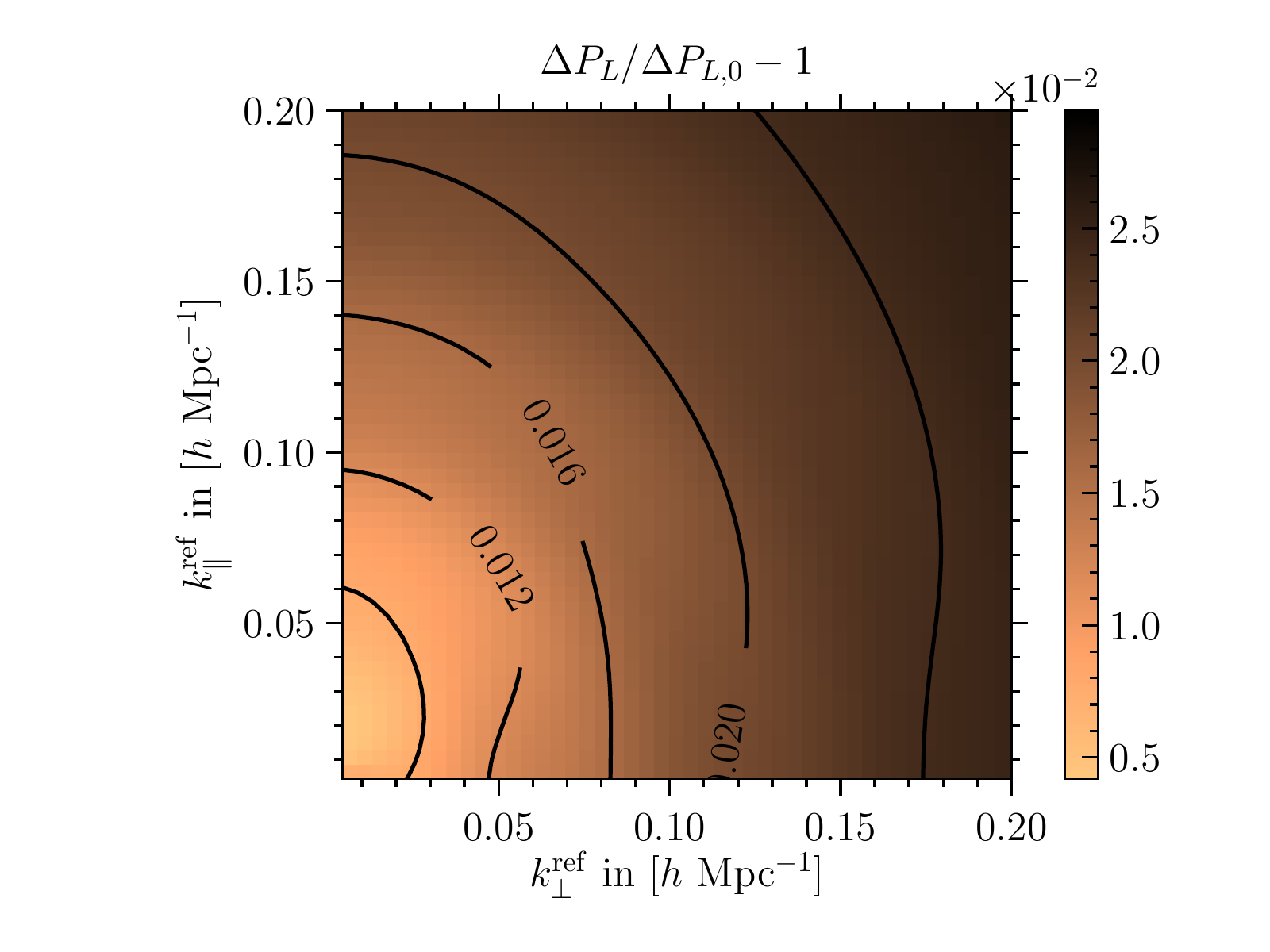}
    \caption{
        Change in the uncertainty of the power spectrum relative to the ideal
        case of zero interloper fractions, i.e., $[\Delta P_L(\vk)/\Delta
        P_{L,0}(\vk)] - 1$, where we marginalize over $\f$, $\g$, and
        $P_O(\vk)$. To first order, the increase in uncertainty is the least
        where the product $\bar n P$ is largest. Here, $\ffid=\gfid=0.01$.
    }
    \label{fig:pkLL_deltachange}
\end{figure}
How accurately can we measure the interloper fractions $\f$ and $\g$ by
requiring only that the true cross-power spectrum must vanish? We address
this question in the most conservative manner, assuming no prior knowledge
about the shape of the galaxy power spectrum; i.e., the case where we measure
the interloper fractions $\f$ and $\g$ along with the amplitude of the
two-dimensional power spectrum $P_L(k_\perp,k_\parallel)$ and
$P_O(k_\perp,k_\parallel)$ from fitting the observed auto- and cross-power
spectra. The expressions for these power spectra are given in \refeq{pkobs_lae}
and \refeq{pkobs_oii} (for the auto-power spectra) and \refeq{pkobs_LO} (for
the cross-power spectrum).

For the HETDEX survey outlined in \refsec{HETDEX}, there are $91^2=8281$
Fourier modes within the maximum wavenumber $k_{\rm max}=0.4~h/{\rm Mpc}$ at
the LAE volume. Thus, there are a total of $2~(\f~{\rm and}~\g)+8281~({\rm
for}~P_L)+8281~({\rm for}~P_O)=16564$ parameters. For each value of
$\f$ and $\g$, we use the Fisher information matrix analysis to calculate the
projected uncertainties.

A cautionary remark is in order here. The analysis in this section only serves
as a proof of concept for the measurement of the interloper fractions $\f$ and
$\g$. The uncertainties found here are a worst case benchmark, because they are
derived from minimal assumptions about the shape of the galaxy power spectrum.
However, we do not advocate such an analysis in practice. In fact, we carried
out a Markov-chain Monte Carlo (MCMC) analysis for the HETDEX case only to find
that the chain does not converge in the 16564-dimensional parameter space. Even
for a simpler analysis of measuring $\f$ and $\g$ by iteration, it is
non-trivial to construct an unbiased estimator. 

\reffig{pLO0_N} presents the projected uncertainties on $\f$ and $\g$ under the
null hypothesis, i.e., when the true interloper fractions are zero. The
shaded ellipse at the center indicates the cosmic-variance limited constraint
without shot noise. The solid ellipse shows the constraint with the fiducial
number density of HETDEX given in \refsec{HETDEX}. The forecast demonstrates
that the cross-correlation constrains both $\f$ and $\g$ to the sub-percent
level. The constraint is better for $\f$ than $\g$ because on large-scales
where the signal-to-noise ratio is the largest (\reffig{pkcross_2d}) the
amplitude of the projected OIIE power spectrum (the contaminant) is much higher
than that for the LAEs; this behavior is clear in the upper two panels in
\reffig{pk2d-with-contam}. In the Figure, the relative contributions of
$P_{\lae}$ and $P_{\oii}^\proj$ to the observed auto-correlation are similar on
large-scales even though the latter is suppressed by $\f^2$.

We also investigate the effect of reducing the number density of LAEs and
OIIEs. The orange dash-dotted ellipse, the green triple-dot-dashed line, and
the red dashed line show the expected constraint when reducing the number of
LAEs, OIIEs, or both galaxy groups, respectively, by \SI{50}{\percent} of what
is predicted. Reducing the number density of galaxies increases the Poisson
shot noise which affects the uncertainties in the power spectrum measurements
on small scales. This change affects the constraint on $\g$ more than $\f$, due
to the different scale-dependence of the $\f$ and $\g$ contributions.
\reffig{pk2d-with-contam} reflected this behavior in the top- and bottom-right
panels.

We next address the situation of non-zero $\f$ and $\g$. As shown in
\refapp{fgplane}, unlike the misidentification fractions $x_\lae$ and $x_\oii$
[defined in \refeqs{numlaeobs}{numoiiobs}] that can take any value between 0
and 1, the true interloper fractions are limited to two regions: one with
$0\leq\ffid\leq\f_\mathrm{lim}$ and
$0\leq\gfid\leq\g_\mathrm{lim}$, and one with
$\f_\mathrm{lim}\leq\ffid\leq1$ and
$\g_\mathrm{lim}\leq\gfid\leq1$. The limiting values $\f_{\rm lim} =
N_\oii^\true/(N_\lae^\true+N_\oii^\true)$ and $\g_{\rm lim} =
N_\lae^\true/(N_\lae^\true+N_\oii^\true)$ occur when $x_\lae+x_\oii=1$. For
HETDEX, $\f_\mathrm{lim}\approx2/3$ and $\g_\mathrm{lim}\approx1/3$. 

\reffig{pLO0_fgplane} presents the projected \SI{68}{\percent} confidence
ellipses of $\f$ and $\g$ for several true interloper fractions spread throughout
the allowed region at intervals of \SI{10}{\percent}. The tightest constraints
are obtained when the interloper fractions are either small or extremely large.
There is a symmetry of exchanging $\f\leftrightarrow1-\g$, which is equivalent
to swapping the two samples. Indeed, a closer inspection of the log-likelihood
given in \refeq{nlnL} reveals that the cross-correlation alone cannot
distinguish which of the two regions a given survey will fall into. However, since we
expect the power spectra of LAEs and OIIEs to have very different shapes on
large scales, inspection of the measured true power spectra should suffice to
discriminate the two regions in cases far from the diagonal $\f+\g=1$. Because
all three observed power spectra are the same, $\f$ and $\g$ are completely
degenerate on the diagonal $\f+\g=1$. In that case, although the non-zero
cross-correlation indicates the existence of interlopers, \refeq{pkobs_lae} and
\refeq{pkobs_oii} cannot be inverted. The cross-correlation alone, therefore,
is insufficient to determine the interloper fraction; this is the reason why
the error ellipse diverges at $(\f_{\rm lim},\g_{\rm lim})$.

Although we plot the projected constraints on $\f$ and $\g$ only in the limited
regions, an analysis with measured data must explore the whole range
of $\f$ and $\g$ between $0$ and $1$. This is because the limits $\f_{\rm lim}$
and $\g_{\rm lim}$ are given by the true ratio $N_\lae^\true/N_\oii^\true$, and
in reality only the observed numbers of galaxies will be known; the true number
of LAEs and OIIEs will be variables that need to be estimated from the observed
numbers and the estimated interloper fractions. Thus, while the
true interloper fractions are restricted to the allowed regions, the
measured interloper fractions $\f$ and $\g$ can have any value
between $0$ and $1$.

How do the uncertainties in measuring the power spectrum change due to the
interlopers? \reffig{pkLL_deltachange} displays the change in the uncertainty
on each mode $P_L(k_\perp,k_\parallel)$ after marginalizing over the interloper
fractions. Here, we use $\ffid=\gfid=0.01$.
\reffig{pkLL_deltachange} shows that the fractional increase in the
uncertainties in the power spectrum closely follows $\bar{n}P$: the power
spectrum uncertainties increase mainly in the shot noise dominated regime
($\bar n_L P_L \ll 1$) and the increase only depends on the interloper
fractions and the number of galaxies.

We can estimate the increase in the uncertainty by inverting \refeq{pkobs_lae}
and \refeq{pkobs_oii} for fixed $\f$ and $\g$, which is valid when $\f$ and
$\g$ are tightly constrained (which is the case for small values, see
\reffig{pLO0_fgplane}). Combining this result with \refeq{Deltapk}, to first
order in $\f$ and $\g$, we obtain the uncertainty $\Delta P_L$ in the LAE power
spectrum
\ba
\label{eq:DeltaPL}
\Delta P_L(\vk)
&\simeq \sqrt{\frac{2}{N_\vk}}\left[
P_L(\vk)
+ \(1+\f+\g\,\frac{\bar n_O}{\bar n_L}\)\frac{1}{\bar n_L}
\right]\,,
\ea
which is the same as \refeq{Deltapk}, except that the shot-noise term is
increased due to the interlopers. The fractional increase of the uncertainty is 
\ba
\label{eq:deltaPLdeltaPL}
\frac{\Delta P_L(\vk)}
{\Delta P_{L,0}(\vk)}-1
&\simeq
\(\frac{\f\,\bar n_L+\g\,\bar n_O}{\bar n_L}\)
\frac{1}{1 + \bar n_L P_L^\mathrm{ref}(\vk_\mathrm{ref})}\,,
\ea
where $\Delta P_{L,0}$ is the uncertainty without interlopers.
\refeq{deltaPLdeltaPL} is consistent with the results seen in
\reffig{pkLL_deltachange}. The estimate in \refeq{deltaPLdeltaPL} reproduces
the numerical result to \SI{\sim3}{\percent} for $\ffid=\gfid=0.01$
and to \SI{\sim30}{\percent} for $\ffid=\gfid=0.1$. This
difference is because, for larger values of $\ffid$ and $\gfid$, the linear
expansion is not accurate and the larger uncertainties in $\f$ and $\g$ also
contribute to $\Delta P_L$.

\subsection{Case~B: knowing full shape of both LAE and OIIE power spectrum}
\label{sec:fullshape}
\begin{figure}
    \centering
    \incgraph[0.49]{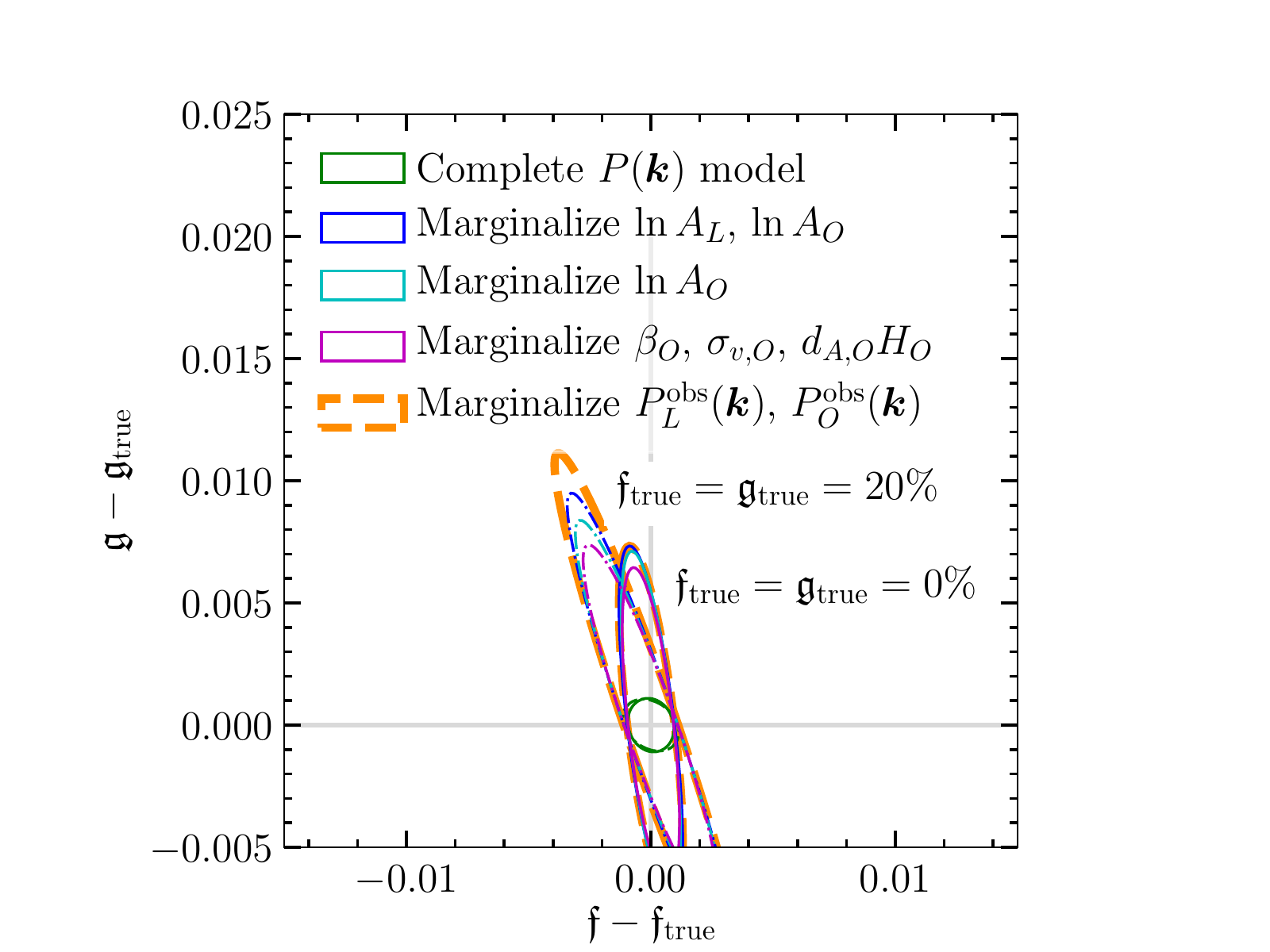}
    \caption{
        $1\sigma$ (68\% C.L.) ellipses on interloper fractions $\f$ and $\g$
        for several models. The first four models correspond to
            \textit{Case~B}, where we assume that the shapes of the power
            spectra are fully known. For the first model in green only the
            interloper fractions are being fit, all other parameters including
        the amplitudes are assumed to be known \emph{a priori}. For the
        second, in blue, we marginalize over the amplitudes. For the third
        (solid cyan), we marginalize over the OIIE amplitude only, and for the
        fourth (solid magenta) over the RSD, FOG, and AP parameters of the
        OIIEs. Finally, the thick dashed orange ellipses correspond to
            the worst-case scenario, \textit{Case~A}, where only the
            cross-correlation is known, marginalizing over both 2D-power
            spectra. We consider two interloper fractions: $\ffid=\gfid=0$ and
            $\ffid=\gfid=0.2$, as labeled in the figure. Most of the
            measurement uncertainty on $\f$ and $\g$ is a result of
            marginalizing over the amplitudes.
    }
    \label{fig:fg_bymodel}
\end{figure}
\begin{figure*}
    \centering
    \incgraph[0.49]{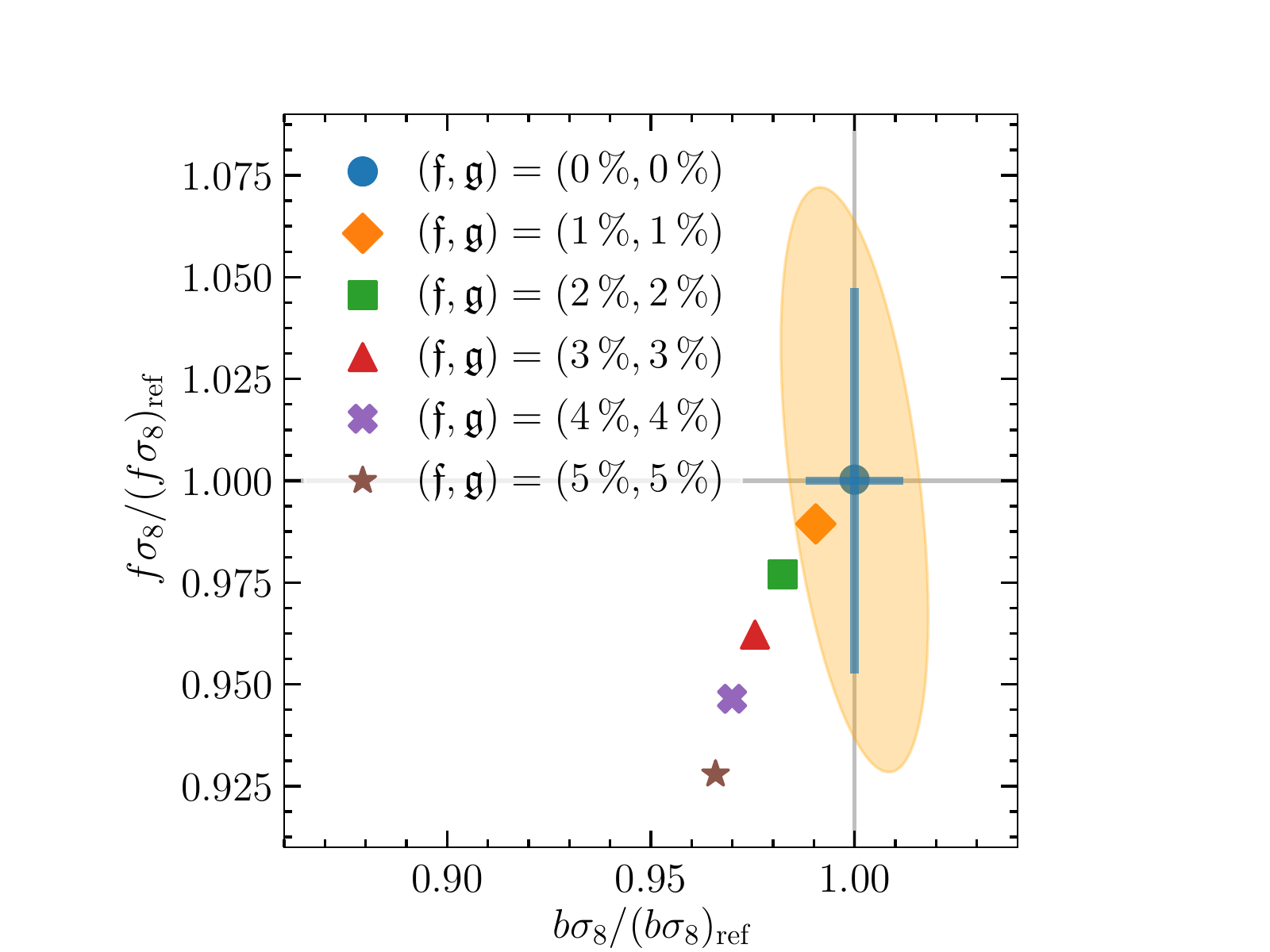}
    \incgraph[0.49]{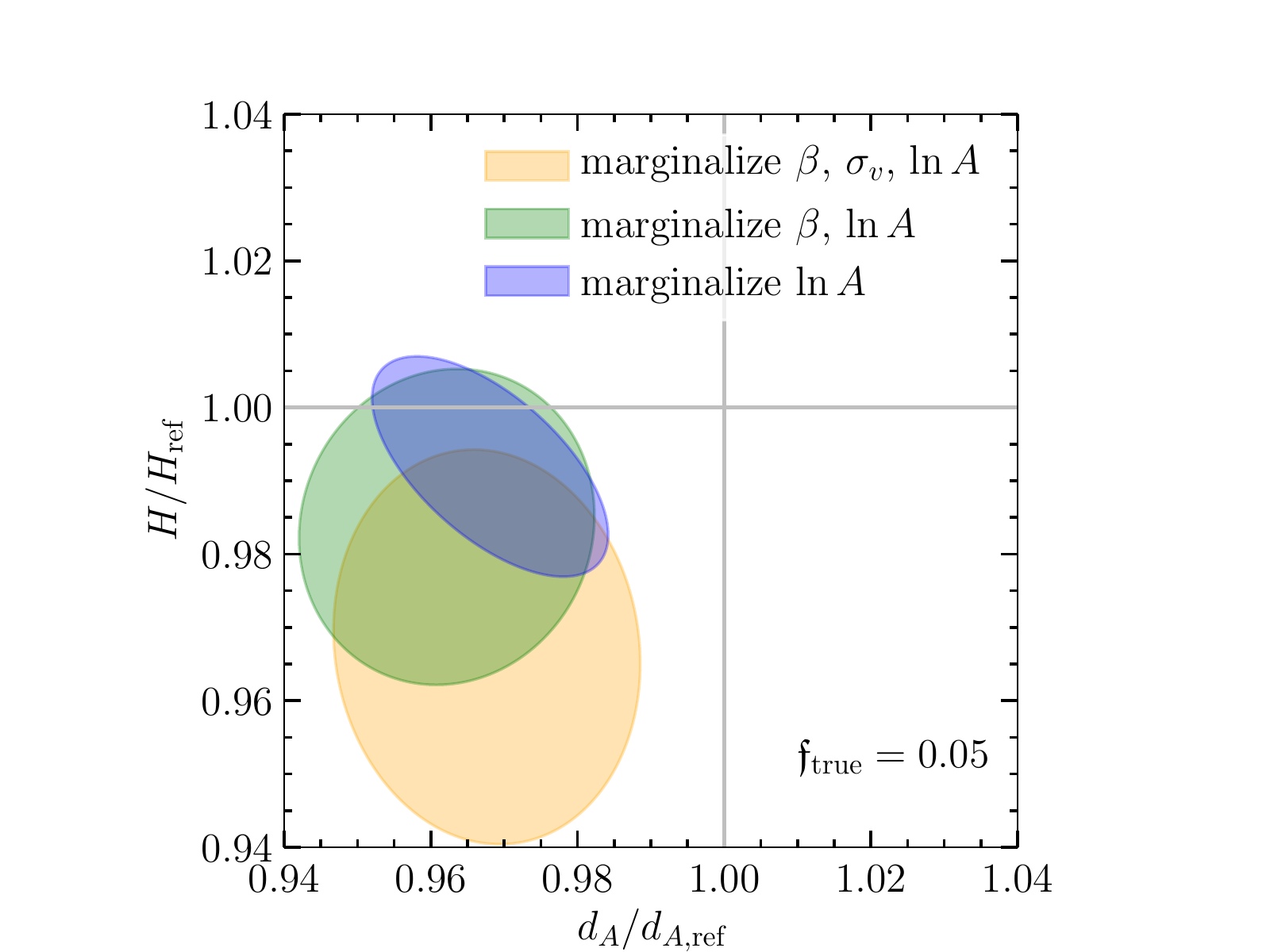}
    \caption{
        \emph{Left:} The interloper bias for $f\sigma_8$ and $b\sigma_8$, for
        uncorrected interloper fractions $\f=0,1,2,3,4,\SI{5}{\percent}$. The
        grey lines show the fiducial value, the ellipse gives the $1\sigma$
        (\SI{68}{\percent} C.L.) uncertainty, and the cross shows the
        marginalized uncertainties.
        \emph{Right:} The interloper bias on distances ($d_A$ and $H$) when
        ignoring the interloper fraction $\f=\SI{5}{\percent}$. The grey lines
        indicate the fiducial value. The bias tends to be more severe when
        marginalizing over more parameters.
    }
    \label{fig:dAH_fg0}
\end{figure*}
\begin{figure*}
    \centering
    \incgraph[0.49]{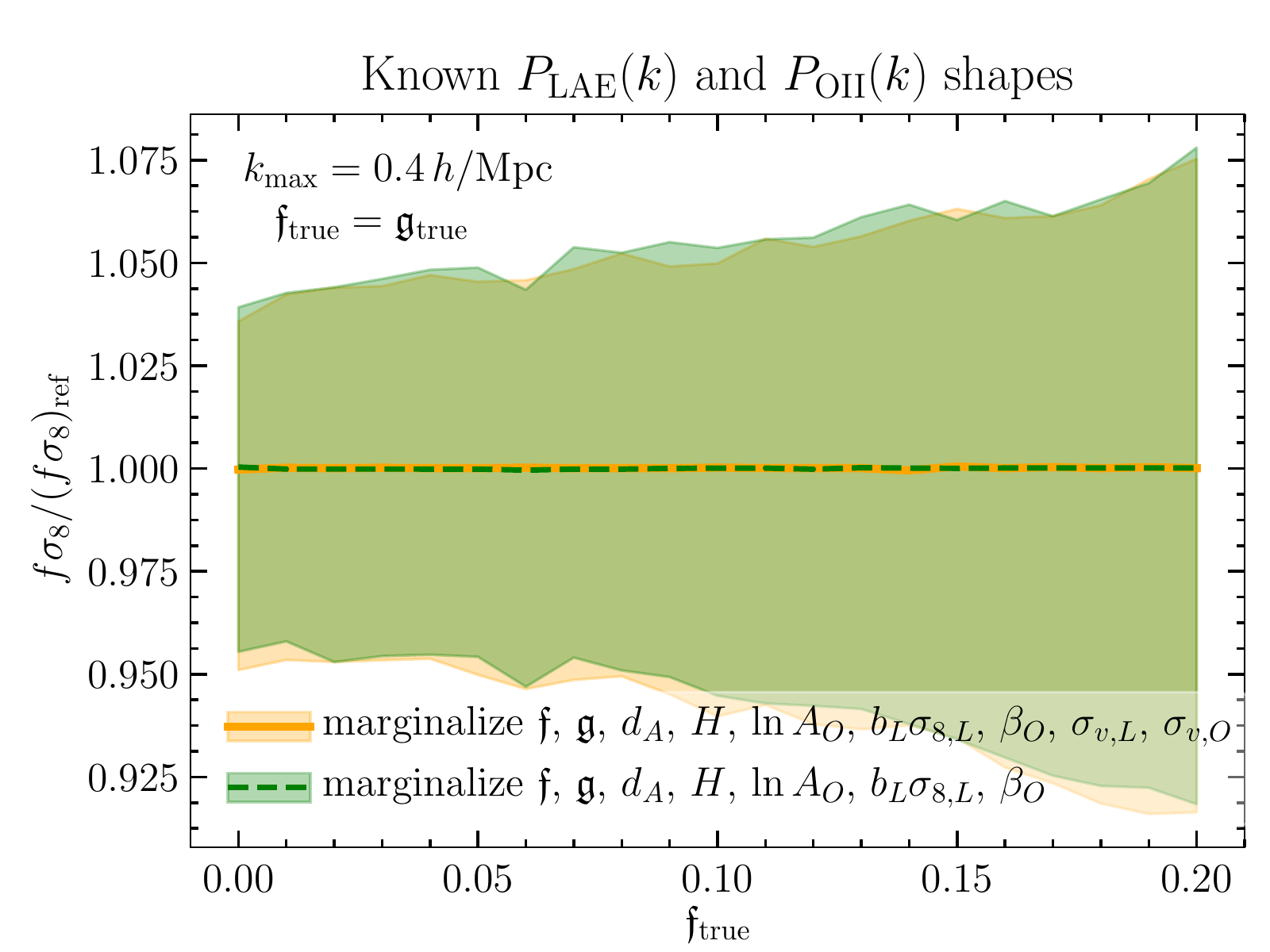}
    \incgraph[0.49]{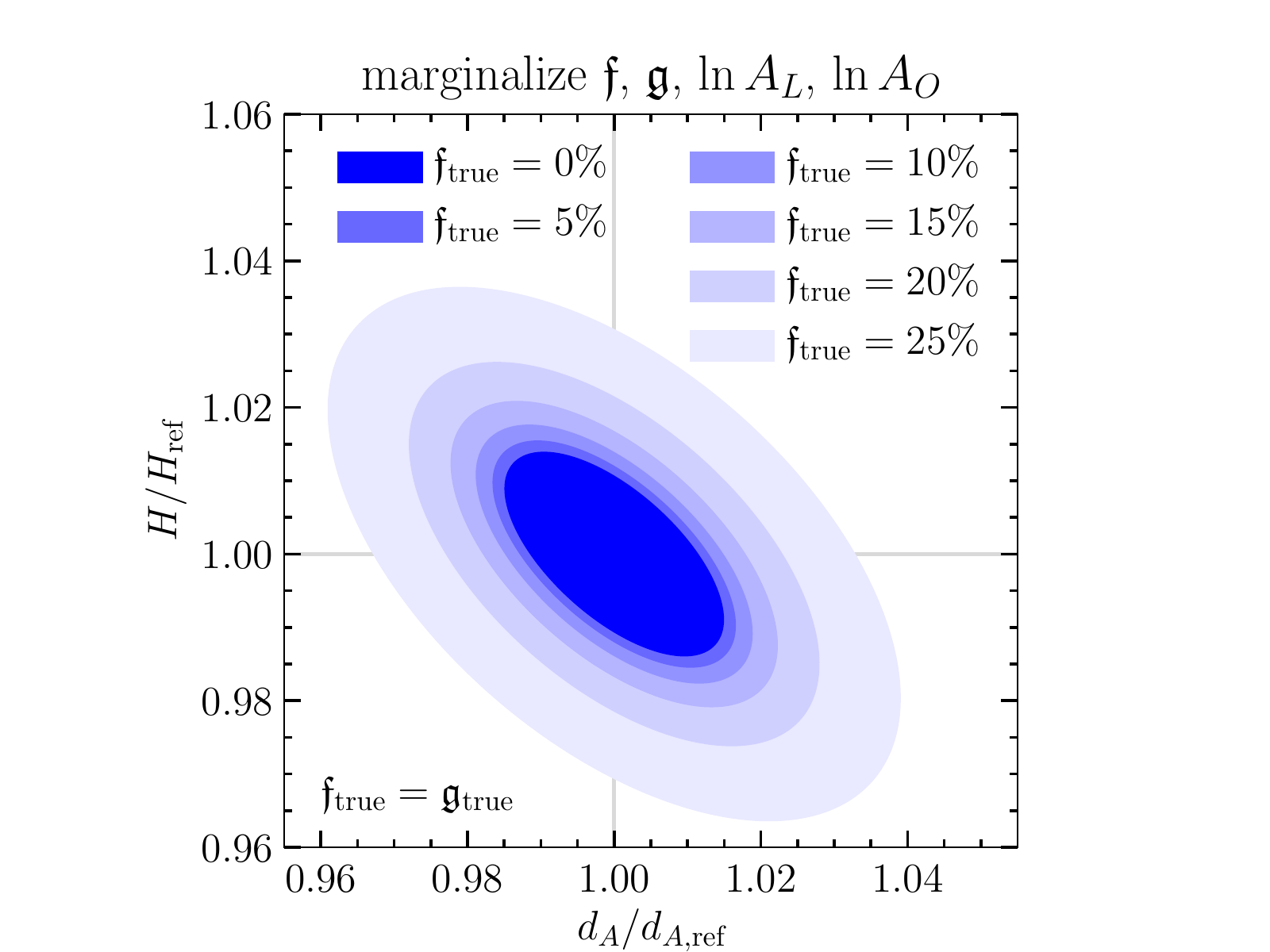}
    \caption{
        \emph{Left:} Results from the joint-analysis method: $1\sigma$
        (\SI{68}{\percent} C.L.) limits on $f\sigma_8$ when the shape of both
        the LAE and OIIE power spectra are known, corresponding to
        \textit{Case~B}. The measurement is unbiased for both sets of
        marginalizations.
        \emph{Right:} Results from the joint-analysis method:
        $1\sigma$-confidence ellipses on the angular diameter distance $d_A$
        and the Hubble expansion rate $H$ at $z_\lae=2.7$. The ellipses
        represent several values of the interloper fractions $\ffid$ and
        $\gfid$ as shown in the legend. The grey lines indicate the fiducial
        $d_A$ and $H$ values. Here, we marginalize over the interloper
        fractions and the amplitudes of the power spectra. Larger interloper
        fractions lead to larger errors.
    }
    \label{fig:fgvwlnA_vw}
\end{figure*}
\begin{figure*}
    \centering
    \incgraph[0.49]{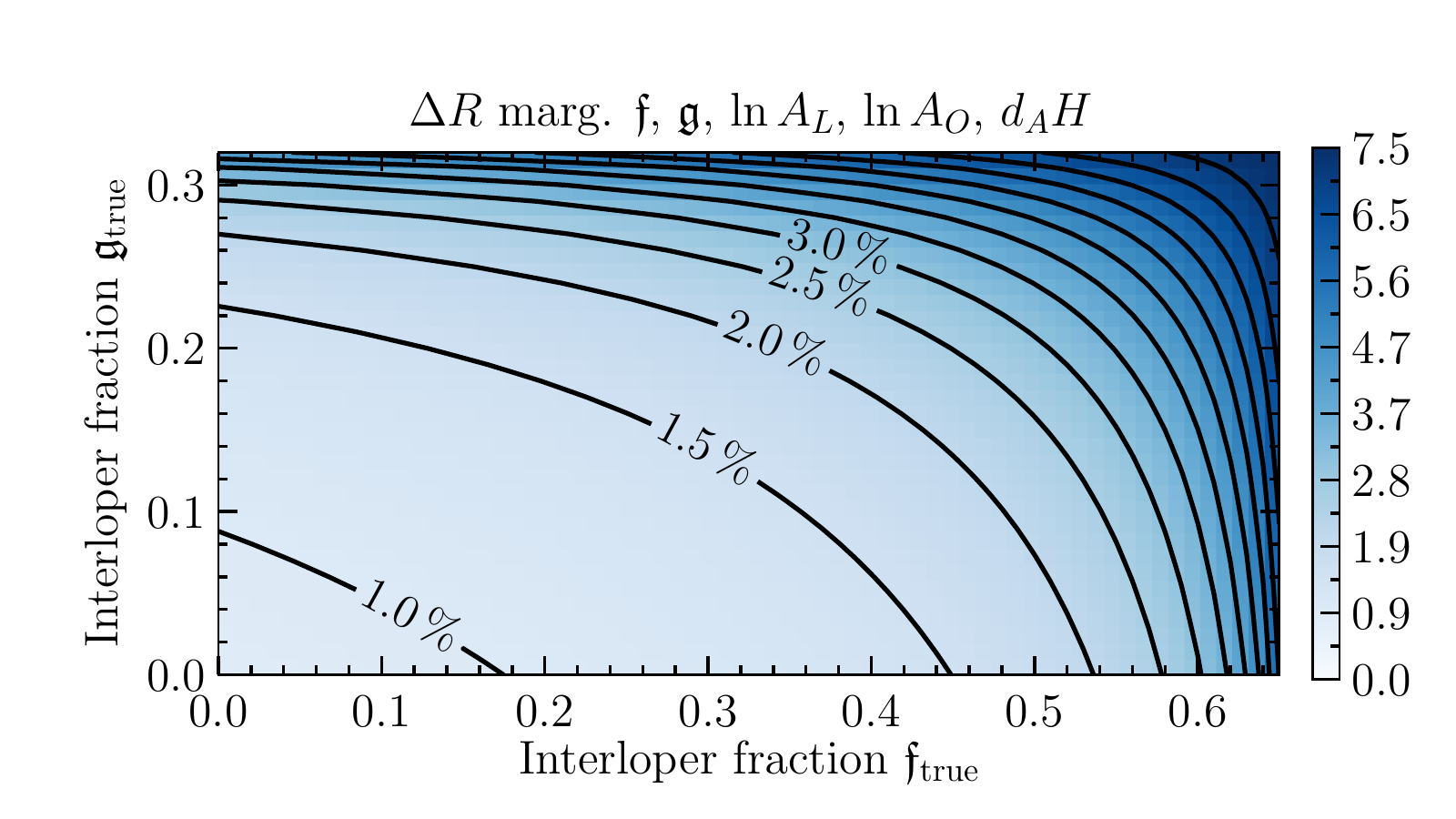}
    \incgraph[0.49]{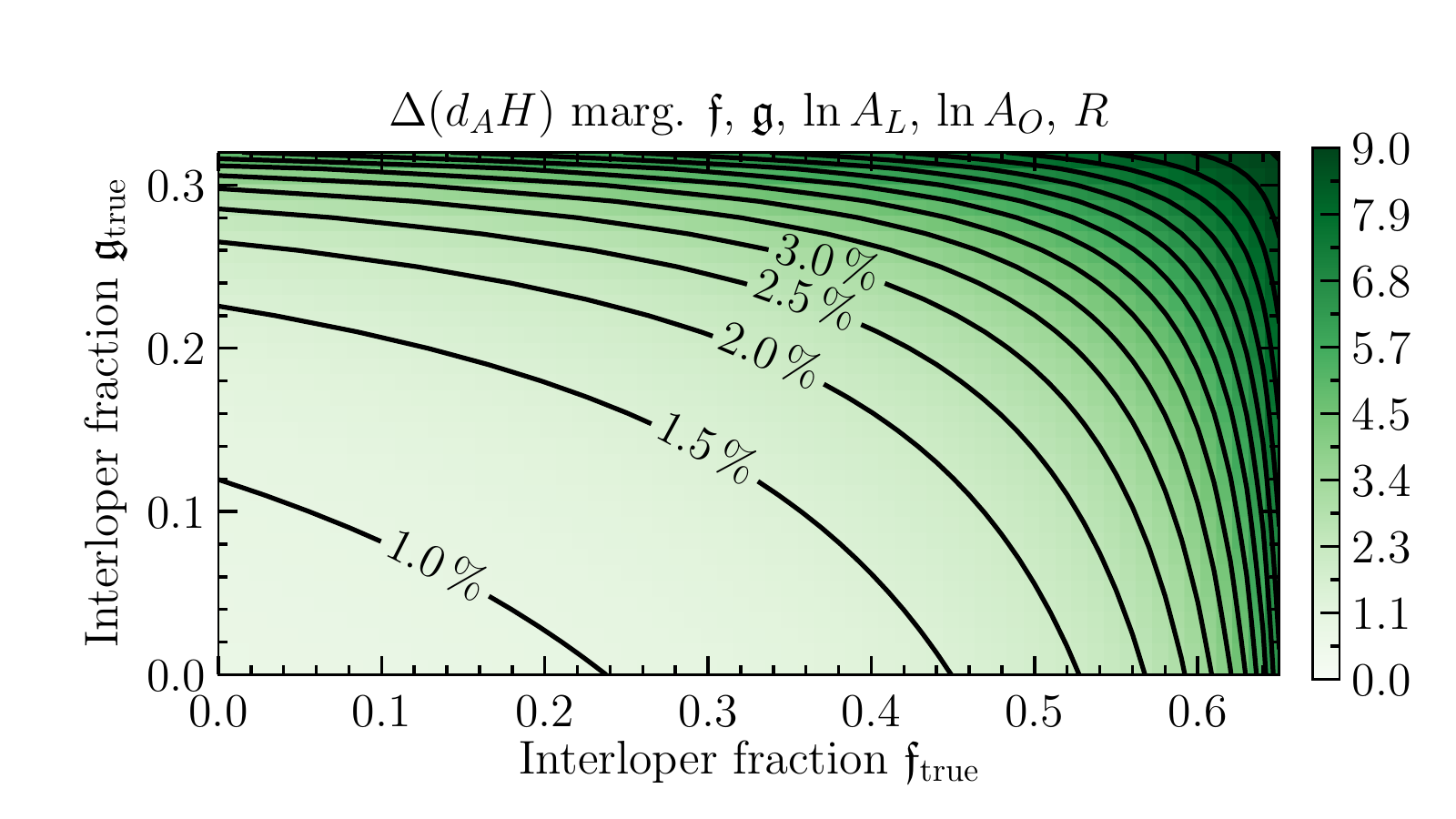}
    \incgraph[0.49]{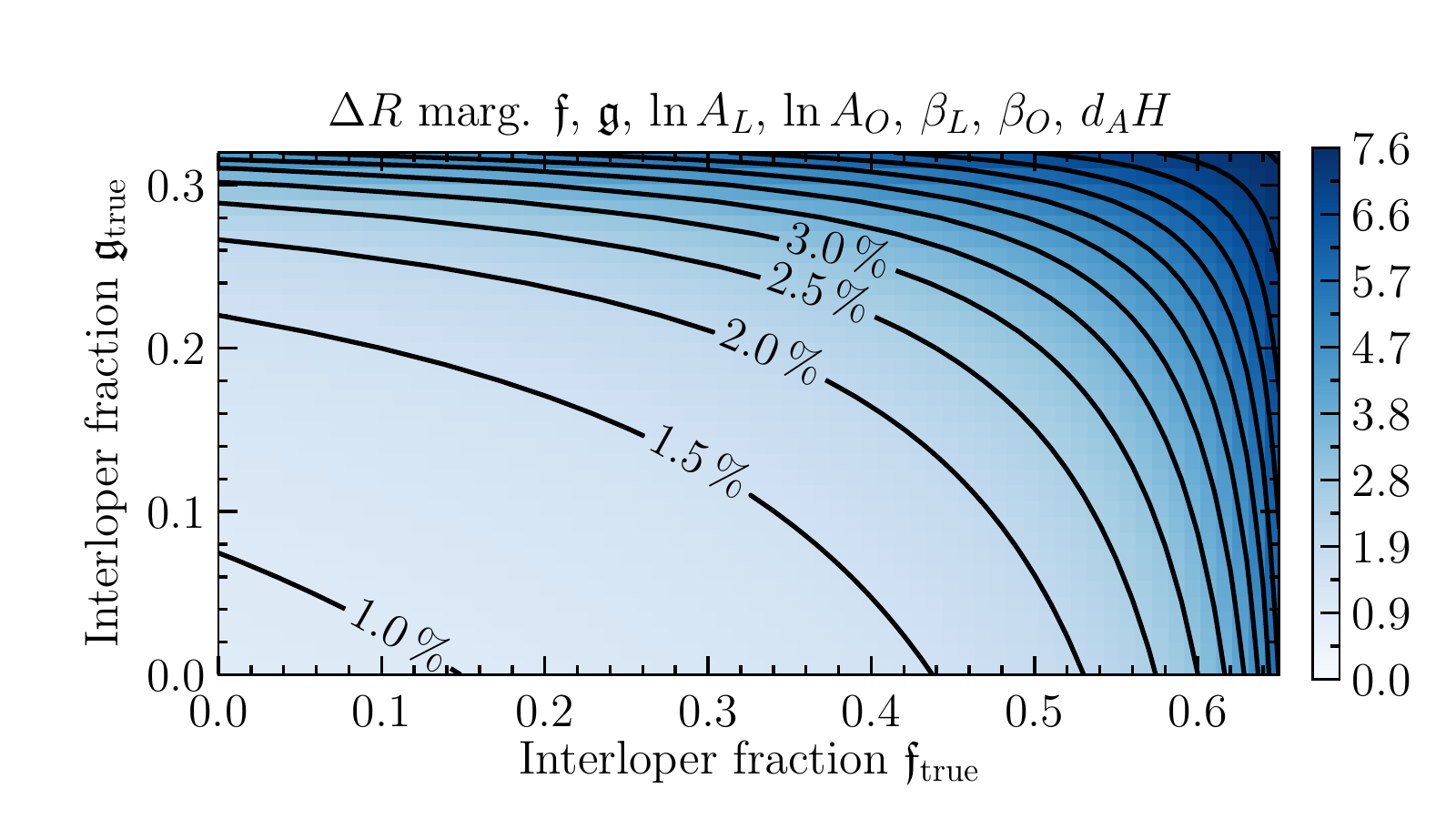}
    \incgraph[0.49]{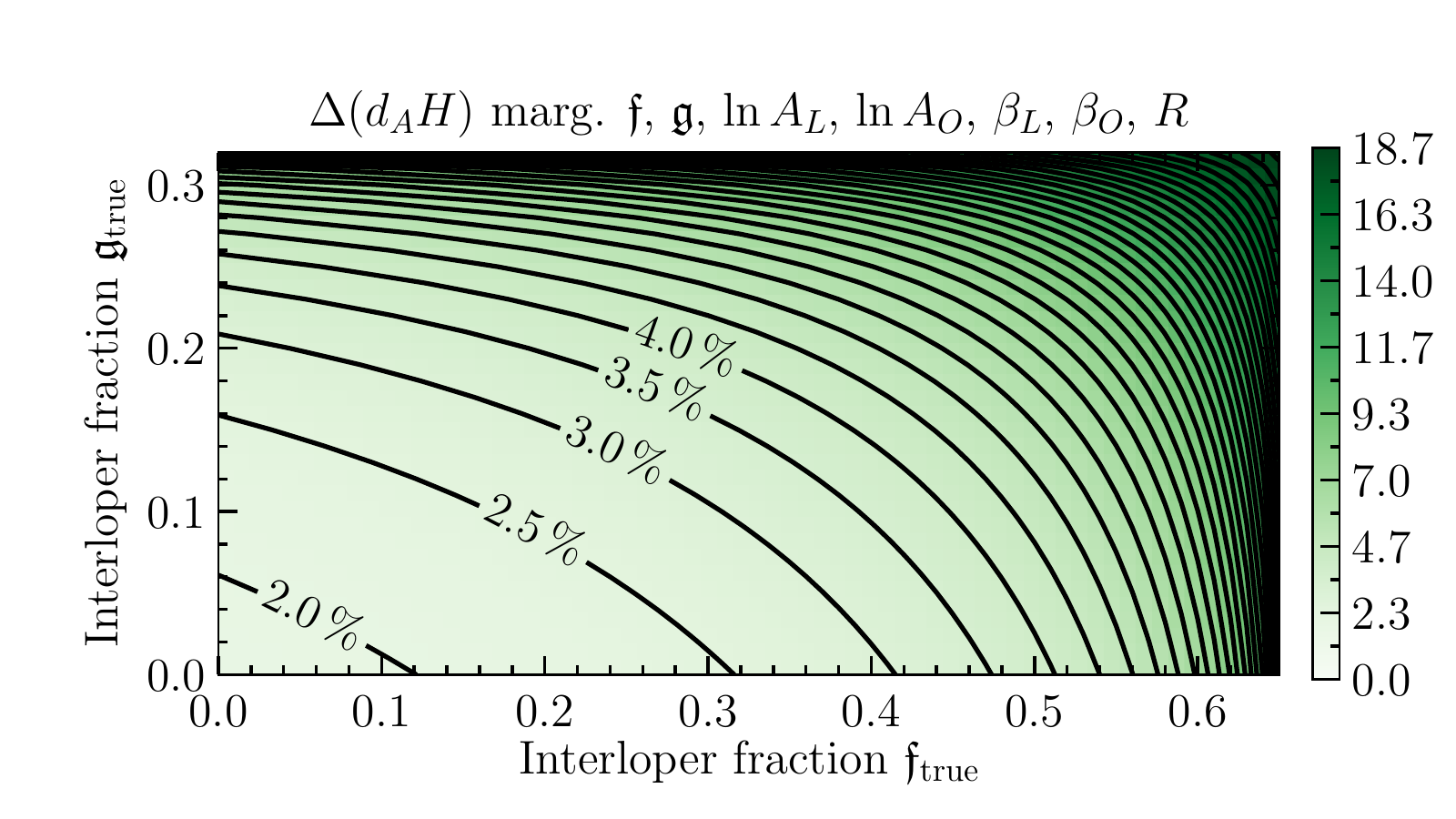}
    \incgraph[0.49]{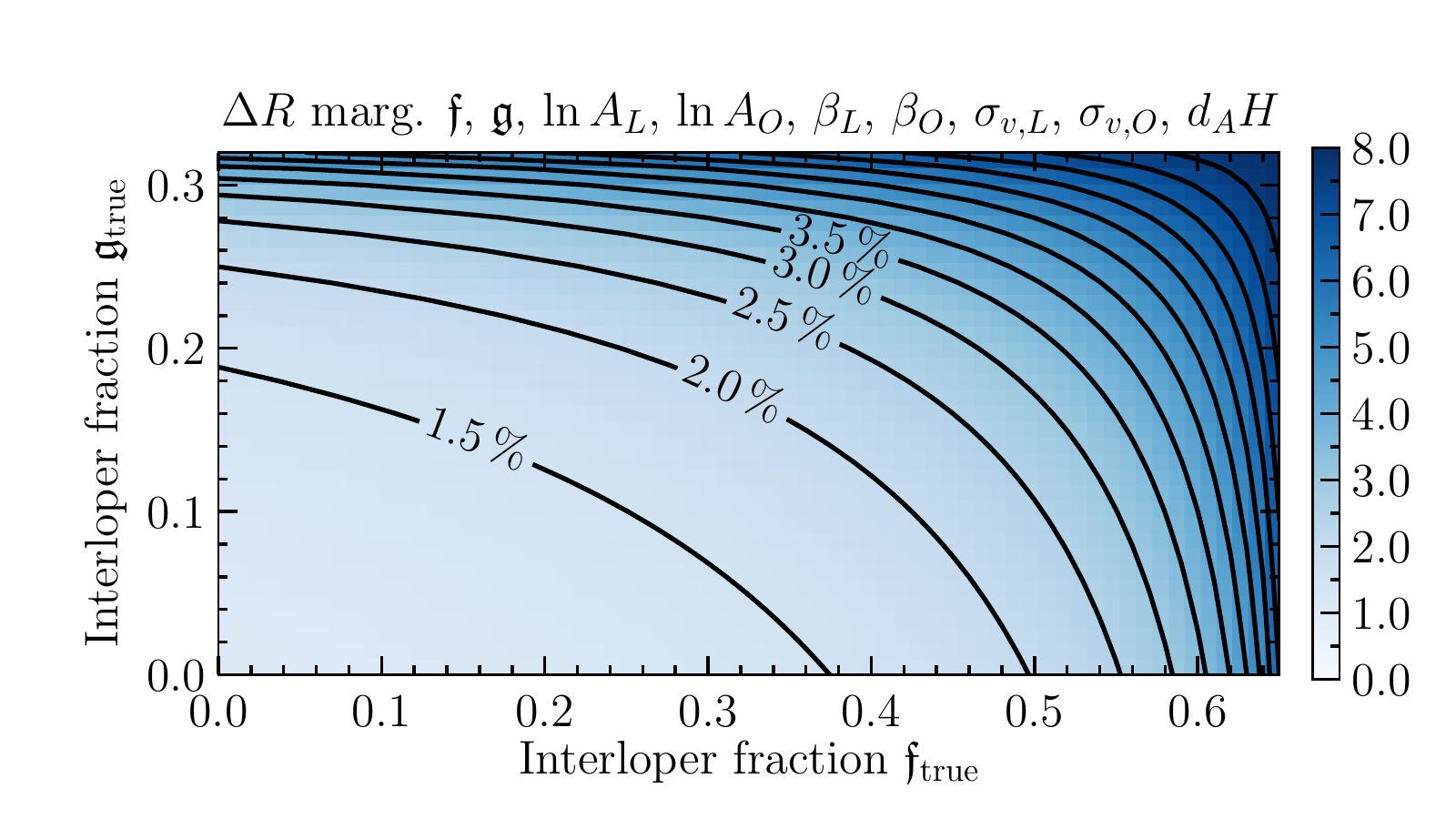}
    \incgraph[0.49]{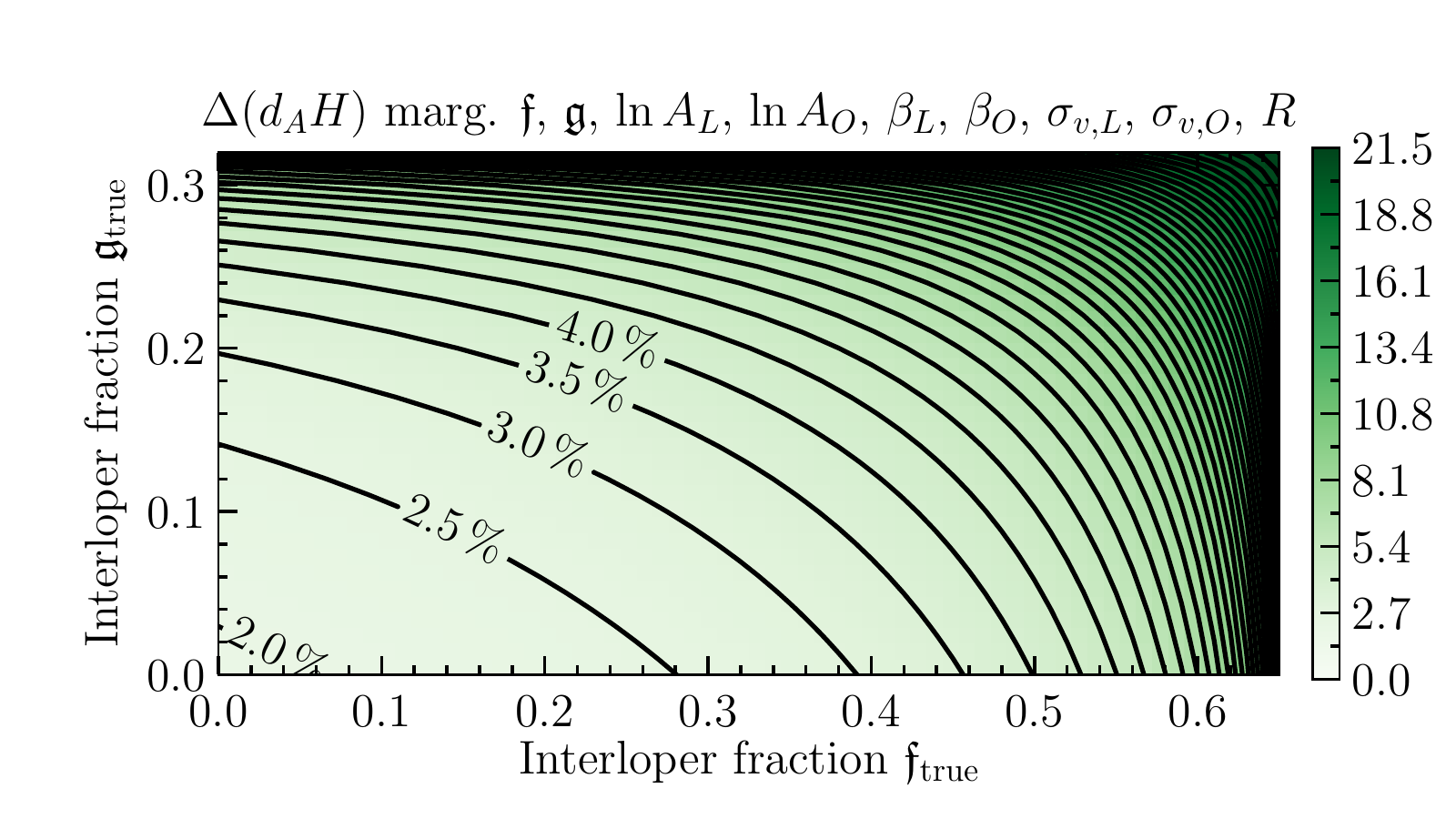}
    \caption{
        Results from the joint-analysis method: the change of $1\sigma$ (68\%
        C.L.) uncertainties on $R=(d_A^2/H)^{1/3}$ (left) and $d_A H$ (right)
        as a function of true interloper fractions $(\ffid, \gfid)$ for
        \textit{Case~B}, when the shapes of the LAE and OIIE power spectra are
        both known. For each point in each panel, we set the true interloper
        fractions as given by the axes and we marginalize over the measured
        interloper fractions and the power spectrum amplitudes. We consider
        three models, from top to bottom:
        1) assume the redshift space distortion parameters are known,
        2) marginalize over $\beta_x=f(z_x)/b_x$, and
        3) additionally marginalize over the FoG velocity dispersion
        $\sigma_{v,x}$. With our method the forecast constraints on $R$ remain
        $\lesssim\SI{1.5}{\percent}$ up to interloper fractions
        \SI{\sim20}{\percent}. The best constraints on $R$ are at the origin;
        from top to bottom, \SI{0.88}{\percent}, \SI{0.9}{\percent}, and
        \SI{1.02}{\percent}. The best constraints on $d_AH$ at the origin are
        \SI{0.79}{\percent}, \SI{1.82}{\percent}, and \SI{1.92}{\percent},
        respectively.
    }
    \label{fig:Rerrbias_fgfnS}
\end{figure*}
\begin{figure*}
    \centering
    \includegraphics[width=0.32\textwidth,trim={52 0 78 0},clip]{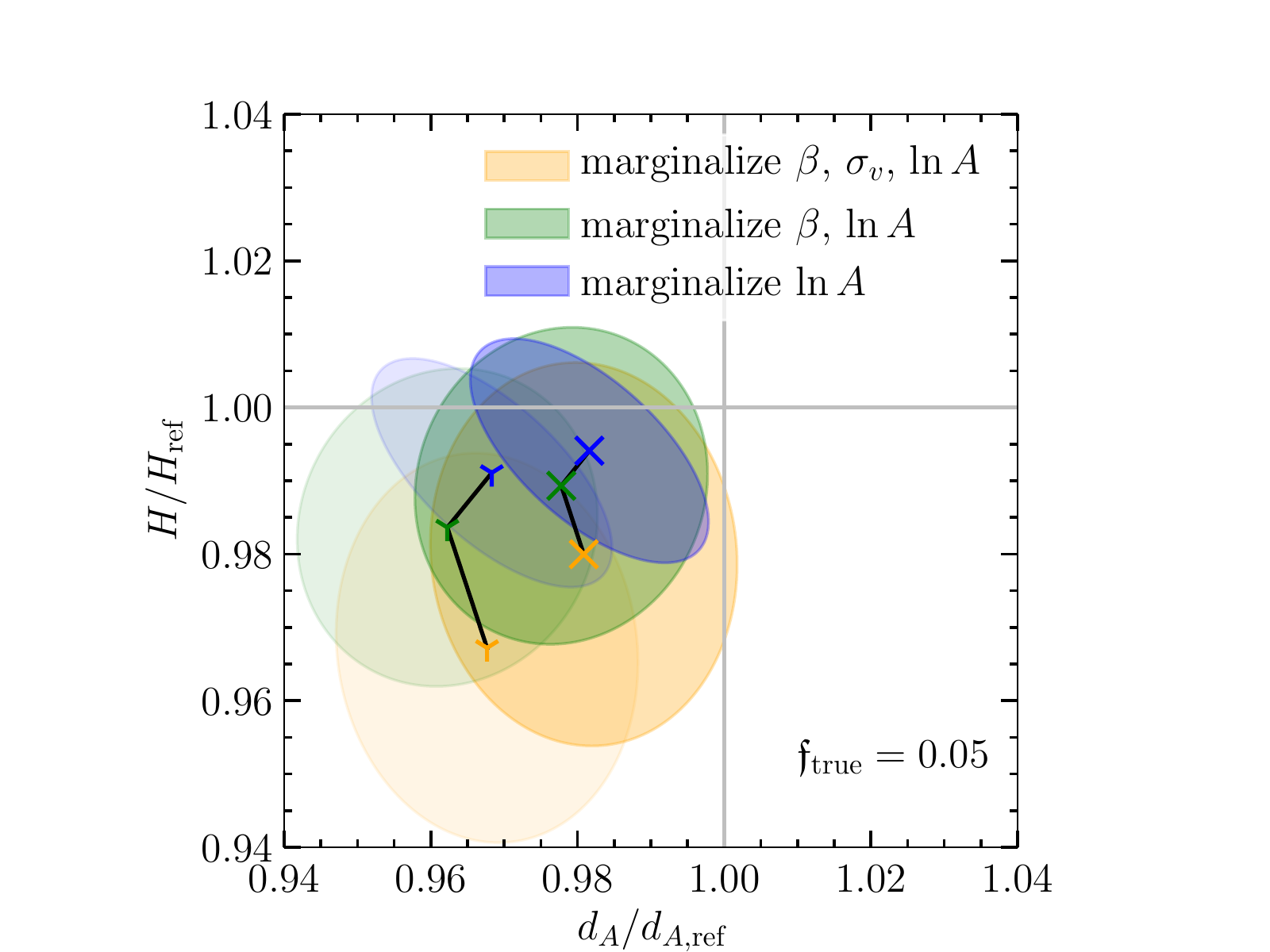}
    \includegraphics[width=0.32\textwidth,trim={52 0 78 0},clip]{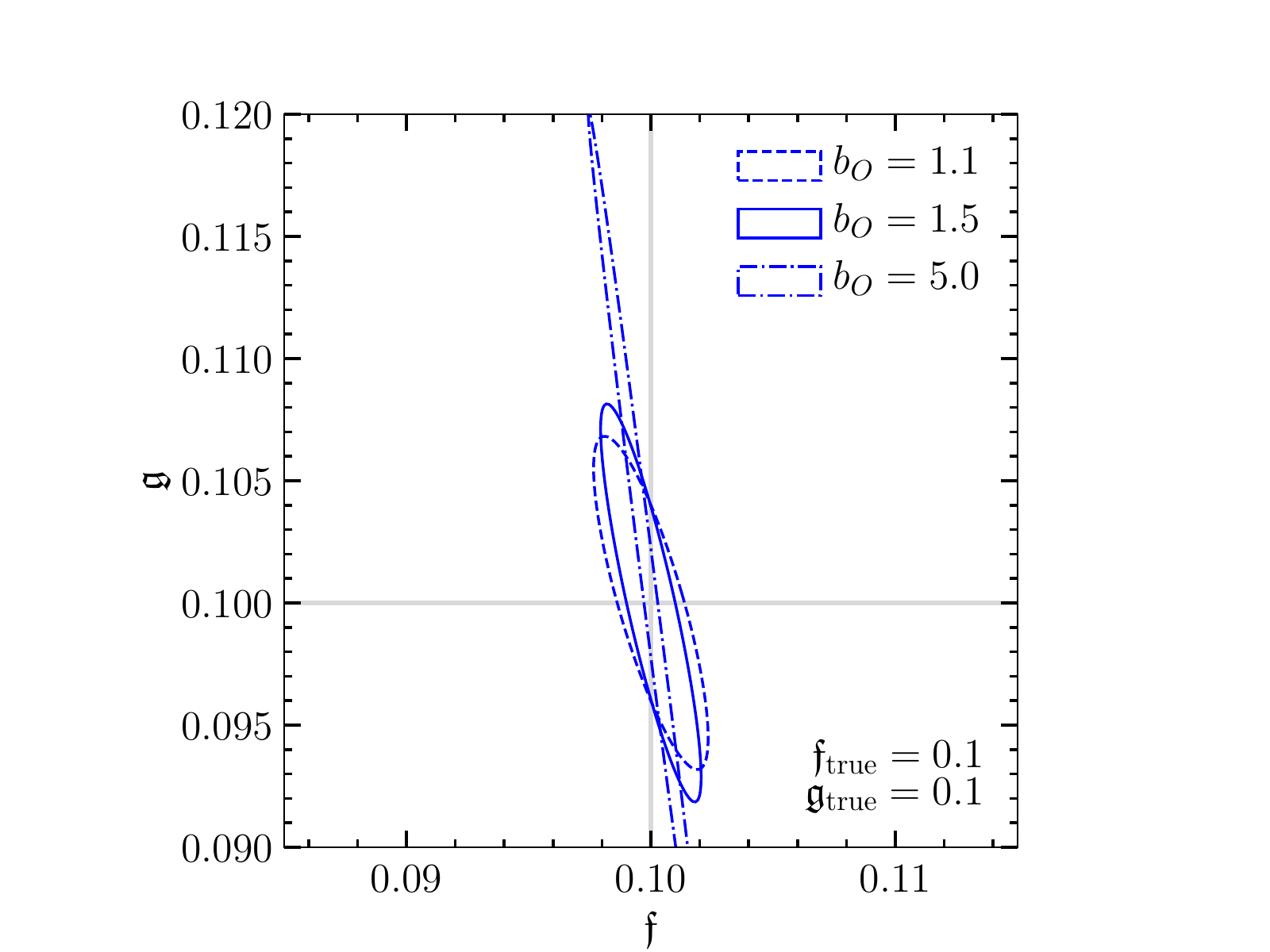}
    \includegraphics[width=0.32\textwidth,trim={52 0 78 0},clip]{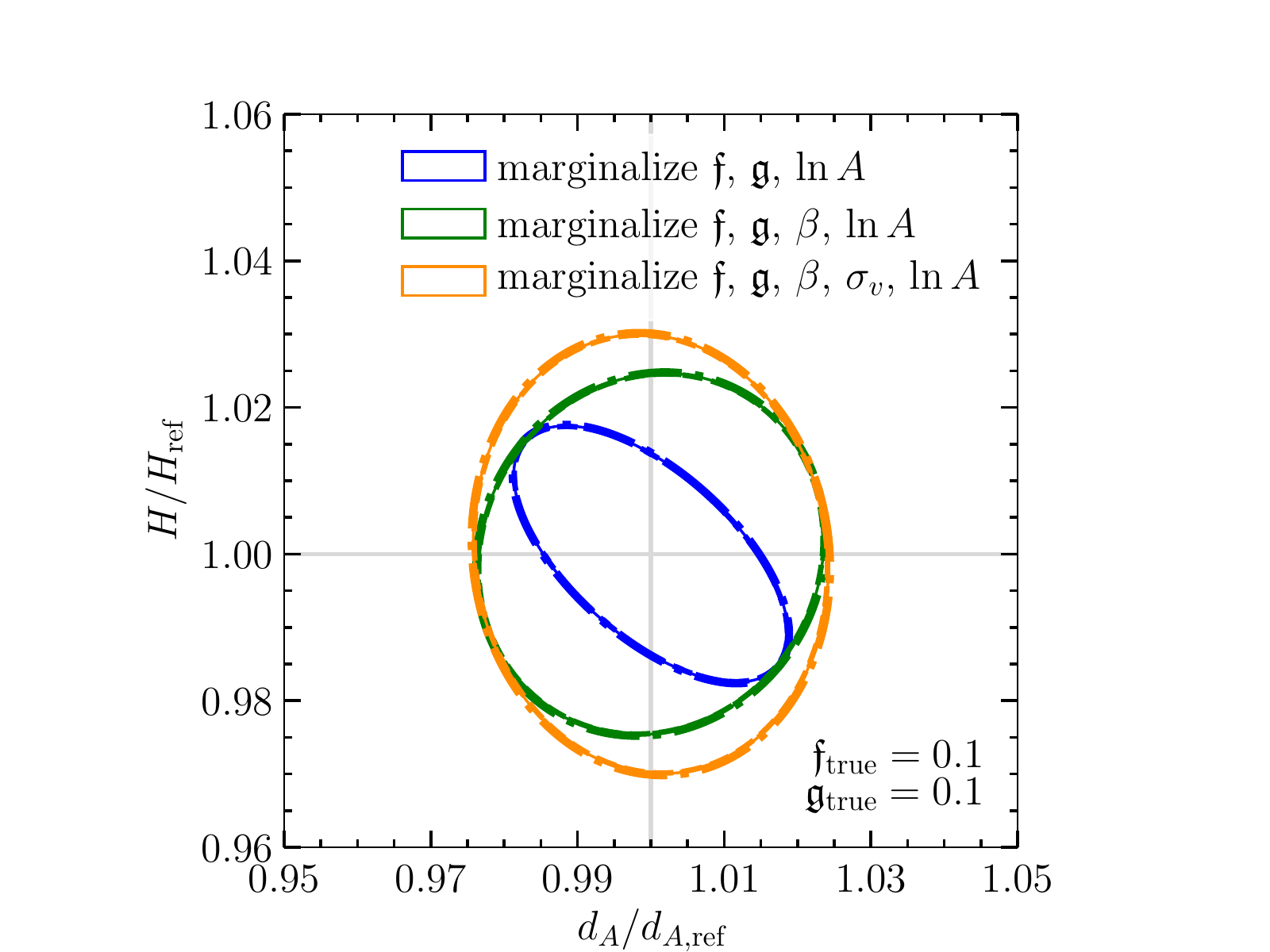}
    \caption{
        Dependence of various parameters on the amplitude of the OIIE power
        spectrum, parametrized by $b_O$. \emph{Left:} The darker ellipses show
        the effect of the interloper bias on $d_A$ and $H$ assuming an OIIE
        linear galaxy bias $b_O=1.1$; the lighter ones are as in
        \reffig{dAH_fg0} for $b_O=1.5$. If $b_O=5$, the bias on $d_AH$ is
        $\SIrange[range-phrase=\textup{--},range-units=single]{\sim20}{30}{\percent}$,
        and
        $\SIrange[range-phrase=\textup{--},range-units=single]{\sim10}{35}{\percent}$
        on $H$. The symbols mark the centers of the ellipses. Here we use
        $\ffid=\gfid=\SI{5}{\percent}$.
        \emph{Center:} Constraints on interloper fractions $\f$ and $\g$ for
        $\ffid=\gfid=\SI{10}{\percent}$ for \textit{Case~B}. Dashed
        lines use a bias of $b_O=1.1$, solid lines assume $b_O=1.5$ (our
        fiducial value throughout the rest of the paper), and dashed-dotted
        use $b_O=5$. A larger OIIE bias reduces the uncertainty on $\f$ and
        increases the uncertainty on $\g$.
        \emph{Right:} The constraints on $d_A$ and $H$, marginalizing over
        parameters as indicated in the legend. The figure demonstrates that
        the OIIE bias does not affect the uncertainty on $d_A$ and $H$, since
        for each of the three marginalizations considered, the three $b_O$
        ellipses lie on top of each other.
    }
    \label{fig:boii_effect}
\end{figure*}

We now address the case of the opposite limit where one can model the full
shape of the galaxy power spectrum for both LAEs and OIIEs using
\refeq{pk-kaiser}. This is perhaps an unrealistically
optimistic case, but it does allow us to set another benchmark point for the
effect of interlopers on the measurement of cosmological parameters such as the
angular diameter distance and the Hubble expansion rate. The list of parameters
that we include in the analysis is: the interloper fractions $\f$ and $\g$, and
for each type of tracer the power spectrum amplitude $\ln A$, the
angular diameter distance $d_A$, the Hubble expansion rate $H$, the redshift
space distortion parameter $\beta=f/b_1$, and the velocity dispersion
$\sigma_v^2$, for a total of up to 12 parameters.

First, we study the projected uncertainties in measuring $\f$ and $\g$.
\reffig{fg_bymodel} shows the Fisher forecasts for
several cases from maximal \emph{a priori} knowledge where we assume we
have complete knowledge of the power spectra and only the interloper
fractions are being fitted, to minimal \emph{a priori} knowledge, where
only the true cross-correlation is known beforehand, i.e.\ our
\textit{Case~A}.
For each case \reffig{fg_bymodel} displays the \SI{68}{\percent}
C.L. ($1\sigma$) contours for two values: $\ffid=\gfid=0$ (inner ellipses) and
$0.2$ (outer ellipses).

Of course, assuming the complete knowledge on the shape of the galaxy power
spectrum enhances the constraint on the contamination fractions. Between the
optimistic case (green, central ellipses) and the pessimistic case (thick,
outer-most ellipses) are three cases in which we marginalize over different
combinations of parameters. Intriguingly, marginalizing over the full 2D power
spectra and marginalizing over just the amplitudes of the power spectra gives
rise to similar constraints on the interloper fractions. Indeed, for $\f=\g=0$,
there is no change between the two, and only at interloper fractions
$\f=\g=20\%$ is the difference apparent. Marginalizing over the parameters
controlling the shape of the redshift-space power spectrum of OIIEs ($\beta_O$,
$\sigma_{v,O}$ and $d_{A,O}H_O$) also changes the $\f$ and $\g$ constraints
significantly. The effects of marginalizing over other parameters are not
as dramatic.

\citet{pullen+2016} demonstrated that ignoring interlopers in the galaxy sample
biases the estimation of cosmological parameters, such as the linear growth
rate $f$ and the galaxy bias $b_g$. Similarly, the left panel of
\reffig{dAH_fg0} presents the bias on $f\sigma_8$ and $b\sigma_8$ induced by
ignoring the interlopers, and the right panel of \reffig{dAH_fg0} shows that
this \emph{interloper bias} also plagues the distance measurement. We simulate
the interloper bias by generating a realization of the LAE power spectrum with
$\ffid=\SI{5}{\percent}$ but ignore the contamination by fixing $\f=0$ in the
analysis. For the analysis, we use an MCMC algorithm with the adaptive
Metropolis sampler \citep{roberts/rosenthal:2006}, and find the interloper bias
by running MCMC on the ensemble-averaged log-likelihood function. We also check
that the result is consistent with the first-order analytical calculation
presented in \refapp{systematicbias}. In the appendix we also justify
the use of the ensemble-averaged log-likelihood.

For each set of parameters (as indicated at the beginning of this
section and the figures), we ran the chain with $\num{1000} \, N_p (N_p+1)$
iterations, where $N_p$ is the number of parameters, updating the
covariance matrix of the proposal distribution every $N_p(N_p+1)$ steps
with $\beta=0.95$ \citep[see Eq.(2.1) in][]{roberts/rosenthal:2006}. The
first half of the iterations are discarded to allow the proposal
distribution to settle, and the analysis is performed on the second half.
We use a flat prior as long as the parameters are within physical limits.
The exceptions are that the auto-power spectra are additionally limited to
\SI{\leq e6}{\per\h\cubed\mega\parsec\cubed}, and in later sections
(\textit{Case~C} and \textit{Case~D}) the cosmological distortion
parameters $v_L$, $w_L$, $v_O$, and $w_O$ are additionally limited so that
the splines of the power spectra are well-defined. These limits are
enforced by setting the likelihood $\Like=0$ outside the bounds. 

The interloper bias disappears when one treats the interloper fractions $\f$
and $\g$ as free parameters and simultaneously analyzes the LAE and OIIE
auto-power spectra and the cross-power spectrum. The left panel of
\reffig{fgvwlnA_vw} displays the results for $f\sigma_8$ from our joint
analysis as a function of $\ffid=\gfid$, after marginalizing over two different
sets of parameters as indicated in the figure legend. In addition to
showing that the interloper bias disappears, the figure also shows that larger
interloper fractions come at the cost of increasing the measurement
uncertainty.

The right panel of \reffig{fgvwlnA_vw} shows the result for the distance
measurements ($d_A$ and $H$) at $z\sim2.7$ (the LAE redshift). Here, we fix the
redshift space distortion parameters $\beta_L$, $\beta_O$, $\sigma_{v,L}$, and
$\sigma_{v,O}$, and marginalize over $\f$, $\g$, $\ln A_{L}$ and $\ln
A_O$\footnote{The uncertainty ellipses here differ from the ones in
\citet{shoji+2009} because we use a smaller galaxy bias parameter ($b_L=2$
versus $b_L=2.5$).}. We consider several interloper fractions with $\f=\g=0$,
$5$, $10$, $15$, $20$, $\SI{25}{\percent}$, and calculate the error ellipses
from the Fisher information matrix. Running MCMC on the ensemble-averaged
log-likelihood function produces the same conclusion: the joint analysis
removes the interloper bias, although the measurement uncertainty increases for
larger interloper fractions.

The primary geometrical observables from the cosmological distortion of the
two-dimensional redshift-space power spectrum are the following combinations of
$d_A$ and $H$,
\be
R \equiv \(d_A^2 / H\)^{1/3}, \quad
AP \equiv d_A H\,.
\ee
These are sensitive to, respectively, the isotropic and anisotropic
stretch/contraction in the \plane{k_\perp}{k_\parallel}
\citep{padmanabhan/white:2008,shoji+2009}, i.e., we measure $R$ from the
isotropic location of the BAO signal and $AP$ from the Alcock-Paczynski test
using the known anisotropies associated with RSD.

\reffig{Rerrbias_fgfnS} presents the uncertainties on $R$ and $AP$ for the
general case $\ffid\neq\gfid$. Since the upper right corner of the
\plane{\f}{\g} (in \reffig{pLO0_fgplane}) should only occur for a catastrophic
failure of line identification, we only present the lower allowed region for
the true interloper fractions. The left three panels display the constraints
for $R$, while the right three are the equivalent figures for $AP\equiv d_A H$.
In the top two panels we only marginalize over $\f$, $\g$, $\ln A_L$, and $\ln
A_O$; in the middle two panels we additionally marginalize over the redshift
space parameters $\beta_L$ and $\beta_O$. In the bottom two panels we include
marginalizations over the Finger-of-God parameters $\sigma_{v,L}$ and
$\sigma_{v,O}$. All plots in \reffig{Rerrbias_fgfnS} have a similar structure:
the measurement uncertainty is lowest near the origin $\f=\g=0$, then increases
slowly at first, then rapidly as one gets closer to the limits
$\f=\f_\mathrm{lim} = N_{\oii}^{\true}/(N_\lae^\true+N_\oii^\true)$ and
$\g=\g_\mathrm{lim} = N_\lae^\true/(N_\lae^\true+N_\oii^\true)$. 

For all cases, the larger interloper fractions degrade the measurement
precision of $R$ and $d_AH$. The largest effect is for the Alcock-Paczynski
parameter $d_AH$ when marginalizing over the RSD parameters because both RSD
and the interlopers contribute to the observed anisotropies in the
two-dimensional power spectrum. For example, in the right panel in
\reffig{Rerrbias_fgfnS}, the uncertainty for the $d_AH$ measurement changes
from \SI{1}{\percent} (top panel) to \SI{\sim2}{\percent} (middle panel) near
$\f=\g=0$, once we marginalize over $\beta_L$ and $\beta_O$. Conversely, the
distance measure $R$ is much less affected by the redshift space distortion
parameters, and the uncertainties at the origin $\f=\g=0$ changes from
\SI{0.88}{\percent} (top) to \SI{1.02}{\percent} (bottom). This behavior arises
because the BAO feature, which dominates the measurement of $R$, is less
affected by RSD \citep{seo/eisenstein:2007,seo/etal:2010,shoji+2009}.

As we have seen in \reffig{fg_bymodel}, the interloper fraction measurement is
most affected by the amplitude of the power spectrum. In addition, since a
larger OIIE power spectrum means a larger contamination amplitude, it also
generates a larger interloper bias, as shown in the left panel in
\reffig{boii_effect}. In the center and right panels of the figure,
we show the effect of a change in OIIE linear galaxy bias on the
measurement of the interloper fractions and the LAE distance measurement.
The figure contains the result for three different values of the OIIE
galaxy bias: $b_O=1.1$, $b_O=1.5$ (this is the value adopted throughout
the paper), and $b_O=5$. Since a larger OIIE bias results in a larger
interloper signal in the LAE and cross power spectra, a large OIIE galaxy bias
results in a tight constraint on $\f$. However, when $\f=\g=0.1$ the LAE
distance measurement is essentially independent of the OIIE bias.

\subsection{Case~C: knowing full shape of LAE power spectrum only}
\label{sec:onlylaeshape}
The \textit{Case~B} presented in \refsec{fullshape} assumes that we can
accurately  model the non-linearities in both the LAE and OIIE power spectra.
For the LAEs at redshift $z\sim2.7$, perturbation-theory based analytical
calculations \citep{PTreview} provide a reliable model for nonlinear evolution
of the density field up to $k\sim\SI{0.4}{\h\per\mega\parsec}$
\citep{jeong/komatsu:2006}. This is also the redshift and wavelength range
where we expect the perturbative bias expansion \citep{PBSreview} to model the
nonlinear galaxy bias. In contrast, the corresponding OIIE power spectrum
extends to $k\sim\SI{2.8}{\h\per\mega\parsec}$ at a mean redshift $z\sim0.2$,
at which point perturbative approaches fail. At this redshift, perturbation
theory can be reliable only for $k\sim\SI{0.1}{\h\per\mega\parsec}$ or larger
scales.

Therefore, it is more realistic to explore the case where we lack any prior
knowledge on the nonlinear 1D power spectrum of OIIEs. That is, we treat the
nonlinear power spectrum for OIIEs as a set of free parameters. We still adopt
the anisotropies due to RSD as in \refeq{pk-kaiser}, and only parameterize the
one-dimensional power spectrum. We shall study the effect of higher order RSD
parameters in \refsec{morersd}.

To measure the 1D OIIE power spectrum $P_O(k)$, we use a cubic spline with
knots linearly spaced at $k<\SI{0.4}{\h\per\mega\parsec}$ (in the original OIIE
volume), and logarithmically spaced above that. \refeq{pkOOref} is then used to
project the OIIE power spectrum onto the LAE volume. Because we perform the
analysis in the projected LAE volume, we determine the minimum and maximum
wavenumbers by scaling the corresponding wavenumbers in the LAE volume, but we
extend the range by \SI{10}{\percent} at each end, in order to provide a buffer
for the cosmological distortion measurement. This procedure effectively sets a
prior on $R$ and $AP$ of \SI{\sim10}{\percent}, because we set a hard prior
outside of the wavenumber range: when the MCMC chain moves outside of the
range, we force the likelihood to be zero.

We find no noticeable difference in the constraint on the interloper fractions
$\f$ and $\g$ between this case and \textit{Case~B} in \refsec{fullshape}.
This result arises because the main information for constraining $\f$ and $\g$
comes from the cross-correlation, and the cross-correlation method works
without knowing the explicit shape of the nonlinear power spectrum (as shown in
\refsec{model_pkcrossonly}).
Since the constraints for \textit{Case~B} and \textit{Case~C} are nearly identical,
we do not duplicate the figures from \refsec{fullshape} for \textit{Case~C}.

Measuring the nonlinear OIIE power spectrum without the shape information means
that it is impossible to measure $R=\(d_A^2/H\)^{1/3}$, unless we specifically
search for the BAO signature. Nevertheless, because the assumed RSD function in
\refeq{pk-kaiser} dictates the angular-dependence of the two dimensional power
spectrum, we can still measure $AP$.
We find that the constraint does not change significantly for
interloper fractions $\f\sim\g\lesssim\SI{20}{\percent}$ because the
contamination from the LAE power spectrum to the OIIE power spectrum is
multiplied by a volume factor $1/(\alpha^2\beta)\sim0.023$ and thus remains
insignificant.

The linear RSD model adopted here is not reliable on scales relevant for the
OIIE galaxy power spectrum ($k<\SI{2.8}{\h\per\mega\parsec}$). To correct for
this effect, a more robust method would be to set the full two-dimensional OIIE
power spectrum completely unconstrained, similar to the analysis in
\refsec{model_pkcrossonly}. However, leaving the 2D OIIE power spectrum free
would require fitting \num{\sim8500} parameters. Given that the measurement
uncertainties in $\f$ and $\g$ hardly change among the three cases examined in
\refsec{model_pkcrossonly}, \refsec{fullshape}, and \refsec{onlylaeshape}, we
expect that the key result would still remain true that (a) joint analysis
removes the interloper bias, and (b) marginalizing over the interloper
fractions increases the uncertainties in the cosmological parameters.

\subsection{Case~D: knowing only anisotropy due to RSD}
\label{sec:noknowledge}
\begin{figure*}[ht]
    \centering
    \incgraph[0.495]{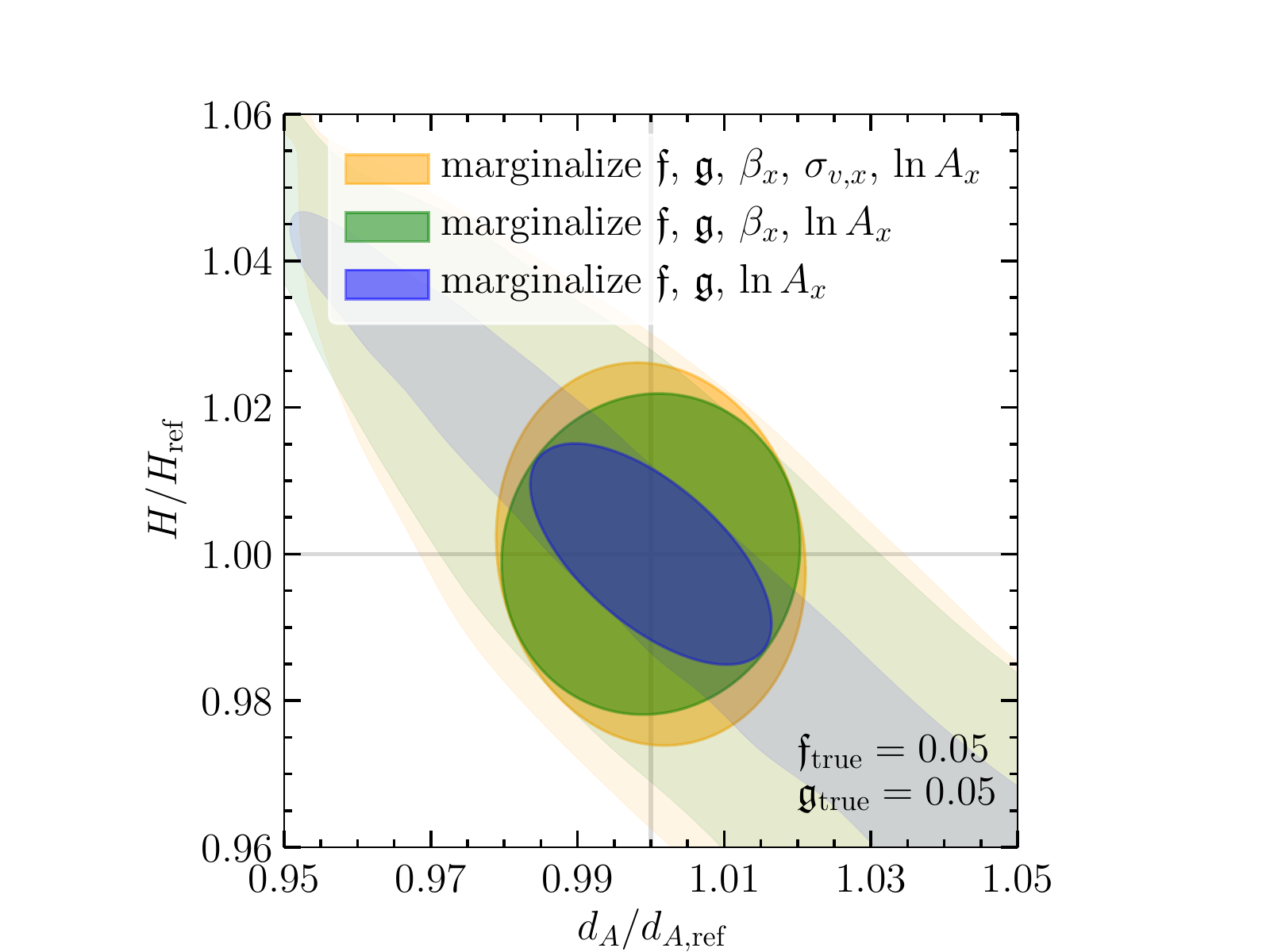}
    \incgraph[0.495]{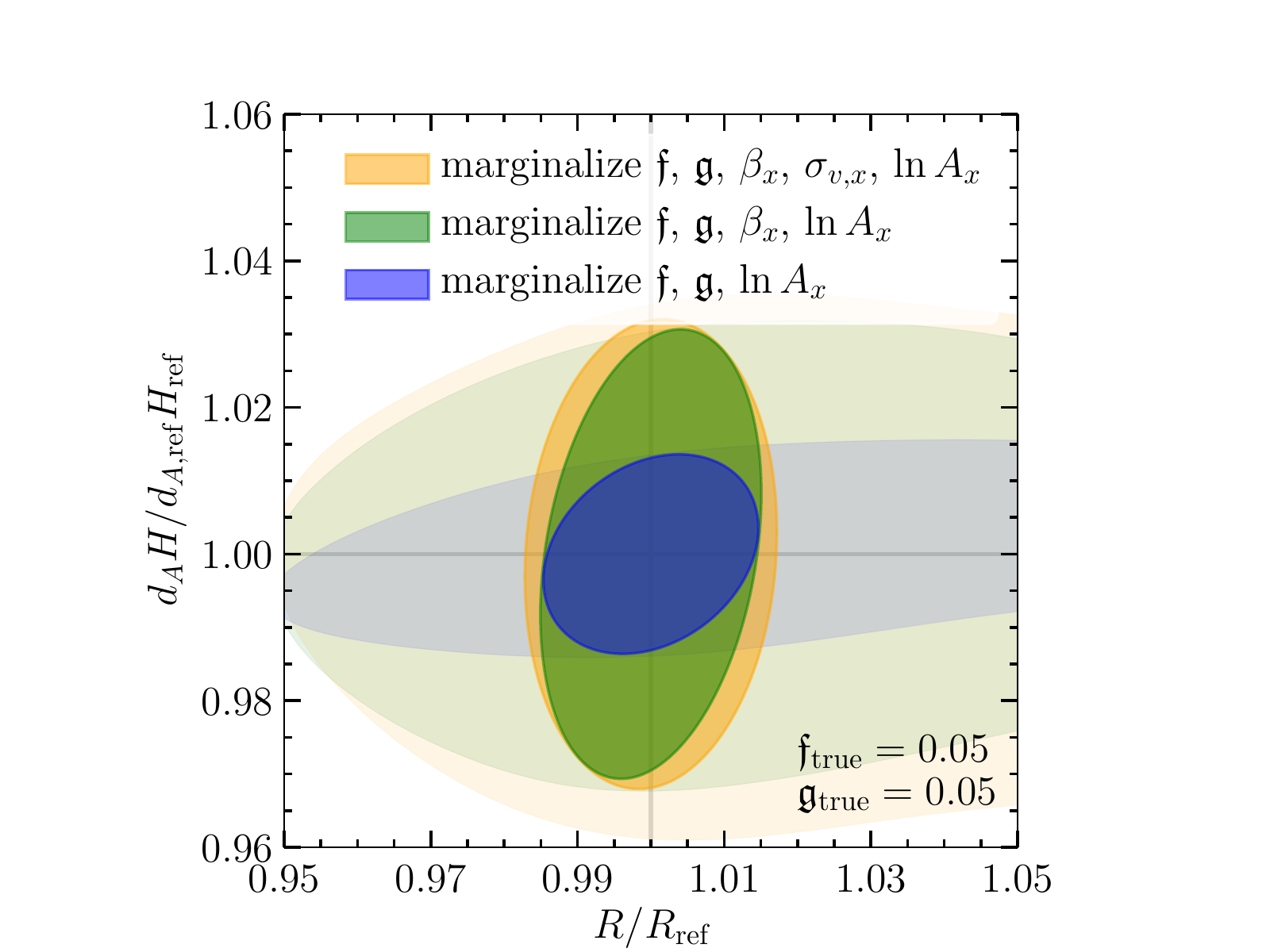}
    \caption{
        Results from the joint-analysis method:
        \textit{Left:} $1\sigma$ (\SI{68}{\percent} C.L.) uncertainties on
        $d_A$ and $H$ at the LAE redshift when $\ffid=\gfid=\SI{5}{\percent}$.
        The strongly shaded ellipses correspond to \textit{Case~C}
        where the shape of the LAE power spectrum is known and we fit for the
        1D OIIE power spectrum. The lightly shaded ellipses are the
        constraints for \textit{Case~D} when fitting both the 1D LAE
        and 1D OIIE power spectra. The color coding indicates which parameters
        are marginalized over, as described in the legend.
        \textit{Right:} Projected constraints for the same cases as the left
        panel, but for $R=(d_A^2/H)^{1/3}$ and $AP=d_A H$.
    }
    \label{fig:M_vw}
\end{figure*}
Now we examine the case when we relax the assumption that the 1D LAE power
spectrum shape is known. Without the shape information, we cannot measure
$R=(d_A^2/H)^{1/3}$. However, the Alcock-Paczynski test still allows
measurement of the parameter $AP = d_A H$ from the anisotropy of the
redshift-space power spectrum. Here, in order to highlight this point, we
exclude the shape information altogether, including the BAO that must enhance
the constraint on $R$.

We model the LAE power spectrum similarly to the way we modeled the OIIE
power spectrum: $P_L(k)$ will be a $3^\text{rd}$-order spline with knots
linearly spaced for $k<\SI{0.04}{\h\per\mega\parsec}$ and logarithmically
spaced for $\SI{0.04}{\h\per\mega\parsec}<k<\SI{0.4}{\h\per\mega\parsec}$.
This approach is adopted to ensure that all major features of the power
spectrum can be represented by the fit. Without a dedicated search for the
BAO \citep{koehler/etal:2007,shoji+2009}, however, we cannot measure $R$.

The left hand side of \reffig{M_vw} shows the projected constraints on $d_A$
and $H$ when $\f=\g=5\%$, after marginalizing over the interloper fraction and
the amplitude (\textit{Case~C}, strongly shaded blue) or the full 1D power
spectrum (\textit{Case~D}, lightly shaded blue). For the green and orange
ellipses we additionally marginalize over the RSD and FoG parameters,
respectively. The figure reveals the degeneracy between $d_A$ and $H$ along
the direction of constant $AP=d_AH$, as shown in the right panel in
\reffig{M_vw}. When the shape of the LAE power spectrum is completely unknown,
HETDEX can still measure $d_A H$ to about \SI{3}{\percent} accuracy.

\subsection{Effect of higher-order RSD}\label{sec:morersd}
\begin{figure}
    \centering
    \incgraph[0.49]{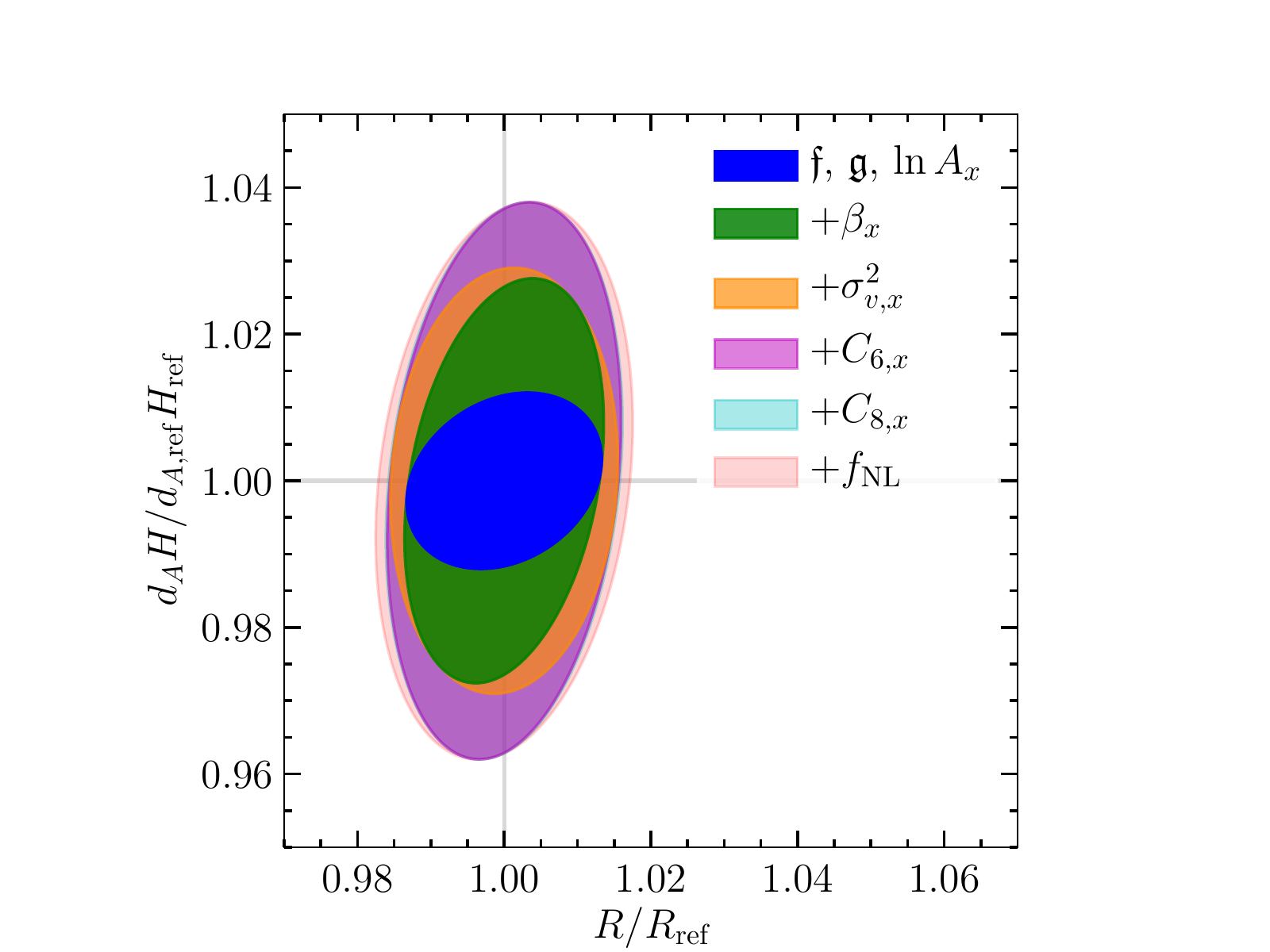}
    \caption{
        Results from the joint-analysis method, \textit{Case~B}:
        $1\sigma$ (68\% C.L.) uncertainty ranges on $R$ and $d_AH$,
        successively marginalizing over more parameters as indicated in the
        legend. Whenever a parameter has an $x$ suffix, it means that we
        marginalize over both the parameter for LAEs and for OIIEs.
    }
    \label{fig:morersd}
\end{figure}
We have used \refeq{pk-kaiser} for modeling the redshift-space distortions,
including the linear theory prediction (Kaiser effect) and the Finger-of-God
suppression. In this section, we study the effect from the non-linear
contribution of redshift space distortion. To fully account for the nonlinear
distortion in a consistent manner, we need to include the full perturbation
theory expression in, for example, \citet{desjacques/etal:pkgs}. For the
purpose of testing the interplay between the interloper fraction and the
nonlinear RSD effect on the distance measurement ($R$ and $d_A H$), however, we
develop an ansatz motivated by the full expression. Specifically, we add
parameters $C_6^L$, $C_6^O$ and $C_8^L$, $C_8^O$ to account for the
higher-order angular dependence, replacing \refeq{pk-kaiser} by
\ba
\label{eq:pk2d}
P_x(k_\perp, k_{\parallel})
&= A_x^2\,A_\mathrm{RSD}(k, \mu)\,A_\mathrm{FoG}(k, \mu)\,P_x^\text{lin}(k)\,,
\ea
where
\ba
\label{eq:pkXXlin}
P_x^\mathrm{lin}(k) &= b_x^2 \, D^2(z_x) \, P_m^\mathrm{lin}(k)
\\
\label{eq:Arsd}
A_\mathrm{RSD}(k,\mu) &= \big[
    1  + 2\beta_x \mu^2
    + \beta_x^2 \mu^4
    \vs&\quad
    + C_6^x b_x^{-2} \sigma_{v,x}^2 k^2 \mu^6
    + C_8^x b_x^{-2} \sigma_{v,x}^2 k^2 \mu^8
\big]
\\
A_\mathrm{FoG} &=
\left(1 + f^2 k^2 \mu^2 \sigma_{v,x}^2\right)^{-1}\,.
\label{eq:Afog}
\ea
This parametrization naturally reduces to the linear Kaiser formula
\refeq{pk-kaiser} in the large-scale limit $k\to 0$. We study the projected
constraints with fiducial values of $C_6=C_8=0$ (under the null hypothesis);
increasing them to $C_6=C_8=1$ does not change the result significantly.

\reffig{morersd} presents the projected constraints on $R$ and $d_A H$ as we
marginalize over successively more parameters, assuming that the shape of both
LAE and OIIE power spectra are known. Including the nonlinear redshift-space
distortion does not affect the projected constraint on the $R$ parameter which
controls the isotropic shift of the wavenumbers because the features in the
monopole galaxy power spectrum such as the BAO provide information orthogonal
to the anisotropies. Conversely, the Alcock-Paczynski test is weakened by the
marginalization over $C_6$. When marginalizing over the $C_8$ parameter,
however, there is no noticeable difference.

\subsection{Interloper bias and primordial non-Gaussianity}\label{sec:fnl}
\begin{figure}
    \centering
    \incgraph[0.49]{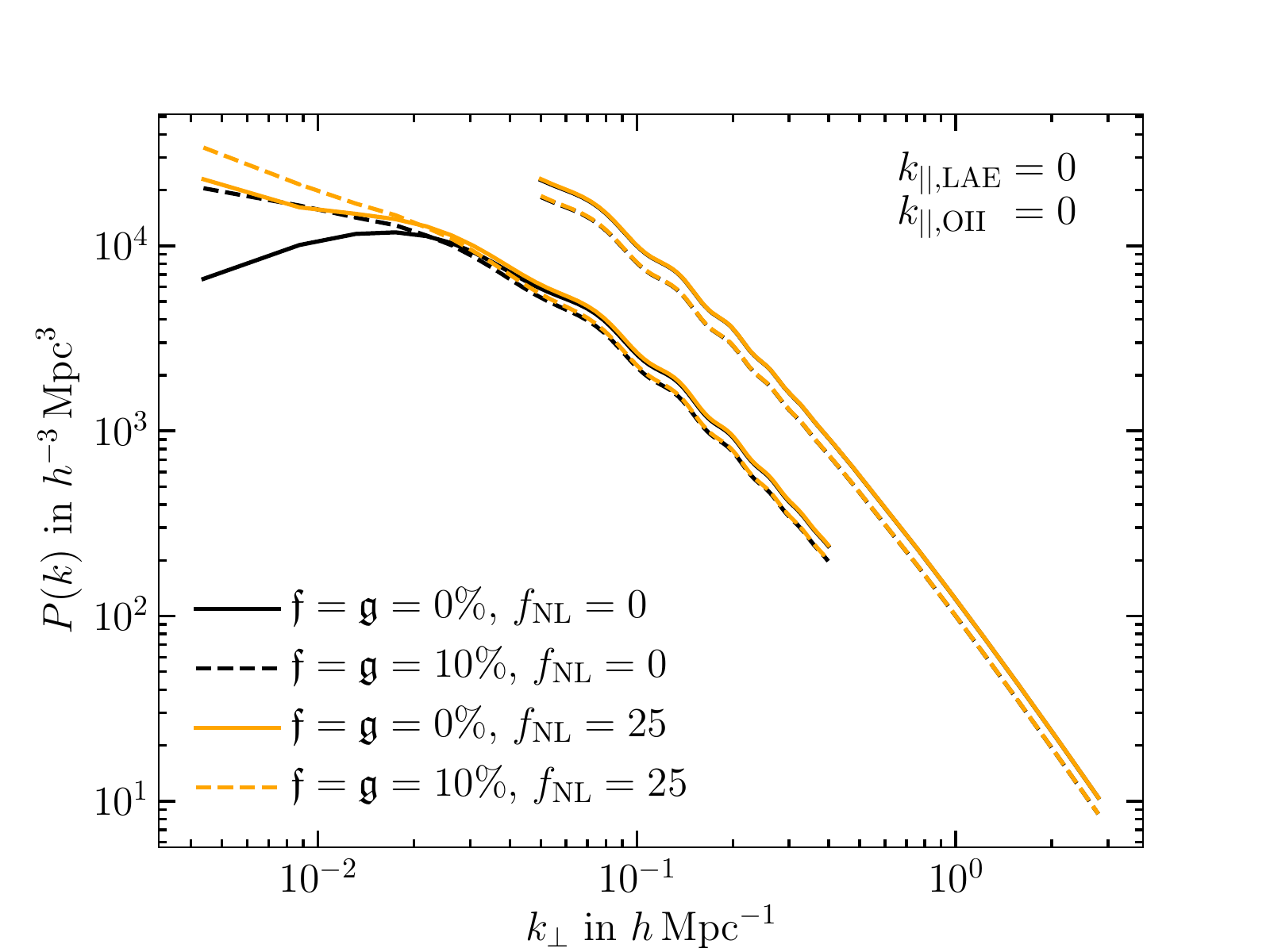}
    \incgraph[0.49]{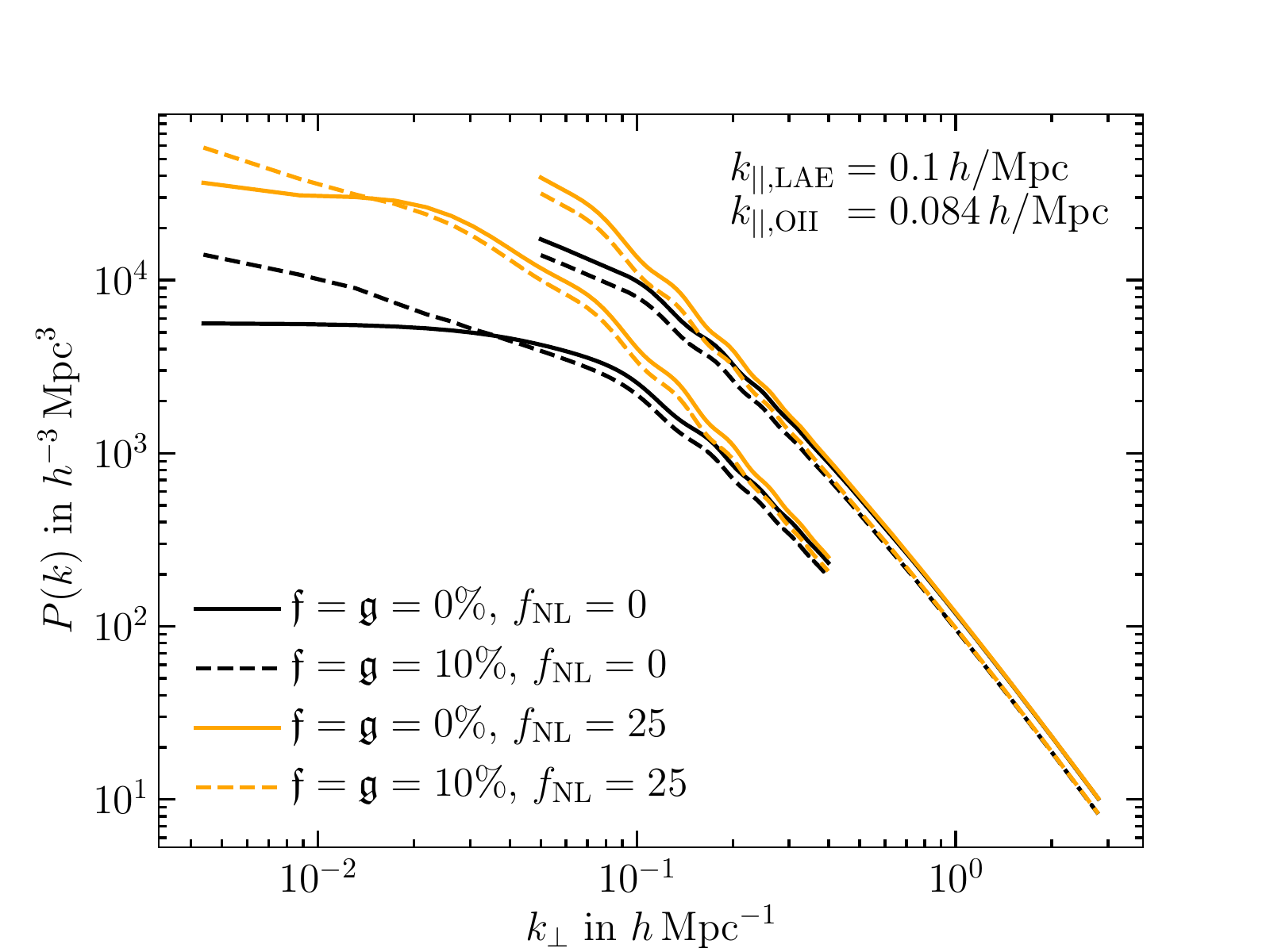}
    \caption{
        Here we show the power spectrum with and without interlopers and with
        and without primordial non-Gaussianity (parameterized by $\fnl$) for
        the HETDEX survey.
        \emph{Top:} At
        $k_{\parallel}=0$ interlopers and $\fnl$ change the power spectrum in
        similar ways. \emph{Bottom:} At $k_{\parallel}\neq0$, the
        non-Gaussianity introduces a $k$-dependence different from interlopers.
    }
    \label{fig:pk_fnl}
\end{figure}
\begin{figure}
    \centering
    \incgraph[0.49]{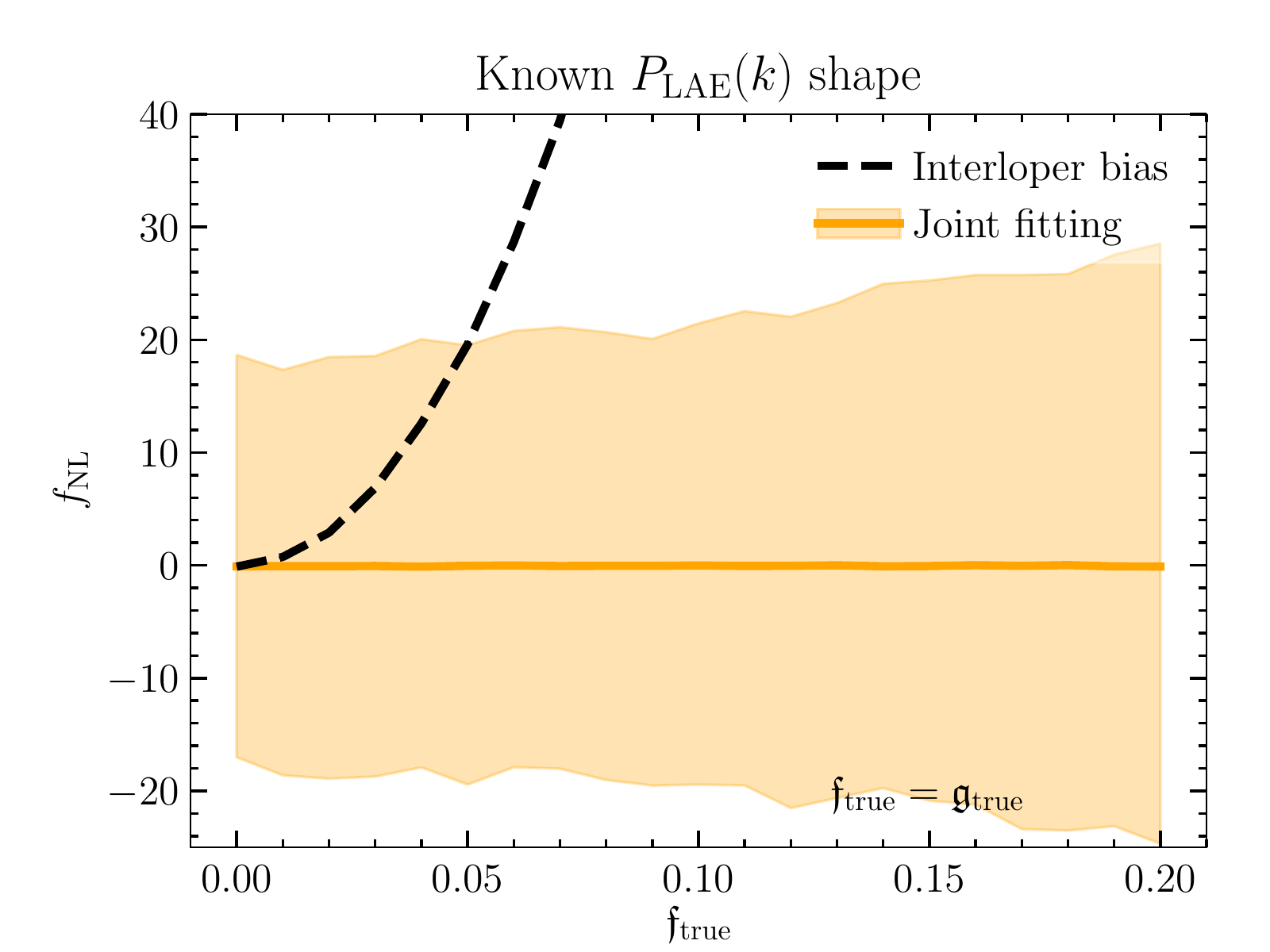}
    \caption{
        The dashed black line indicates the interloper bias in $\fnl$ when
        interlopers are ignored, while the orange line is the result of our
        joint fitting method. In both cases, the fiducial $\fnl=0$. The shaded
        area represents the $1\sigma$ (\SI{68}{\percent} C.L.) range calculated
        from the MCMC chain using the galaxy power spectrum. We use our
        \textit{Case~C} calculation (\refsec{onlylaeshape}), marginalizing over
        the interloper fractions ($\f$ and $\g$), angular diameter distance
        ($d_A$), Hubble expansion rate ($H$), amplitude of LAE power spectrum
        ($\ln A_L$), 1D OIIE power spectrum, and RSD and FoG parameters.
    }
    \label{fig:fnl_f}
\end{figure}
Inspection of the interloper effect on the monopole and quadrupole power
spectra in \reffig{pk_with_contam} suggests that at large scales the lower
redshift interlopers generically add significant power to the power spectrum of
the main higher-redshift sample. This behavior arises because the small-scale
interloper power spectrum is boosted and added to the main sample power
spectrum. Also, the scale-dependent addition to the power spectrum on large
scales is the characteristic feature of the scale-dependent bias from
primordial non-Gaussianities \citep{dalal/etal:2008,PBSreview}. In this
section, we study the effect of interlopers on measuring the primordial
non-Gaussianity parameter $\fnl$ of local type
\citep{salopek/bond:1990,komatsu/spergel:2001}.

The scale-dependent bias generated from local-type primordial non-Gaussianity
adds to the linear bias $b_x$ ($x=L$ and $x=O$ for, respectively, LAEs and
OIIEs) as
\ba
\label{eq:b1k}
\Delta b_{x}(k) = \fnl\,2\,\delta_c\,(b_x-1)\,\frac{3\Omega_m H_0^2}{2k^2 T(k) D_{\rm md}(z)}\,,
\ea
where $\fnl$ is the nonlinearity parameter, $\delta_c=1.686$ is the critical
density contrast in the spherical collapse model, $\Omega_m$ is the present-day
matter density parameter, $H_0$ is the Hubble constant, $T(k)$ is the transfer
function, and $D_{\rm md}(z)$ is the linear growth function normalized to the
scale factor $a$ during the matter-dominated epoch.

As is clear in \refeq{b1k}, for a positive nonlinearity parameter $\fnl$, the
scale-dependent bias increases the power at large scales; this behavior is
similar to the effect of interlopers. \reffig{pk_fnl} demonstrates this point
by comparing the LAE power spectrum without interlopers (solid lines) and with
\SI{10}{\percent} interlopers (dashed lines). The black lines are for $\fnl=0$
while the orange lines are for $\fnl=25$. The top and bottom panels differ in the value of $k_\parallel$ that is fixed, as indicated in the figure.
The top panel of \reffig{pk_fnl} reveals that interlopers have a
similar effect as a positive $\fnl$. The bottom panel of \reffig{pk_fnl},
however, shows that the scale-dependence from non-Gaussianities are distinct
from the scale-dependence from interlopers for different $k_{\parallel}$,
i.e., the scale-dependence of the interloper contamination has a distinctive
angular dependence from the isotropic scale-dependence of the local primordial
non-Gaussianity.

We first test the effect of primordial non-Gaussianities on the distance
measurement in \reffig{morersd}. The projected uncertainties on $R$ and $d_AH$,
after marginalizing over the nonlinearity parameter $\fnl$, slightly increase
along the $R$ direction.

The fact that the interloper effect and primordial non-Gaussianity produce
similar scale-dependencies in the galaxy power spectrum causes larger
interloper bias. \reffig{fnl_f} shows the interloper bias in $\fnl$ when
ignoring the interlopers (dashed black line) and compares it with the result
from the joint fitting method (solid orange). Thus, non-Gaussianity
can be distinguished from interlopers, and, once again, \reffig{fnl_f}
clearly demonstrates that the joint fitting method removes the interloper
bias.

\section{Statistical analysis for WFIRST}\label{sec:wfirst}
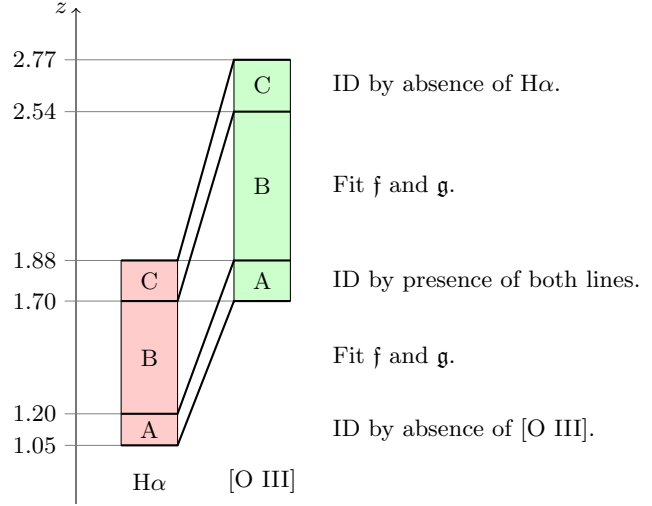
\begin{figure}
    \centering
    \begin{tikzpicture}[scale=3]
        \draw[<-] (0.3,3) node[left]{$z$} -- (0.3,0.8);
        \draw[help lines] (0.25,1.06) node[left,black]{$1.05$} -- (0.5,1.06);
        \draw[help lines] (0.25,1.2) node[left,black]{$1.20$} -- (0.5,1.2);
        \draw[help lines] (0.25,1.7) node[left,black]{$1.70$} -- (1,1.7);
        \draw[help lines] (0.25,1.88) node[left,black]{$1.88$} -- (1,1.88);
        \draw[help lines] (0.25,2.54) node[left,black]{$2.54$} -- (1,2.54);
        \draw[help lines] (0.25,2.77) node[left,black]{$2.77$} -- (1,2.77);
        \draw[fill=red!20] (0.5,1.06) rectangle (0.75,1.88);
        \draw[fill=green!20] (1,1.7) rectangle (1.25,2.77);
        \draw[thick] (0.5,1.06) -- (0.75,1.06) -- (1,1.7) -- (1.25,1.7);
        \draw[thick] (0.5,1.2) -- (0.75,1.2) -- (1,1.88) -- (1.25,1.88);
        \draw[thick] (0.5,1.7) -- (0.75,1.7) -- (1,2.54) -- (1.25,2.54);
        \draw[thick] (0.5,1.88) -- (0.75,1.88) -- (1,2.77) -- (1.25,2.77);
        \node at (0.625,0.9) {H$\alpha$};
        \node at (1.125,0.9) {[O~III]};
        \node at (0.625,1.13) {A};
        \node at (0.625,1.45) {B};
        \node at (0.625,1.79) {C};
        \node at (1.125,1.79) {A};
        \node at (1.125,2.21) {B};
        \node at (1.125,2.66) {C};
        \node[right] at (1.4,1.13) {ID by absence of [O~III].};
        \node[right] at (1.4,1.45) {Fit $\f$ and $\g$.};
        \node[right] at (1.4,1.79) {ID by presence of both lines.};
        \node[right] at (1.4,2.21) {Fit $\f$ and $\g$.};
        \node[right] at (1.4,2.66) {ID by absence of H$\alpha$.};
    \end{tikzpicture}
    \caption{
        A schematic indicating how we divide the H$\alpha$ and [O~III]
        samples to avoid overlap in our joint fitting method applied to the
        WFIRST mission. We split the spectrum into three wavelength ranges 
        A ($1.35\,{\rm \mu m}<\lambda_\obs<1.44\,{\rm\mu m}$),
        B ($1.44\,{\rm \mu m}<\lambda_\obs<1.77\,{\rm\mu m}$), and 
        C ($1.77\,{\rm \mu m}<\lambda_\obs<1.89\,{\rm\mu m}$), corresponding
        to the redshift ranges for H$\alpha$ and [O~III] as shown in the
        figure. In region A, an [O~III] line is identified by the
        simultaneous presence of H$\alpha$ in the spectrum. Thus, in region A
        the absence of a second line identifies H$\alpha$. Similarly, in
        region C we either expect to see both lines, or else it must be
        [O~III]. In region B only one of the two lines is present at a time,
        and thus we expect misidentification to be a potential problem where
        our method fitting for $\f$ and $\g$ may be needed.
    }
    \label{fig:wfirst_redshift_ranges}
\end{figure}
\begin{figure*}
    \centering
    \incgraph[0.49]{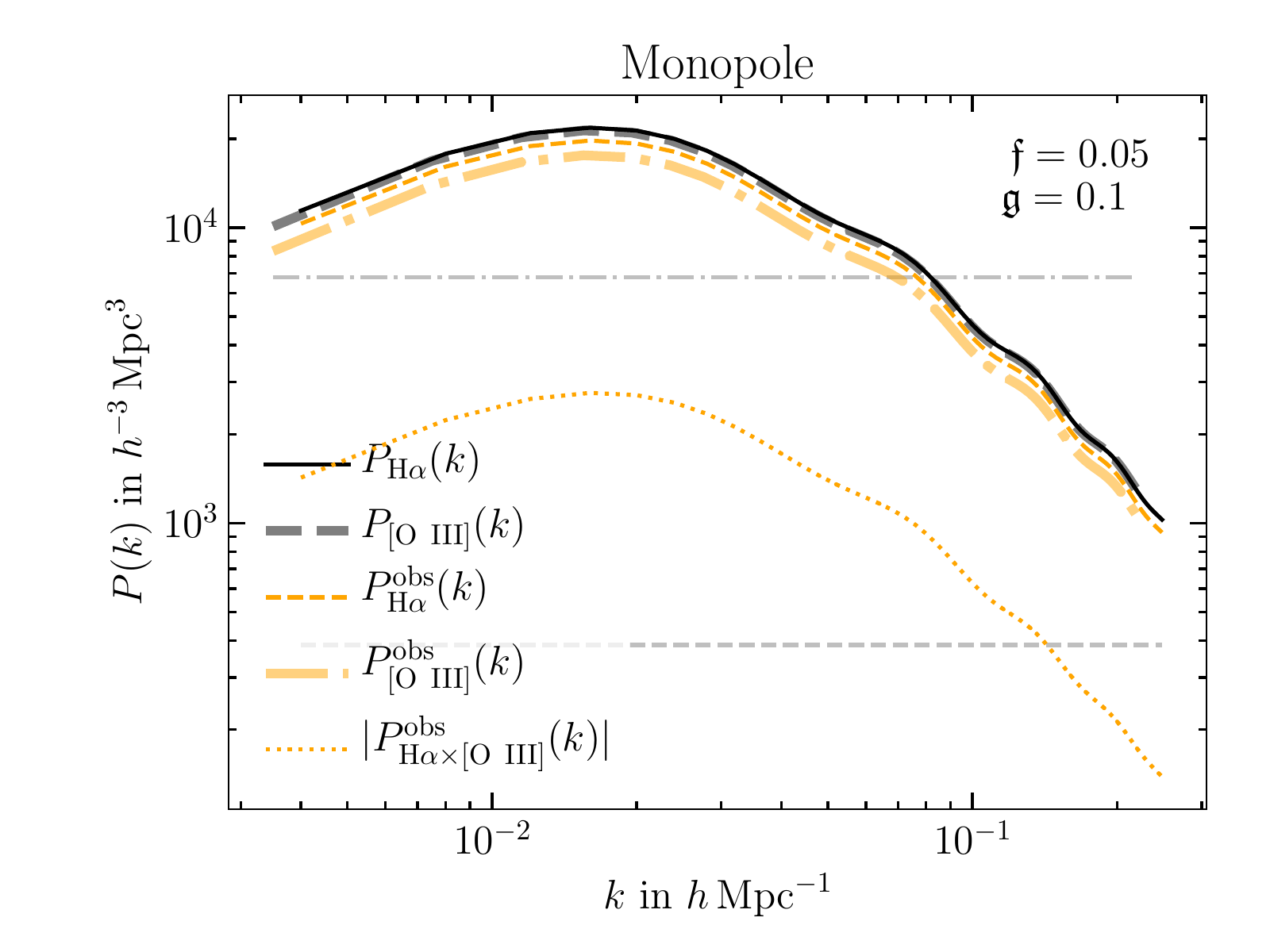}
    \incgraph[0.49]{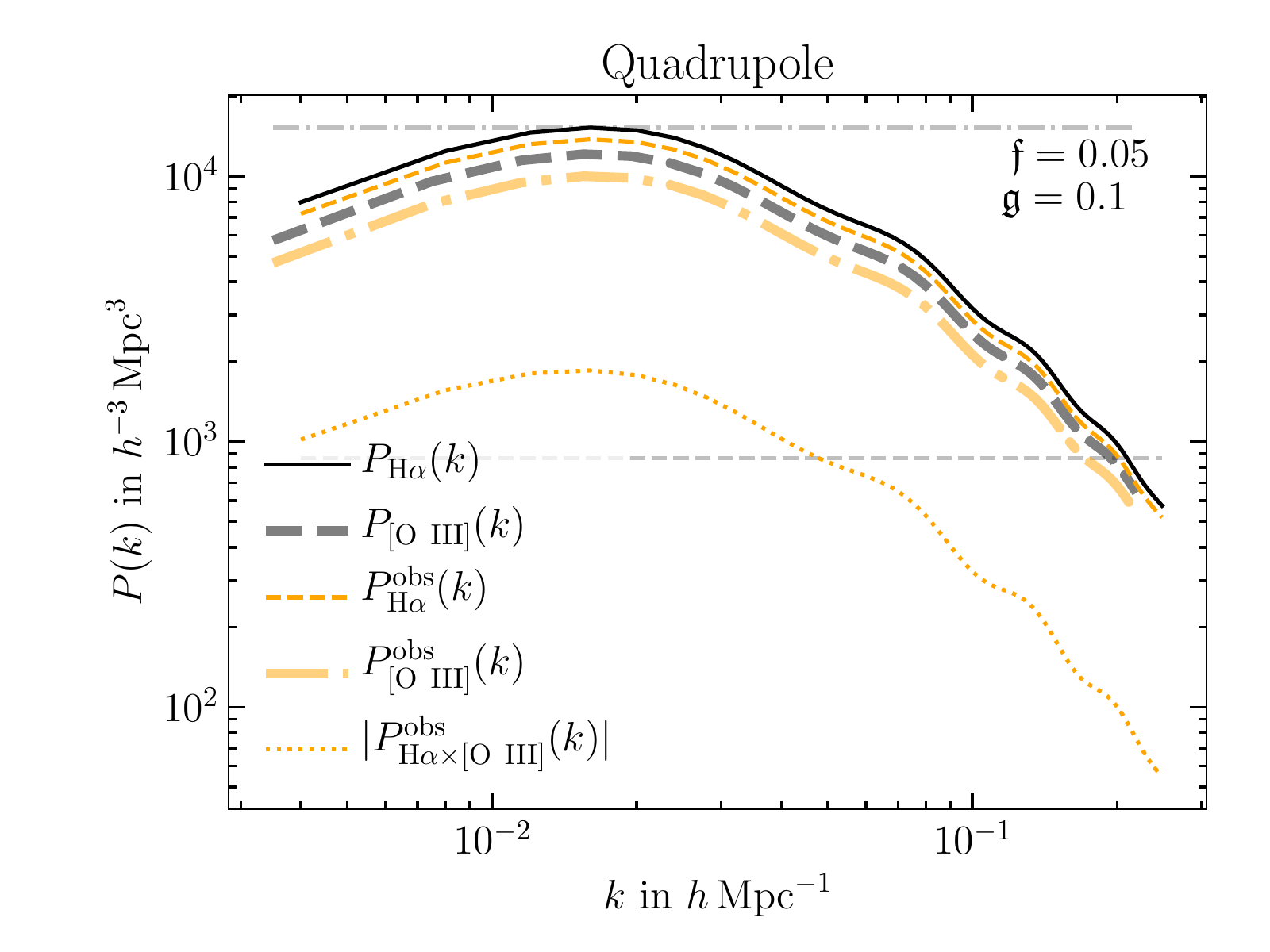}
    \caption{
        Monopole (left) and quadrupole (right) for the WFIRST galaxy samples.
        The black solid line represents the H$\alpha$ power spectrum; the thick
        grey dashed shows the [O~III] power spectrum without interlopers. The
        thin orange dashed line indicates the H$\alpha$ samples with $\f=0.05$
        interlopers, while the thick orange dash-dotted line shows [O~III] with
        $\g=0.1$ interlopers. The horizontal grey dashed line shows the shot
        noise for the H$\alpha$ sample, the dashed-dotted line is the same for
        the [O~III] sample. The orange dotted line is the cross-correlation in
        the presence of interlopers. In all cases the main effect from
        interlopers is to reduce the power by a constant factor.
    }
    \label{fig:wfirst_multipole}
\end{figure*}
In this section, we apply the joint fitting method to the planned High
Latitude Spectroscopic Survey of NASA's Wide-Field Infrared Survey Telescope
(WFIRST) mission \citep{WFIRST}. WFIRST is an emission-line galaxy survey
using slitless grism spectroscopy in the infrared wavelength range between
\SIrange[range-phrase=\textup{--},range-units=single]{1.35}{1.89}{\micro\meter} with spectral resolution of $R\equiv
\lambda/\Delta\lambda \simeq \numrange[range-phrase=\textup{--}]{620}{870}$. The total sky area coverage of the
survey is \SI{2200}{deg\squared} ($f_{\rm sky}\simeq 0.05$) for which we can
safely apply the Fourier-based analysis method in the previous section.

Focusing on the two largest emission line samples, we consider the main galaxy
sample of H$\alpha$ ($\lambda$6563\AA) emitters (HAEs) contaminated by [O~III]
($\lambda$5007\AA) emitters (OIIIEs). With the wavelength coverage of the
survey, the observed HAEs will be in the redshift range $1.05 < z < 1.88$, and
the observed OIIIEs will be at $1.70 < z < 2.77$. 

We assume that the line intensities for both H$\alpha$ and [O~III] are
strong for all emission-line galaxies at the redshift range of WFIRST. This is
motivated by \citet{Bowman/etal:2018}. Using the 3D-HST grism data
\citep{3dHST,skelton+:2014,momcheva+:2016}, they find that both lines are
strong at $z\sim2$. However, in the local universe, [O~III] is often weak.
Thus, our assumption may turn out to be an optimistic one. A proper solution
would take into account the sample of [O~II] emitters and other lines that may
be present in the spectrum. We leave this to a future investigation.

The assumption of both lines being strong implies two important
points for the line identification. First, in the overlapping redshift range
$1.7<z<1.88$, both H$\alpha$ and [O~III] lines will be observed and the HAE
sample coincides with the OIIIE sample; the presence of a second
line unambiguously identifies the sample. Second, for the lower redshift
($1.05<z<1.20$) HAEs, the interpretation is unambiguous; if
we identified the line as [O~III], the corresponding H$\alpha$ must be
detected as well. Similarly, the line identification for the higher redshift
($2.54<z<2.77$) OIIIEs is also unambiguous. 

The relation between the H$\alpha$ and [O~III] redshift bins is illustrated in
\reffig{wfirst_redshift_ranges}, where we identify three regions in the
spectrum; for the main HAE sample: $1.05<z<1.2$, $1.2<z<1.7$, and $1.7<z<1.88$;
for the interloper OIIIE sample: $1.7<z<1.88$, $1.88<z<2.54$ and $2.54<z<2.77$.
For both cases, we assume no interloper for the first and the third bins. For
the middle bins ($1.2<z<1.7$ for HAEs and $1.88<z<2.54$ for OIIIEs) we apply
our method of measuring the interloper fractions $\f$ (OIIIEs contaminating
HAEs) and $\g$ (HAEs contaminating OIIIEs) from the cross-correlation. The HAEs
and OIIIEs coincide in the overlapping region ($1.7<z<1.88$) that we analyze
only once.

For the Fourier analysis, we use the central redshifts of each of the bins to
calculate the geometrical quantities. For HAEs, the survey volume for each bin
is $V_\text{survey}^\text{lo}=\SI{0.92}{\per\h\cubed\giga\parsec\cubed}$ (for
the low-$z$ bin),
$V_\text{survey}^\text{mid}=\SI{3.87}{\per\h\cubed\giga\parsec\cubed}$ (for the
middle-$z$ bin), and
$V_\text{survey}^\text{hi}=\SI{1.53}{\per\h\cubed\giga\parsec\cubed}$ (for the
high-$z$ bin). For OIIIEs, they are
$V_\text{survey}^\text{OIIIE,lo}=\SI{1.67}{\per\h\cubed\giga\parsec\cubed}$,
$V_\text{survey}^\text{OIIIE,mid}=\SI{5.77}{\per\h\cubed\giga\parsec\cubed}$,
and
$V_\text{survey}^\text{OIIIE,hi}=\SI{2.09}{\per\h\cubed\giga\parsec\cubed}$.
For the number of samples, we use 16.4~million HAEs and 1.4~million OIIIEs and
adopt the linear galaxy bias of $b_\hae=1.5$ and $b_\oiii=2$ \citep{WFIRST}. We
spread the galaxies uniformly over the survey volume to calculate the number
densities for HAEs and OIIIEs. Just like for HETDEX, we project the OIIIEs onto
the HAE redshift for the Fourier analysis. The scaling factors in our reference
cosmology are $\alpha=0.78$, $\beta=1.10$. We use all Fourier modes below the
maximum wavenumber
$k_{\perp,\max}=k_{\parallel,\max}=\SI{0.25}{\h\per\mega\parsec}$ at the HAE
redshift. Because the HAEs and OIIIEs are at high redshifts, we assume that
both HAE and OIIIE galaxy power spectra are well modeled by a theoretical
template. This assumption corresponds to the \textit{Case~B} of
\refsec{fullshape}. As for the baseline model, we use the expression in
\refeq{pk-kaiser}. 

Finally, after the analysis for each bin, we combine the result by adding the
Fisher information matrices. The combined center redshifts are $z_{\hae}=1.47$
and $z_{\rm OIIIE}=2.32$. We count the galaxies in the overlapping range
($1.70<z<1.88$) as a part of the main HAE sample. For our baseline analysis,
we marginalize over the amplitudes of the HAE and OIIIE power spectra, the
angular diameter distances and the Hubble expansion rates at $z_\hae$ and
$z_\oiii$, the RSD parameters ($\beta_x$), and the FoG velocity dispersions
($\sigma_{v,x}^2$).

\subsection{Power spectra}
The monopole and quadrupole of the observed galaxy power spectra along with
the contribution from contaminants (for $\f=0.05$ and $\g=0.1$) are presented
in \reffig{wfirst_multipole}. The interloper contribution is quite different
from the case for HETDEX (shown in \reffig{pk_with_contam}) because the WFIRST
HAEs (the main sample) and OIIIEs (the interlopers) are at similar redshifts,
which makes the projection parameters $\alpha\sim0.78$ and $\beta\sim1.1$
close to unity.

The main effect is, therefore, to suppress the observed power spectra of HAEs
and OIIIEs by the factors of $(1-\f)^2$ and $(1-\g)^2$ [see \refeq{pkobs_lae}
and \refeq{pkobs_oii}, and replace LAE with HAE and OII with OIII]. Since the
contributions from the contaminants are suppressed by the square of the
contamination fraction and the volume factor $\alpha^2\beta=0.67$ is
of order unity, the observed power spectra are mainly affected by a change in
their amplitude. The products of the linear galaxy bias and growth factor for
the HAE and OIIIE samples are quite similar, so their uncontaminated monopole
power spectra lie nearly on top of each other.

Because the contamination effect is quite minor in the auto-power spectra of
HAEs and OIIIEs, we expect that the cross-power spectrum is the main driver
for the measurement of the contamination fractions $\f$ and $\g$ (see
\reffig{wfirst_fg} below).

\subsection{Interloper bias}
\begin{figure}
    \centering
    \incgraph[0.49]{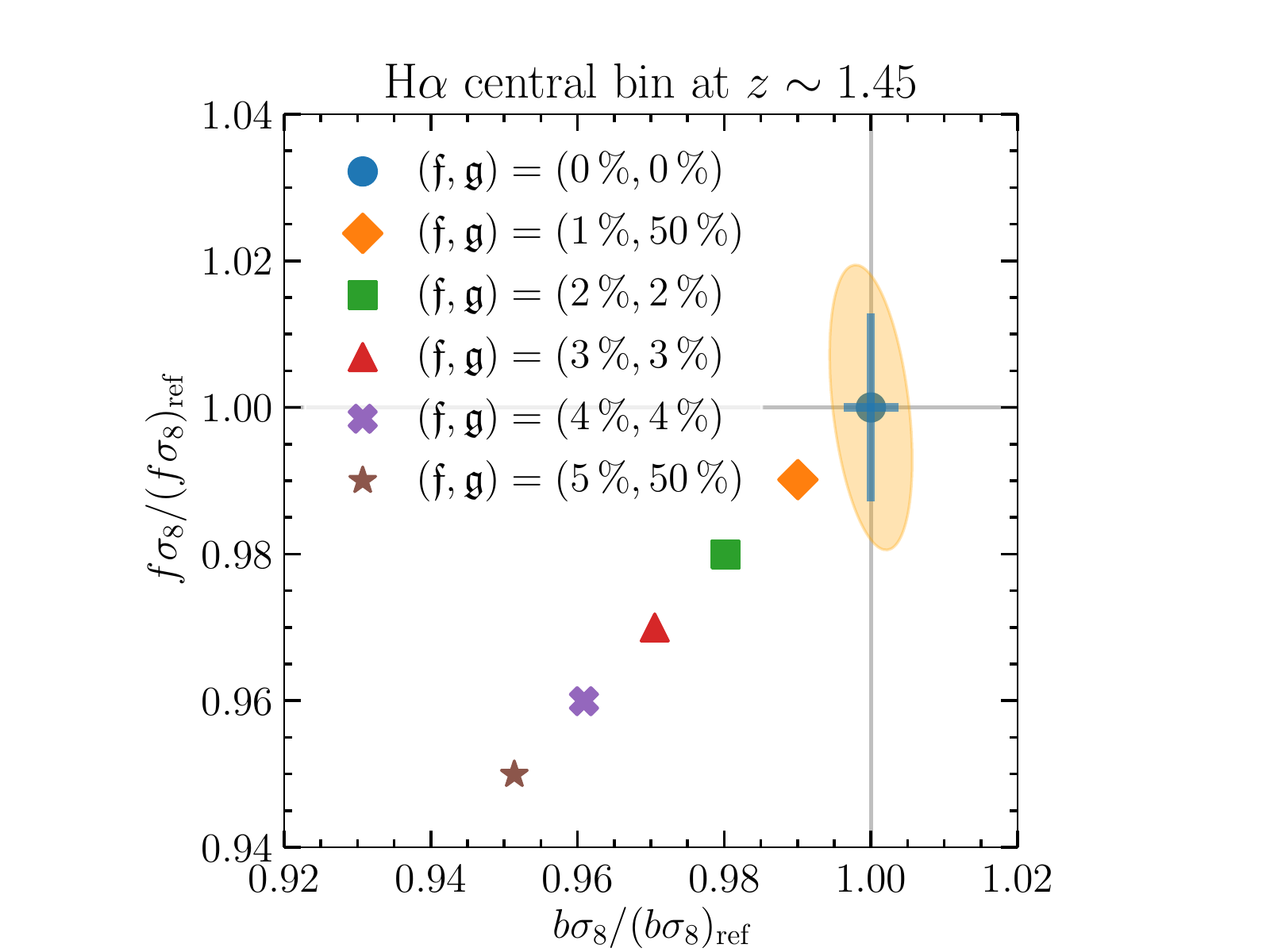}
    \incgraph[0.49]{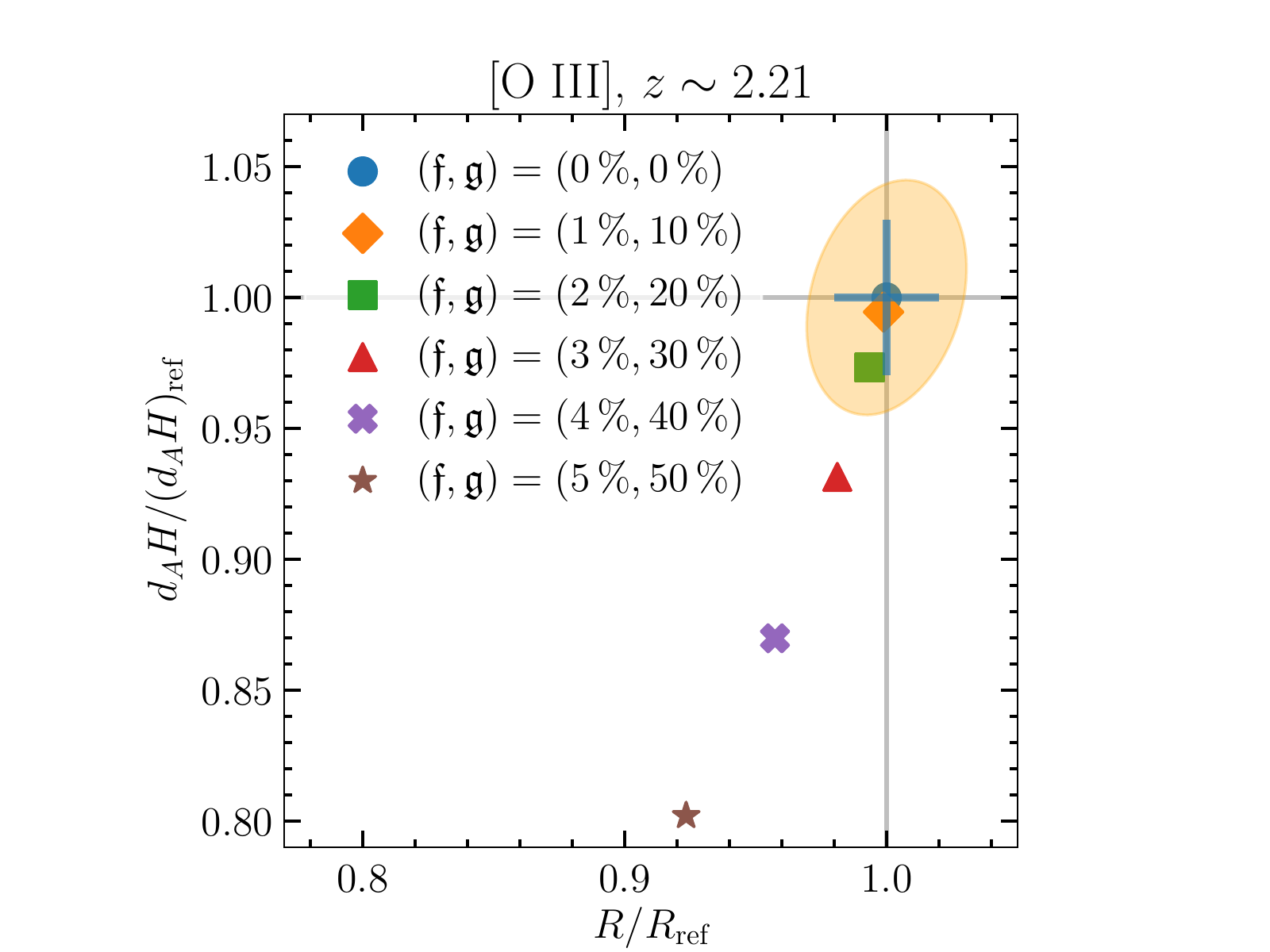}
    \caption{
        \emph{Top:} When the presence of interlopers is ignored, the
        measurements of $f\sigma_8$ and $b\sigma_8$ will be biased. This is
        primarily due to the reduction in power from the factor $(1-\f)^2$ in
        \refeq{pkobs_lae}, as applied to WFIRST. To show that the interloper
        bias does not depend on the fraction $\g$, we set
        $\g=\SI{50}{\percent}$ for the $\f=\SI{1}{\percent},\SI{5}{\percent}$
        cases.
        \emph{Bottom:} Although there is no interloper bias for the HAE
        distance measurements, the distances $d_AH$ and $R$ measured from the
        OIIIEs (at $z\sim2.21$) will be biased without a proper account of the
        HAE interloper effect.
    }
    \label{fig:wfirst_bias}
\end{figure}

We estimate the systematic changes in the maximum-likelihood value of
cosmological parameters $p_i$ due to the interloper contamination by using 
\ba
\label{eq:systematicbias}
\Delta p_i
&= \bar F^{-1}_{ij} \sum_\vk \frac{N_\vk}{2}\,
\frac{P_{,j}(\vk;\bar p_i)}{P(\vk;\bar p_i)}\,
\frac{\Delta\hat P(\vk)}{P(\vk;\bar p_i)}\,,
\ea
where $\bar F_{ij}$ is the Fisher information matrix, and $\Delta\hat P(\vk)$
is the change in the power spectrum due to interlopers. We give the derivation
of \refeq{systematicbias} in \refapp{systematicbias}. 

Let us first consider the interloper bias for the main HAE sample. From
\refeq{pkobs_lae} this is
\ba
\label{eq:wfirst_deltapk}
\Delta\hat P_{\rm HAE}(\vk)
&= -\f\,(2-\f) P_\mathrm{HAE}(\vk) + \f^2 P_\oiii^\proj(\vk)\,.
\ea
Since the OIIIE power spectrum gets projected into a smaller volume
($\alpha^2\beta \sim 0.67$) and the interloper fraction $\f$ is smaller than
$\f_{\rm lim} = \bar{N}_{\oiii}/(\bar{N}_\hae + \bar{N}_\oiii) = 0.079$ (see
\refapp{fgplane}), the second term in \refeq{wfirst_deltapk} must be negligible
compared to the first term. The main interloper effect on the HAE power
spectrum, therefore, is to change the observed amplitude. We have also
indicated this effect in \reffig{wfirst_multipole}.

Because the interloper contamination does not distort the shape of the power
spectrum, we forecast that there will be no significant interloper bias for the
measurement of the angular diameter distance and the Hubble expansion rate at
the HAE redshift.

On the other hand, both $f\sigma_8$ and $b\sigma_8$ (two direct observables
from the dynamical measurement of redshift-space distortion) would be
systematically biased if the presence of interlopers is ignored. We show this
in the top panel of \reffig{wfirst_bias} for five different values for the
interloper fractions. This figure is similar to Fig.~(4) in
\citet{pullen+2016}, but for the direct observables from the two-dimensional
galaxy power spectrum. From the Figure it is apparent that the interloper bias
in the \plane{b\sigma_8}{f\sigma_8} is quite strongly correlated, and the
correlation is due to the bias in the amplitude of the observed galaxy power
spectrum.

For the OIIIE samples, the story is quite different. Because the contamination
fraction can be as high as $g_{\rm lim} = \bar{N}_{\hae}/(\bar{N}_\hae +
\bar{N}_\oiii) = 0.92$, a small leakage of HAEs into the OIIIE sample can
generate significant interloper bias for the distance measurement. We show this
in the bottom panel of \reffig{wfirst_bias}. Here, we fix the ratio $\f/\g=0.1$
to reflect the sample size ratio between HAEs and OIIIEs.

\subsection{Joint fitting}
\begin{figure}
    \centering
    \incgraph[0.49]{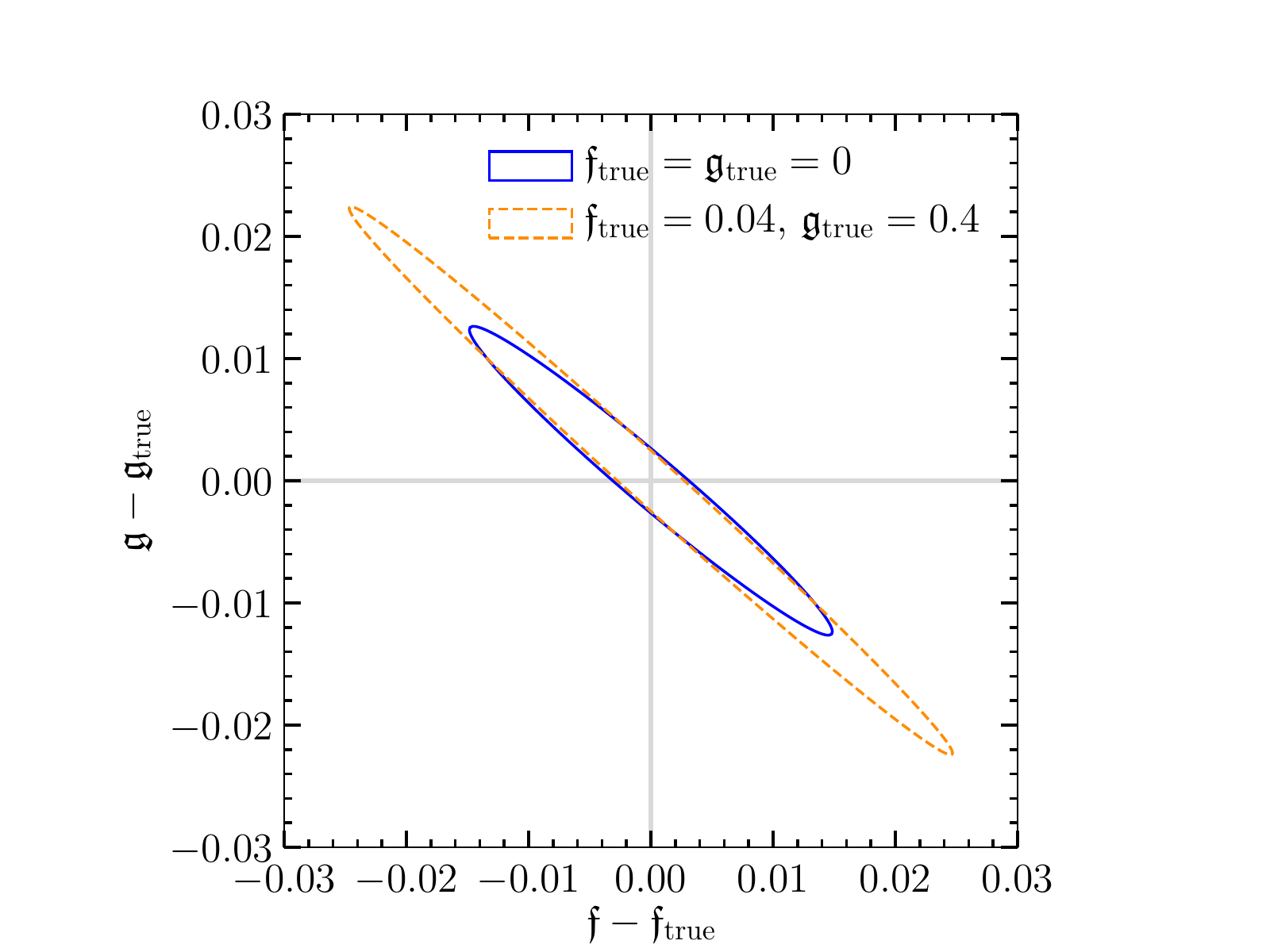}
    \caption{
        The result of the joint-analysis method: $1\sigma$ (68\% C.L.)
        constraints for WFIRST on the interloper fractions $\f$ and $\g$ for
        $\ffid=\gfid=0$ (solid blue ellipse) and
        $(\ffid,\gfid)=(\SI{4}{\percent},\, \SI{40}{\percent})$ (dashed orange
        ellipse). We assume \textit{Case~B}.
    }
    \label{fig:wfirst_fg}
\end{figure}
\begin{figure}
    \centering
    \incgraph[0.49]{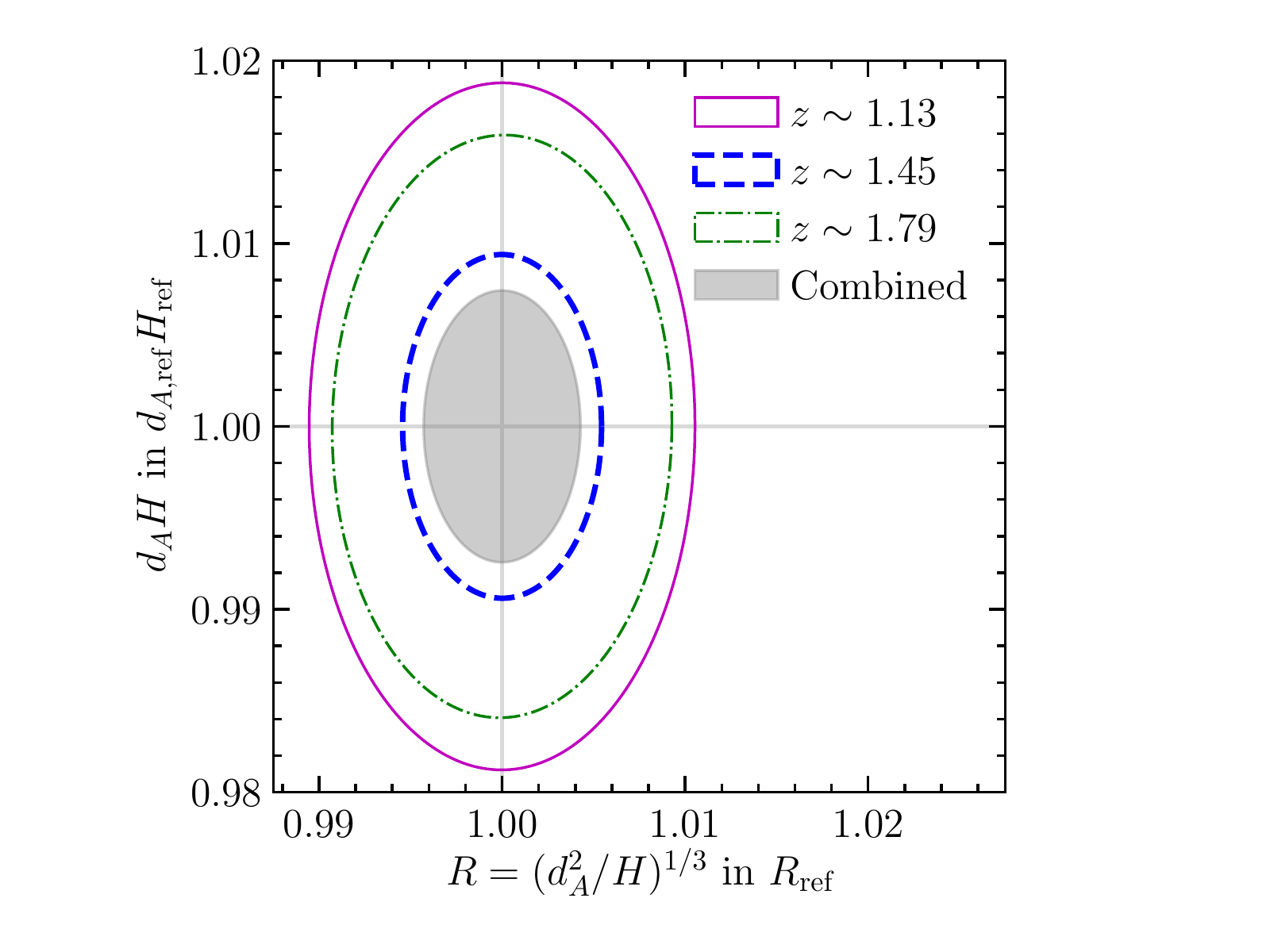}
    \caption{
        Results from the joint-analysis method: projected $1\sigma$ (68\%
        C.L.) range for $R$ and $d_A H$ for WFIRST, assuming
        \textit{Case~B}. The solid magenta ellipse shows the confidence
        interval from the lowest redshift bin assuming no misidentification,
        the thick dashed blue ellipse from the center redshift bin
        marginalizing over interloper fractions, and the dash-dotted green
        ellipse shows the confidence interval from the highest redshift bin
        again assuming no misidentification. The grey shaded ellipse shows the
        combined constraint from all three bins. Here, $\ffid=\gfid=0$.
    }
    \label{fig:wfirst_RdAH}
\end{figure}
\begin{figure*}[htb]
    \centering
    \incgraph[0.32]{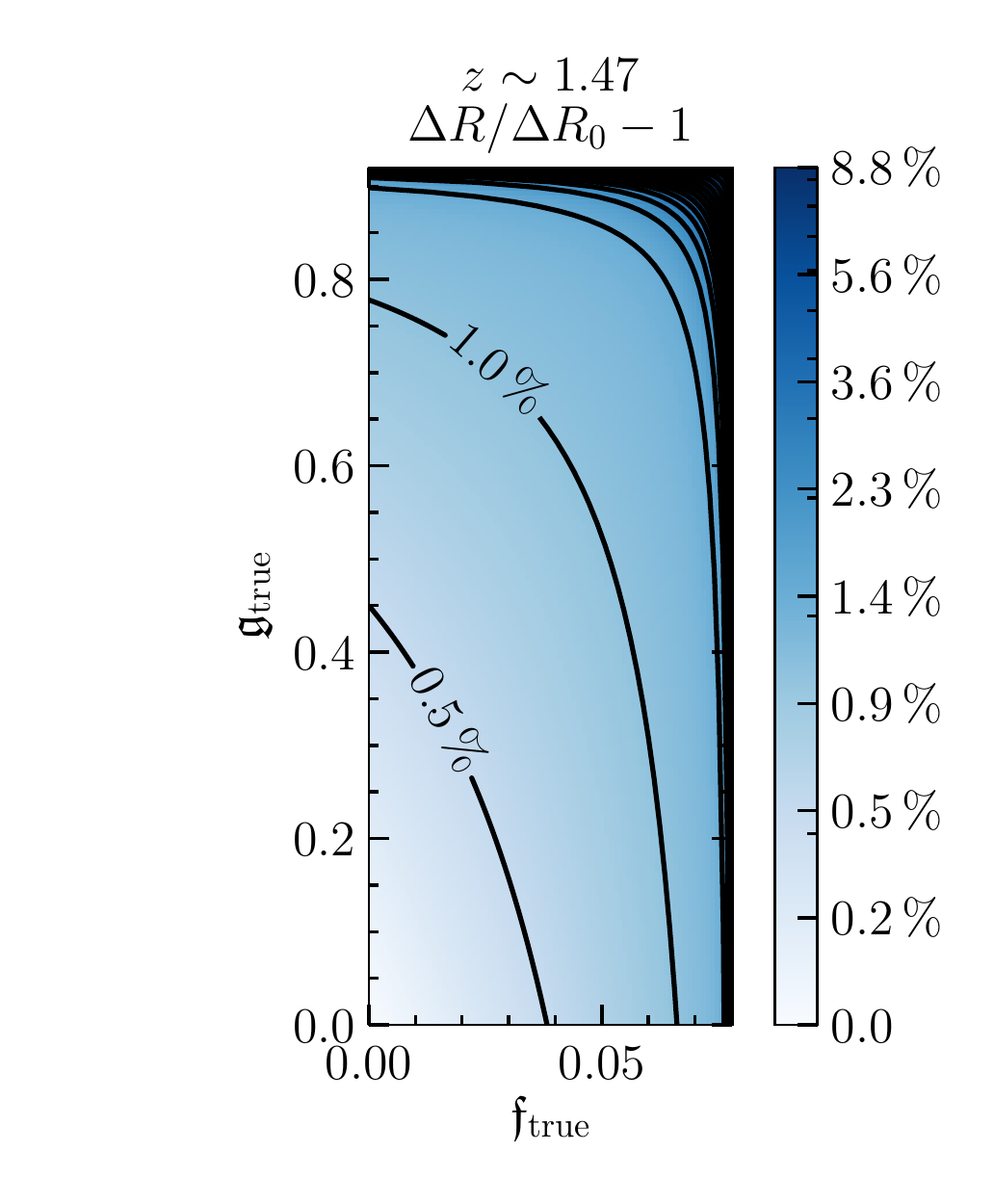}
    \incgraph[0.32]{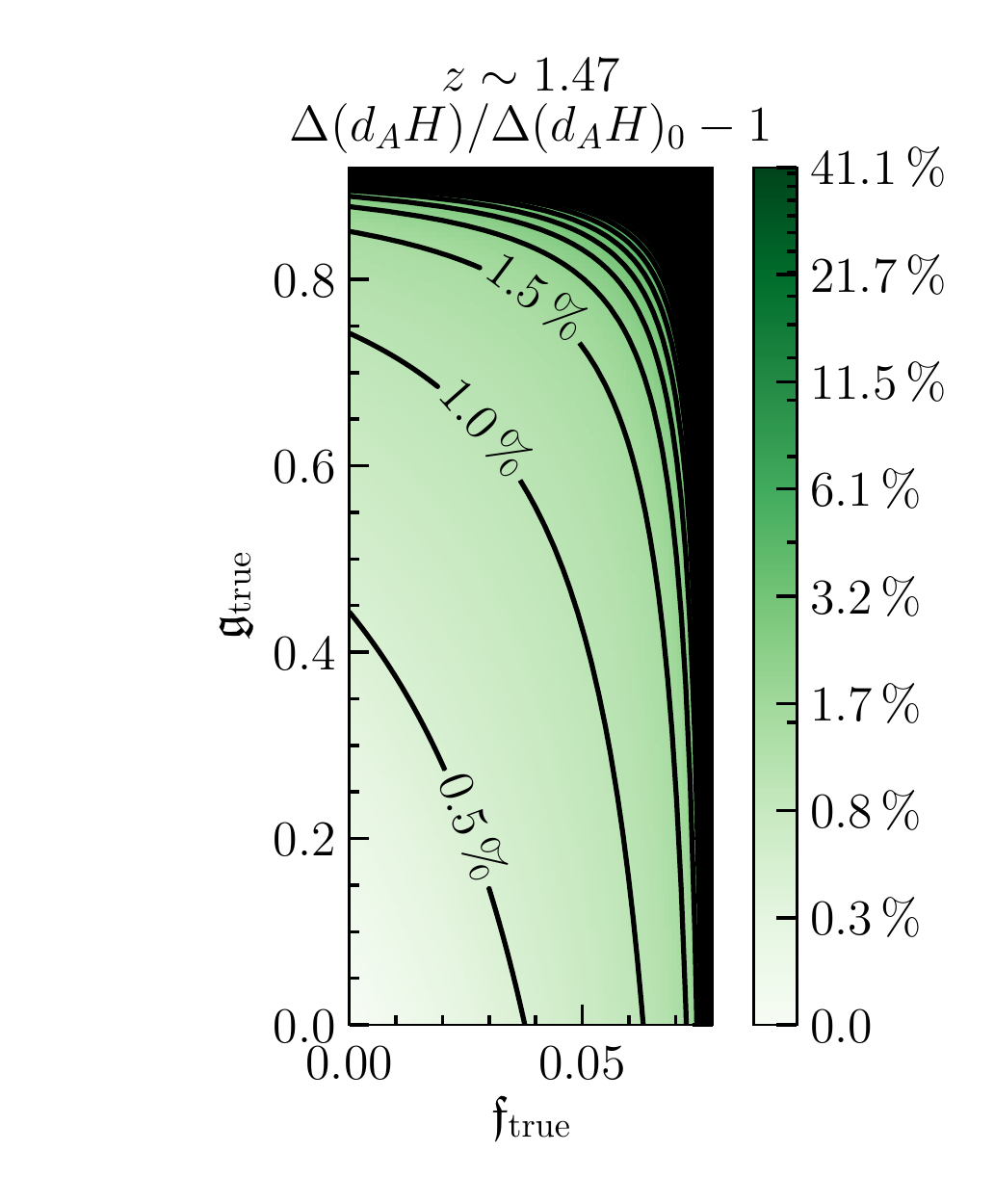}
    \incgraph[0.32]{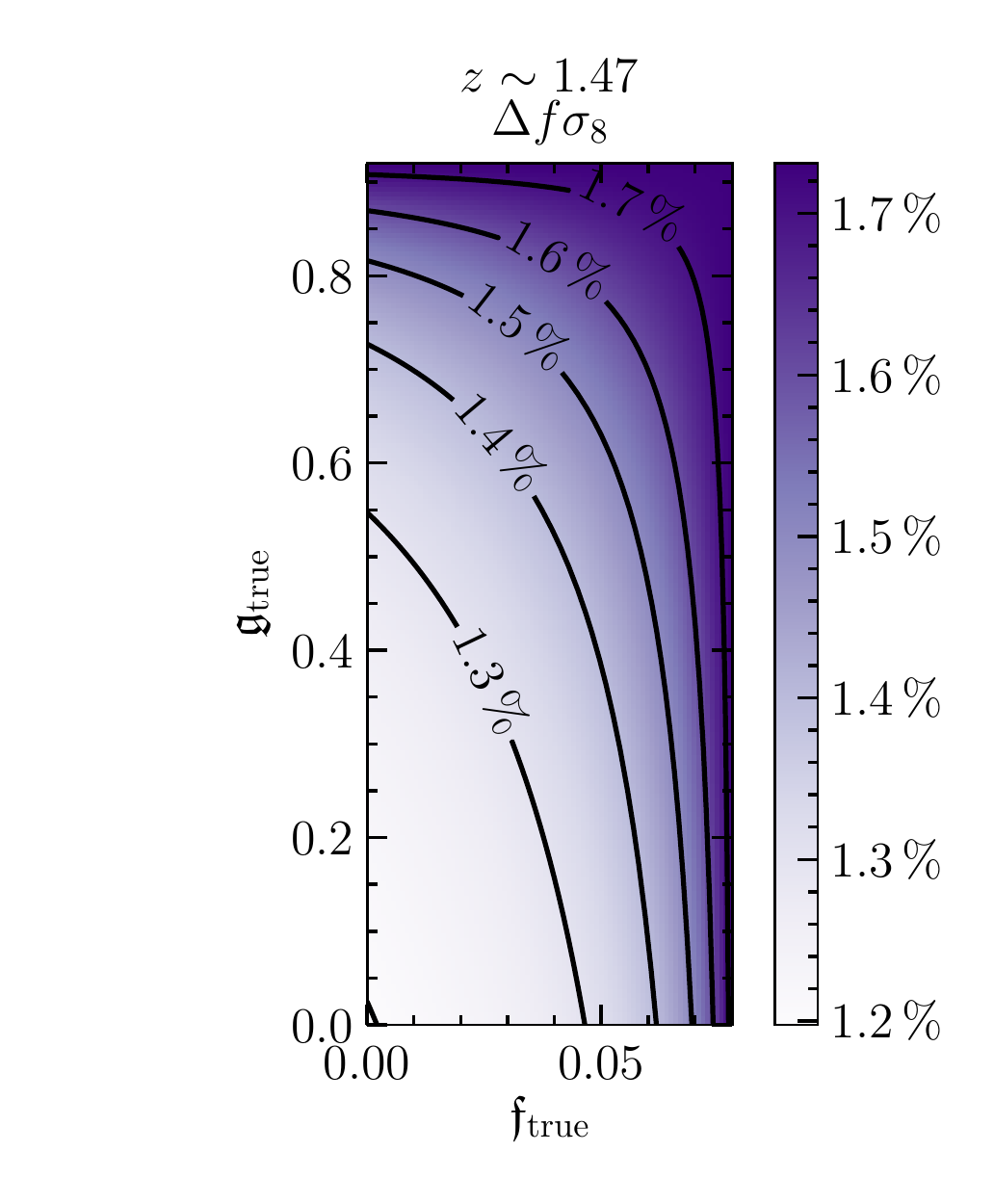}
    \caption{\emph{Left:} Results from the joint-analysis method: 
        the forecast for the {\it relative change} in the $1\sigma$ (68\% C.L.)
        uncertainty on $R$ at the H$\alpha$ redshift as a function of the true
        $\f$ and $\g$ for WFIRST. Since the expected ratio of HAEs to OIIIEs is
        \num{\sim12}, the limiting interloper fractions are
        $\f_\mathrm{lim}=0.079$ and $\g_\mathrm{lim}=0.92$. Note that for
        easier display the abscissa is on a larger scale than the ordinate. We
        forecast the best constraint to be $\Delta R_0=0.28\%$ at $\f=\g=0$.
        The forecast measurement uncertainty $\Delta R$ changes by less than a
        percent over most of the plot.
        \emph{Center:} Similar to the left plot, but for the change in the
        uncertainty on the $AP$ parameter $d_AH$, starting with
        $\Delta(d_AH)_0=\SI{0.49}{\percent}$.
        \emph{Right:} Here we show the expected constraints on $f\sigma_8$, the
        best constraints being $\Delta(f\sigma_8)_0=\SI{1.2}{\percent}$ at
        $\f=\g=0$.
        In alls plots we assume \textit{Case~B}.
    }
    \label{fig:wfirst_delta_Ha}
\end{figure*}
\begin{figure*}[htb]
    \centering
    \incgraph[0.32]{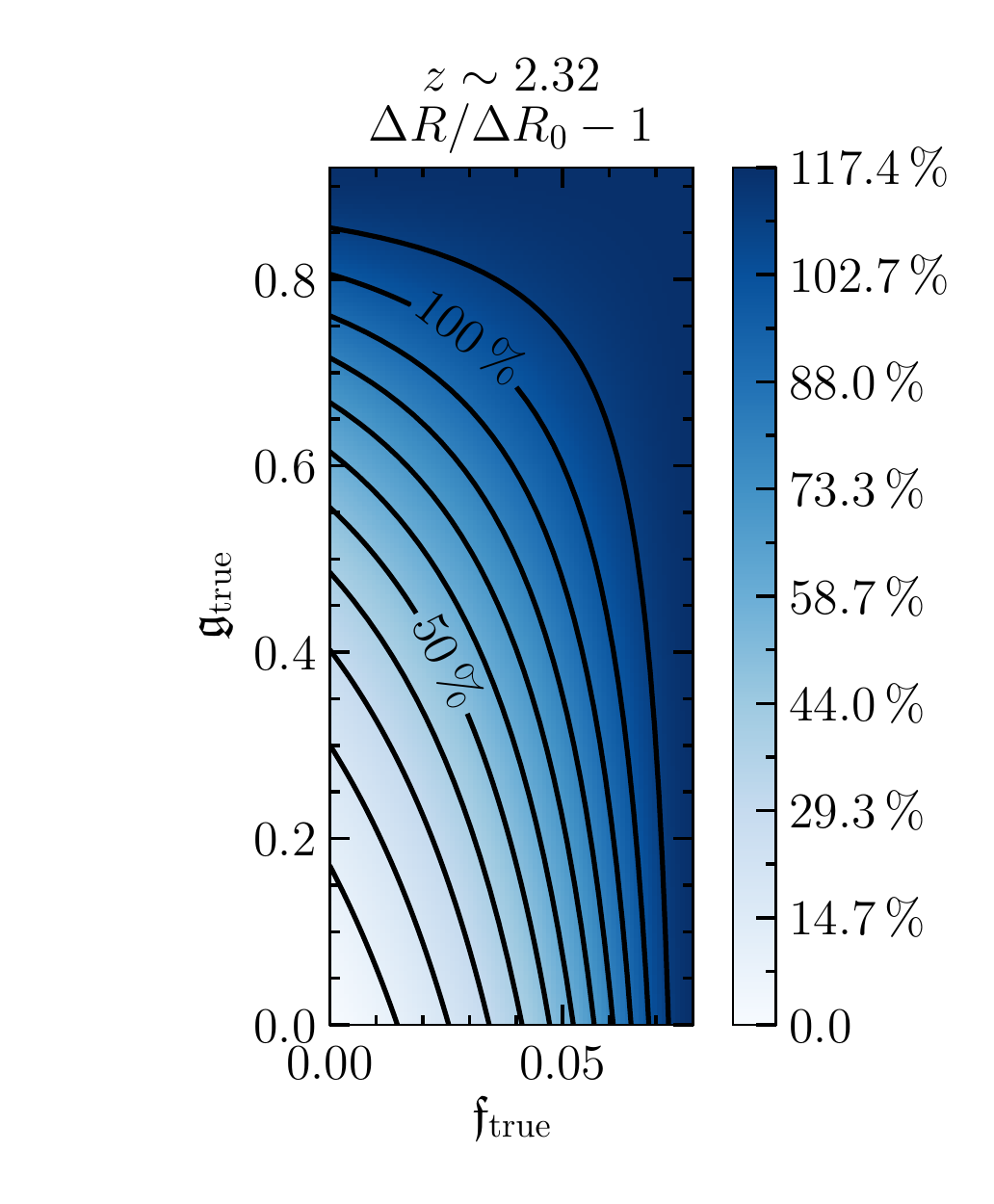}
    \incgraph[0.32]{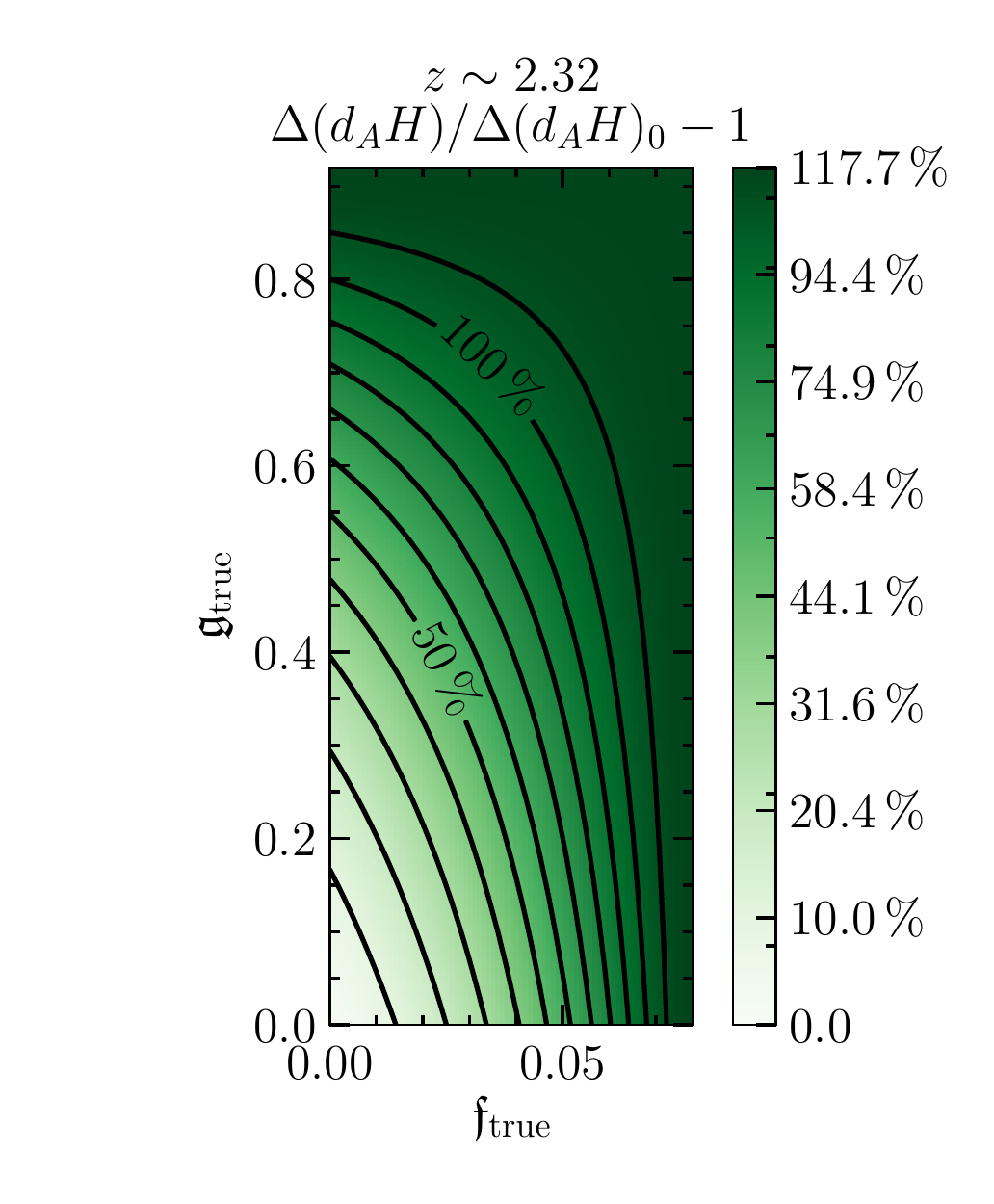}
    \incgraph[0.32]{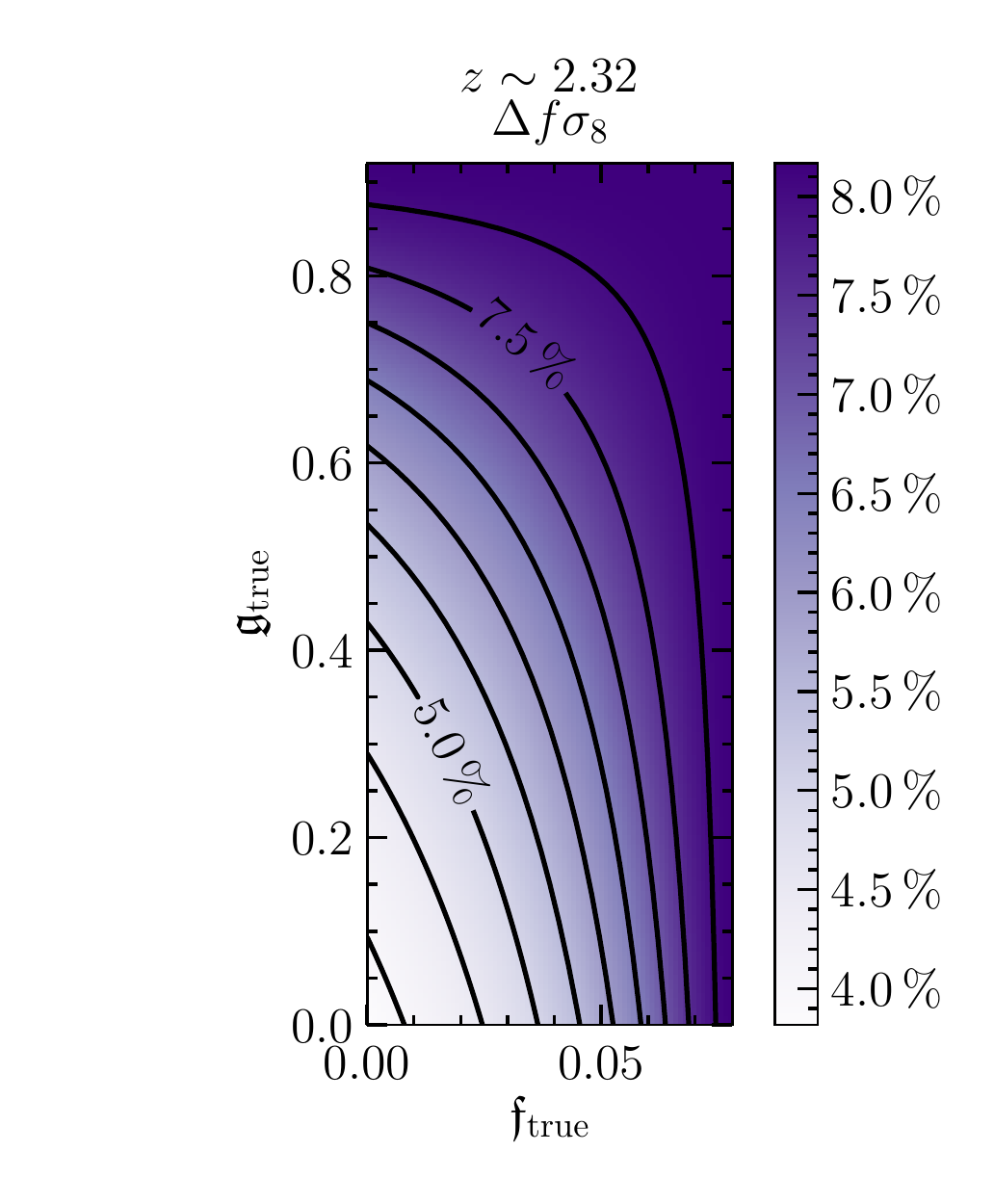}
    \caption{Similar to \reffig{wfirst_delta_Ha} here we show
        our forecast for the relative change in the uncertainty on $R$ (left)
        and $dA_H$ (center) as a function of the true $\f$ and $\g$ for
        WFIRST, and the uncertainty on $f\sigma_8$ (right) at the redshift
        of the [O~III] emitters. We forecast the best constraint to be
        $\Delta R_0=0.89\%$ at $\f=\g=0$, 
        $\Delta(d_AH)_0=\SI{1.8}{\percent}$, and
        $\Delta(f\sigma_8)_0=\SI{3.82}{\percent}$.
        In alls plots we assume \textit{Case~B}.
    }
    \label{fig:wfirst_delta_OIII}
\end{figure*}
\begin{figure}
    \centering
    \incgraph[0.32]{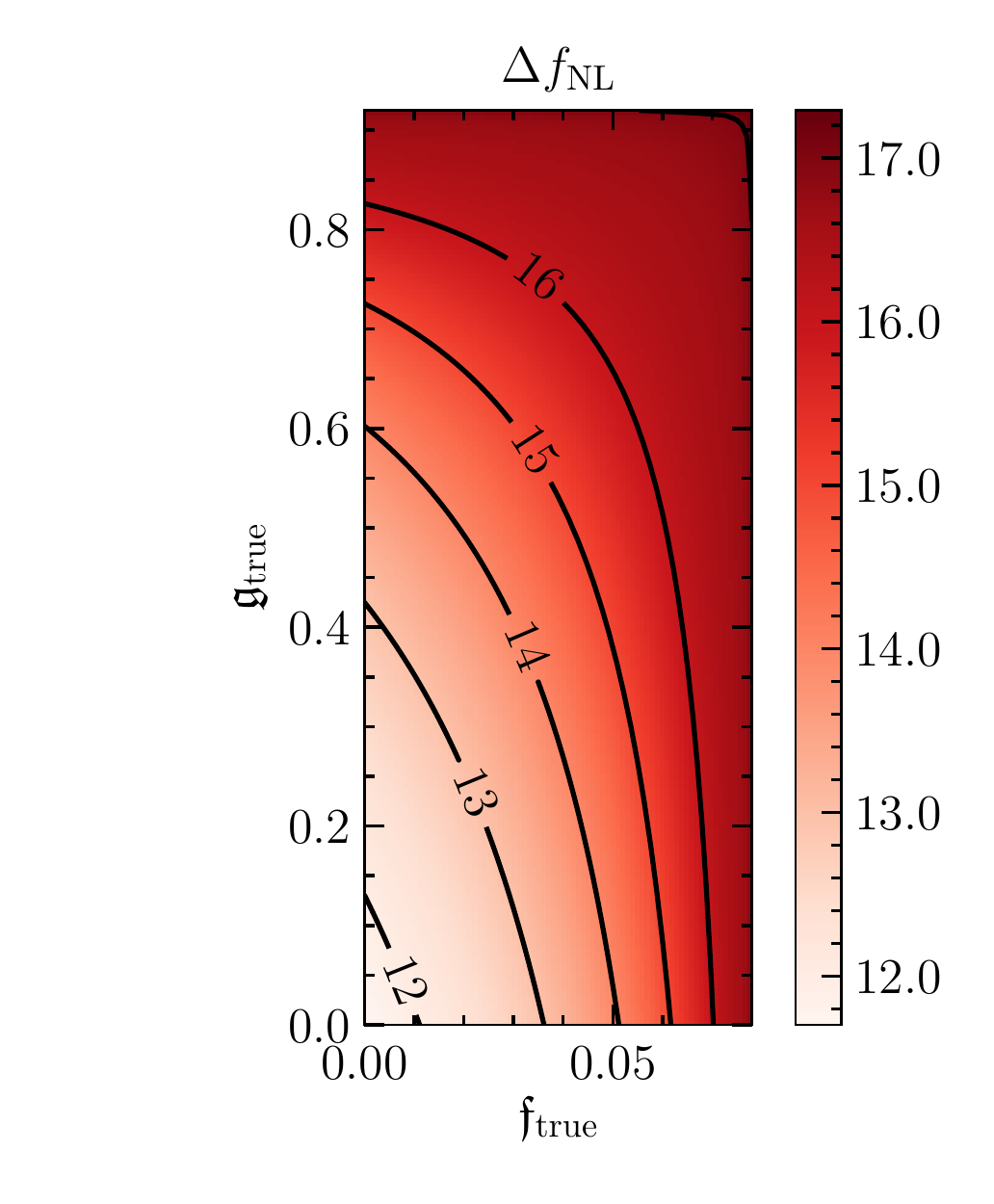}
    \caption{
        Results from the joint-analysis method: $1\sigma$ (68\% C.L.) range on
        $\fnl$ as a function of interloper fractions ($\ffid$, $\gfid$). To
        constrain $\fnl$ we only consider the galaxy power spectrum, combining
        five bins: the three H$\alpha$ bins, and the upper two [O~III] bins.
        We assume \textit{Case~B}.
    }
    \label{fig:wfirst_fnl}
\end{figure}
In this section, we apply the joint analysis technique to WFIRST. That is, we
use all three power spectra (HAE power spectrum, OIIIE power spectrum and
HAE-OIIIE cross power spectrum) as observables to estimate the cosmological
parameters, and show that the parameters estimated from the joint analysis are
unbiased.

First, \reffig{wfirst_fg} shows the projected constraints for the interloper
fractions $\f$ and $\g$ for two sets of interloper fractions. Compared
to HETDEX (\reffig{fg_bymodel}), the measurement uncertainty on the interloper
fractions $\f$ and $\g$ is larger and they are more highly correlated. This is
explained by the fact that the two samples are closer in redshift, and so the
two power spectra have less distinct signals in the cross-correlation. This
plot shows that we can measure a percent level interloper fraction from the
cross power spectrum of WFIRST .

With our joint-analysis method the interloper bias on $f\sigma_8$ and
the distance measures shown in \reffig{wfirst_bias} is removed. For the
H$\alpha$ sample we forecast \SIrange{\sim1.5}{2}{\percent} constraints, and
for the [O~III] sample \SIrange{\sim4.5}{7}{\percent} for interloper fraction
$\gfid\lesssim\SI{30}{\percent}$, and higher for larger interloper fractions.

In \reffig{wfirst_RdAH} we show the projected constraints on $R$ and $d_AH$ for
all three bins when the true interloper fractions vanish. The bins are
labeled in the legend by their central redshifts. We marginalize over the
amplitude, RSD, and FoG parameters for all three bins. For the central bin we
additionally marginalize over $\f$, $\g$, and the amplitude, RSD, and FoG
parameter of the corresponding OIIIE sample. The constraints largely reflect
the size of the survey volume. The weakest constraints come from the bin with
the smallest volume, the most tight constraints from the bin with the largest
volume. The grey shaded ellipse shows the constraints combining all three bins
assuming they are statistically independent. Combined, we get
\SI{\sim0.28}{\percent} uncertainty on $R$. This is more optimistic than
\citet{WFIRST} since we model the full shape of the power spectrum, including
the broadband shape.

On the left of \reffig{wfirst_delta_Ha} we show how the projected uncertainty
on $R$ measured from the HAE sample relative to the best case (without
contamination, at $\f=\g=0$) changes as a function of the contamination
fractions $\f$ and $\g$. Since the fiducial ratio of OIIIEs to HAEs is
\num{\sim12}, the limiting interloper fractions are
$(\f_\mathrm{lim},\g_\mathrm{lim})=(0.079,0.92)$. Note that to better show the
lower allowed region, the abscissa has been stretched compared to the
ordinate. We neglect the upper allowed region that corresponds to catastrophic
misidentification. At $(\ffid,\gfid)=(0,0)$ the constraint is $\Delta
R_0=\SI{0.28}{\percent}$. The constraint changes by less than \SI{1}{\percent}
for most of the lower allowed region before increasing rapidly near the
limiting fractions. Similarly, the center panel of \reffig{wfirst_delta_Ha}
shows the change in the projected uncertainties for the $AP$ parameter $d_AH$
as a function of $\f$ and $\g$. Here, the best constraint is
$\Delta(d_AH)_0=\SI{0.49}{\percent}$. The right panel of
\reffig{wfirst_delta_Ha} shows the uncertainty on $f\sigma_8$ for the HAE
sample. The best constraint we get with marginalization over the interloper
fractions is \SI{\sim1.2}{\percent}.

\reffig{wfirst_delta_OIII} shows the same combining the upper two [O~III]
bins. Note that we do not include the lowest [O~III] bin, as we assume that
the OIIIEs of the lowest redshift bin are already included with the HAEs in
their highest redshift bin. On the left of the figure, we show the change of
the uncertainty relative to the best case $\Delta R_0=\SI{0.89}{\percent}$, in
the center, the change relative to $\Delta(d_AH)_0=\SI{1.8}{\percent}$, and on
the right the uncertainty on $f\sigma_8$, which we forecast to be $\Delta
f\sigma_8\sim\SI{3.8}{\percent}$ in the ideal case.

Finally, in \reffig{wfirst_fnl} we show forecasts for the non-Gaussianity
parameter $\fnl$. We only consider the constraint from the galaxy power
spectrum, combining the information from both HAEs and OIIIEs. From the galaxy
power spectrum alone, we forecast
$\Delta\fnl\sim\numrange[range-phrase=\textup{--}]{12}{16}$ for WFIRST after
marginalizing over all nuisance parameters, with a best-case $\Delta
\fnl=11.7$ at zero interloper fractions.

\section{Conclusion}\label{sec:conclusion}

In this paper, we study the effects of interlopers in the cosmological analysis
based on the power spectrum measured from emission-line galaxy surveys. In
particular, the paper focuses on the two relatively narrow field surveys HETDEX
and WFIRST.

For HETDEX, we define the interloper fraction $\f$ of the Lyman-$\alpha$
emitter (LAE) sample and the interloper fraction $\g$ of the [O~II] emitter
(OIIE) sample in \refeqs{f}{g}. For WFIRST, we define $\f$ to be the interloper
fraction in the H$\alpha$ sample, and $\g$ the interloper fraction in the
[O~III] sample. We then derive the effect of interlopers on the power spectrum
in \refeq{pkobs_lae}, \refeq{pkobs_oii}, and \refeq{pkobs_LO}. The change in
the power spectrum is given in terms of two geometrical factors defined in
\refeq{scaling_parameters}: the direction perpendicular to the line-of-sight
gets scaled by the factor $\alpha$, while the direction parallel scales by the
factor $\beta$. For the two surveys that we study, $\alpha\neq\beta$, and thus
the projection introduces anisotropies in the two-dimensional galaxy power
spectrum, in addition to redshift-space distortions. The volume factor
$\alpha^2\beta$ also multiplies the contamination contribution from the
interloper power spectrum. In \refapp{shotnoise}, we also provide the rigorous
derivation of the shot noise under the assumption that the galaxies are a
Poisson sample of the underlying continuous galaxy density field, and the
interloper fraction plays the role of the probability of having contamination.

We then investigate the joint-analysis method including auto-power spectra of
both samples as well as the cross-power spectrum as observables. We show that
the joint analysis yields robust measurements of the interloper fractions and
it removes the \textit{interloper bias}, a systematic shift of the best-fitting
cosmological parameters when ignoring the interlopers. Although measuring and
marginalizing the interloper fractions increases the measurement uncertainties
in cosmological parameters, it does not bias their maximum likelihood values.
We explicitly show this for the geometrical parameters (angular diameter
distance and Hubble expansion parameter) as well as the dynamical parameters
(linear growth rate and $\sigma_8$), higher-order RSD parameters
(\refsec{morersd}), and non-Gaussianity (\refsec{fnl}).

For the joint-analysis, we investigate several models for the power spectra of
the main survey galaxies and the interlopers.
We consider four cases:
\begin{description}
    \item[\textit{Case~A}] This case makes minimal assumptions, only assuming
        that the true cross-correlation vanishes, see
        \refsec{model_pkcrossonly}.
    \item[\textit{Case~B}] This case makes maximal assumptions, assuming that
        the redshift-space distortions and the shapes of the 1D isotropic
        power spectra are well-modeled by theory, see \refsec{fullshape}.
    \item[\textit{Case~C}] Similar to \textit{Case~B}, but we only asssume to
        be able to model the isotropic auto-correlation power spectrum of the
        higher-redshift sample, see \refsec{onlylaeshape}.
    \item[\textit{Case~D}] Here we assume we can model only the redshift-space
        distortions of the two samples, see \refsec{noknowledge}.
\end{description}
By doing a Fisher analysis, we show that the constraints on the interloper
fractions $\f$ and $\g$ are essentially the same for the two extreme cases
\textit{A} and \textit{B} (see \reffig{fg_bymodel}), confirming that the
information comes primarily from the cross-correlation.

Naturally, there is a large continuum of intermediate cases, and it needs to be
assessed on a survey-by-survey basis which one to use.
For HETDEX, while the
main LAE samples probe the quasi-linear scales at high redshift, the
interlopers are at low redshift and probe scales deep into the nonlinear
regime. Thus, we assume \textit{Case~C} as the baseline, where we marginalize
over the 1D OIIE power spectrum. For WFIRST, we take \textit{Case~B} as the
baseline, because both the main sample and interloper sample probe the
quasi-linear scales at high redshifts.

This paper shows that the better the line classification, the tighter
we can constrain the cosmological parameters. For the astrophysical methods of
line classification, our joint-analysis method provides an estimate of the
total contamination fractions. The usual approach to compliment the emission
line surveys is to have follow-up imaging data in the same survey footprint.
For HETDEX, \citet{leung+2017} developed a Bayesian framework taking into
account the equivalent width distribution of LAEs and OIIEs, and searching for
other lines that may be present in the spectrum. For WFIRST,
\citet{pullen+2016} investigated the use of sensitive photometric data. All of
these methods work for the classification of individual emission-line galaxies.
The estimated interloper fractions from the joint analysis then provide a
global figure of merit, based on which we can modify the line identification
criteria to reach smaller interloper fractions. The interplay between the two
methods will provide a way to optimize the analysis pipeline for the
emission-line galaxy surveys.

Although we have not investigated further, one can also incorporate
cross-correlations with external datasets. For example, for HETDEX,
low-redshift galaxy samples from SDSS \citep{eisenstein+:2011} that correlate
with the OIIE sample should provide an extra constraint on the contamination
fraction $\g$ when cross-correlating the low-redshift galaxies with the
high-redshift LAEs.

Although we have not investigated further, one can also incorporate
cross-correlations with external datasets. For example, for HETDEX,
low-redshift galaxy samples from SDSS (Eisenstein et al. 2011), or radio
catalogs from LOFAR \citep{shimwell+:2019}, or APERTIF \citep{adams+:2018}
that correlate with the OIIE sample should provide an extra constraint on the
contamination fraction g when cross-correlating the low-redshift galaxies with
the high-redshift LAEs.

The results of this paper required two key simplifying assumptions: a flat-sky
approximation and no redshift evolution in the interloper fractions, the galaxy
bias, or the linear growth rate. To address these caveats in the future and to
apply the joint-analysis method to wide-angle galaxy surveys including Euclid
and SPHEREx, we must extend the method to spherical harmonic space while
incorporating astrophysically motivated assumptions about the redshift
evolution of key parameters.  For example, \citet{leung+2017} predict the
interloper fractions, $\f$ and $\g$, as a function of redshift for the LAEs and
OIIEs, which could be incorporated in the future.

\acknowledgements

The authors thank Charles Bennett and Graeme Addison for useful discussion and
comments. We thank the anonymous referee for useful comments that
helped to clarify the paper. This work was supported at Pennsylvania State
University by NSF grant (AST-1517363) and NASA ATP program (80NSSC18K1103).

\bibliography{references}

\appendix   

\section{Transformation between misidentification and interloper fractions}
\label{app:fgplane}
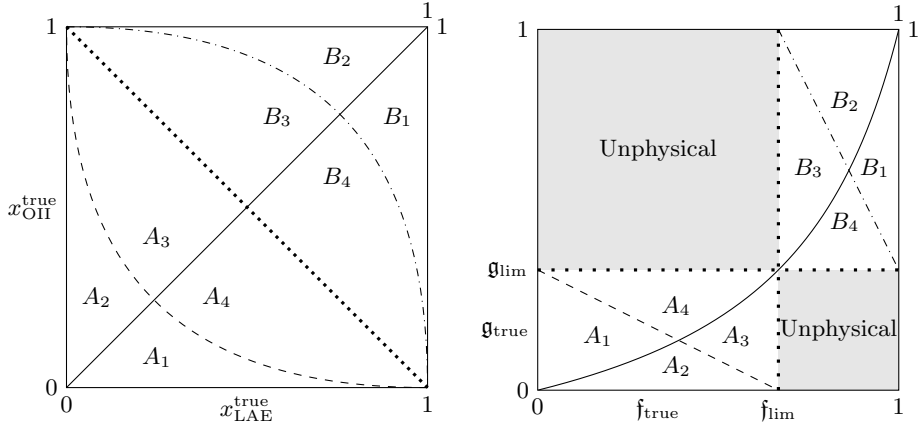
\begin{figure*}
    \centering
    \begin{tikzpicture}[scale=0.8]
        \draw (0,0) node[below]{0} node[left]{0}
        -- (6,0) node[below]{1}
        -- (6,6) node[right]{1} node[above]{1}
        -- (0,6) node[left]{1} -- cycle;
        \node[below] at (3,0) {$x_\lae^\mathrm{true}$};
        \node[left] at (0,3) {$x_\oii^\mathrm{true}$};
        \draw[dotted,very thick] (0,6) -- (6,0);
        \draw (0,0) -- (6,6);
        \draw[domain=0:0.6666,samples=50,dashed] plot(
        {6*1/3*(1-\x/(2/3))/(1-\x-1/3*(1-\x/(2/3))) * (-\x + (1-\x)*2)},
        {6*\x/(1-\x-1/3*(1-\x/(2/3))) * (-1/3*(1-\x/(2/3)) + (1-1/3*(1-\x/(2/3)))/2)});
        \draw[domain=0.6667:1,samples=50,dashdotted] plot(
        {6*(7/3-2*\x)/(1-\x-7/3+2*\x) * (-\x + (1-\x)*2)},
        {6*\x/(1-\x-7/3+2*\x) * (-7/3+2*\x + (1-7/3+2*\x)/2)});
        \node at (1.5,0.5) {$A_1$};
        \node at (0.5,1.5) {$A_2$};
        \node at (1.5,2.5) {$A_3$};
        \node at (2.5,1.5) {$A_4$};
        \node at (5.5,4.5) {$B_1$};
        \node at (4.5,5.5) {$B_2$};
        \node at (3.5,4.5) {$B_3$};
        \node at (4.5,3.5) {$B_4$};
    \end{tikzpicture}
    \begin{tikzpicture}[scale=0.8]
        \fill[black!10] (0,2) rectangle (4,6);
        \fill[black!10] (4,0) rectangle (6,2);
        \draw (0,0) node[below]{0} node[left]{0}
        -- (6,0) node[below]{1}
        -- (6,6) node[right]{1} node[above]{1}
        -- (0,6) node[left]{1} -- cycle;
        \node[below] at (2,0) {$\f_\mathrm{true}$};
        \node[left] at (0,1) {$\g_\mathrm{true}$};
        \draw[loosely dotted, very thick] (4,0) node[below]{$\f_\mathrm{lim}$} -- (4,6);
        \draw[loosely dotted, very thick] (0,2) node[left]{$\g_\mathrm{lim}$} -- (6,2);
        \draw[domain=0.001:1, samples=50] plot({6/(1 + (1-\x)/\x/2)},
        {6/(1 + (1-\x)/\x*2)});
        \draw[dashed] (0,2) -- (4,0);
        \draw[dashdotted] (4,6) -- (6,2);
        \node at (2.3,1.4) {$A_4$};
        \node at (3.3,0.9) {$A_3$};
        \node at (2.3,0.4) {$A_2$};
        \node at (1.0,0.9) {$A_1$};
        \node at (5.6,3.7) {$B_1$};
        \node at (5.1,4.8) {$B_2$};
        \node at (4.5,3.7) {$B_3$};
        \node at (5.1,2.8) {$B_4$};
        \node at (2,4) {Unphysical};
        \node at (5,1) {Unphysical};
    \end{tikzpicture}
    \caption{Transformation from $(x_\lae^\mathrm{true},x_\oii^\mathrm{true})$ to
    $(\ffid,\gfid)$ for fixed
    $r_\true=N_\oii^\true/N_\lae^\true$. On the left we show the $x_\lae
    x_\oii$-space, on the right the $\f\g$-space. Since counts of galaxies must
    be positive, the greyed-out regions are physically impossible, as explained
    in \refapp{fgplane}. The limiting values $\f_\mathrm{lim}$ and
    $\g_\mathrm{lim}$ are given in \refeqs{flim}{glim}.
    When fixing the observed ratio $r_\obs$ [\refeq{robs}] instead of
    $r_\true$, then $\f$ and $\g$ are unrestricted, and instead $x_\lae$ and
    $x_\oii$ are limited, as elaborated in the text.
    }
    \label{fig:fgplane}
\end{figure*}
In this appendix, we seek a better understanding of the mapping between the
misidentification fractions $x_\lae$ and $x_\oii$, and the interloper fractions
$\f$ and $\g$. We introduced $\f$ and $\g$ since they are relevant for relating
the true and observed power spectra. However, $x_\lae$ and $x_\oii$ are a more
natural choice for the process of misidentification. We will assume that the
misidentification of LAEs as OIIEs is independent from the misidentification of
OIIEs as LAEs. That is, we assume that there are two degrees of freedom
describing interlopers.

We assume that when performing a forecast for a survey such as HETDEX, the
fiducial values for the true numbers of galaxies $N_\lae$ and
$N_\oii$ are given, e.g., from previously measured luminosity functions
\citep[e.g.][]{ciardullo/etal:2012}. This implies that the true interloper
fractions $\ffid$ and $\gfid$ cannot take on arbitrary
values. For example, for HETDEX, we expect there to be about twice as many OIIEs
as LAEs. Therefore, the fraction of OIIEs in the LAE sample cannot exceed 2/3
unless some LAEs are misidentified as well. Hence, it is not possible to have
more than 2/3 OIIEs in the LAE sample at the same time as an uncontaminated
OIIE sample, given the true numbers of galaxies.

To find which interloper fractions are physical, we consider when either the
observed LAE number density or the observed OIIE number density vanishes. Then,
\refeqs{f}{g} implies that the limiting cases are on the two lines
$\ffid=\f_\mathrm{lim}$ and $\gfid=\g_\mathrm{lim}$, where
\ba
\label{eq:flim}
\f_\mathrm{lim} &= \frac{N_\oii}{N_\lae + N_\oii}\,,
\\
\label{eq:glim}
\g_\mathrm{lim} &= \frac{N_\lae}{N_\lae + N_\oii}
= 1 - \f_\mathrm{lim}\,,
\ea
where the numbers are the true numbers of galaxies.
The lines $\ffid=\f_\mathrm{lim}$ and $\gfid=\g_\mathrm{lim}$ mark boundaries
between allowed and disallowed (unphysical) regions for the interloper
fractions. By requiring positive number densities, we find that interloper
fractions $\ffid<\f_\mathrm{lim}$ are unphysical unless
$\gfid<\g_\mathrm{lim}$. Similarly, interloper fractions
$\ffid>\f_\mathrm{lim}$ are unphysical unless $\gfid>\g_\mathrm{lim}$. Thus, we
have two allowed regions and two unphysical regions, as shown in the right
panel of \reffig{fgplane}, where we use $N_\oii/N_\lae=2$. Thus,
the true interloper fractions must be chosen from the allowed regions $A_i$ or
$B_i$ shown in the figure.

Fixing the ratio of true numbers $N_\oii/N_\lae$, we derive the non-linear
transformation between interloper fractions and misidentification fractions
from \refeqs{numlaeobs}{g} to get
\ba
\label{eq:fxx}
\ffid
&= \(1 + \frac{1-x_\lae^\mathrm{true}}{x_\oii^\mathrm{true}}\,\frac{N_\lae}{N_\oii}\)^{-1}
\,,\\
\label{eq:gxx}
\gfid
&= \(1 + \frac{1-x_\oii^\mathrm{true}}{x_\lae^\mathrm{true}}\,\frac{N_\oii}{N_\lae}\)^{-1}\,.
\ea
To gain a better understanding of the transformation, we show the possible
values for $x_\lae^\mathrm{true}$ and $x_\oii^\mathrm{true}$ in the left panel
of \reffig{fgplane}. The points $(0,0)$ and $(1,1)$ in $x_\lae x_\oii$-space
become $(0,0)$ and $(1,1)$ in $\f\g$-space. The point $(0,1)$ becomes the line
$\ffid=\f_\mathrm{lim}$ (since $\g$ is undetermined due to the observed number
of OIIEs vanishing in this case) and $(1,0)$ becomes $\gfid=\g_\mathrm{lim}$
(since the observed number of LAEs vanishes). The points on the diagonal
$x_\lae^\mathrm{true}+x_\oii^\mathrm{true}=1$ map onto the single point
$(\f_\mathrm{lim},\g_\mathrm{lim})$, because the LAE and OIIE samples have the
same fraction of LAEs in each, and thus the observed auto power spectra are the
same. Furthermore, we have added three more lines: the diagonal
$x_\lae^\mathrm{true}=x_\oii^\mathrm{true}$ becomes the curved line in
$\f\g$-space, and the dashed and dash-dotted lines, which are curved in $x_\lae
x_\oii$-space, become diagonals of the allowed regions in $\f\g$-space.
Finally, for clarity, the sub-regions $A_i$ and $B_i$ have been labeled in
the two panels of \reffig{fgplane} correspondingly.

The reason two regions appear in $\f\g$-space is due to fixing the
true ratio $N_\oii/N_\lae$. However, when measuring the interloper
fractions $\f$ and $\g$, this ratio cannot be assumed to be known. Rather, it
must be viewed as a parameter to be determined in the fit. However, the ratio
\ba
\label{eq:robs}
r_\obs &\equiv \frac{N_\oii^\obs}{N_\lae^\obs}
\ea
will be known, so that the transformation \refeqs{fxx}{gxx} now become
\ba
\label{eq:xlaefg}
x_\lae
&= \(1 + \frac{1-\f}{\g} \, r_\obs^{-1}\)^{-1}
\,,\\
\label{eq:xoiifg}
x_\oii
&= \(1 + \frac{1-\g}{\f} \, r_\obs\)^{-1}
\,,
\ea
where now $r_\obs$ is fixed instead of $r_\true=N_\oii/N_\lae$. This
transformation has the same form as \refeqs{fxx}{gxx}, provided that we switch
$x_\lae\leftrightarrow\f$, $x_\oii\leftrightarrow\g$, and $r_\true\to r_\obs$.
Thus, the picture is reversed: when only $r_\obs$ is known, the full plane
$0\leq\f\leq1$ and $0\leq\g\leq1$ is allowed. Indeed, since $r_\true$ is
allowed to vary it is possible to find that the measured values $\f,\g$ will be
within the unphysical regions. Finally, we note that now there are
restrictions on the physically-allowed values for $x_\lae$ and $x_\oii$ similar
to those for $\ffid$ and $\gfid$ in the right panel of
\reffig{fgplane}.

\section{The statistics of galaxy samples contaminated with interlopers}
\label{app:shotnoise}
In the main text, we have derived the galaxy two-point correlation functions of
the contaminated galaxy samples based on the underlying, continuous density
fields and the relations between them [\refeqs{deltaobs}{deltaobsoii}]. In this
appendix, we shall extend the derivation including the discrete, point-like
nature of the observed galaxy distribution. This analysis clarifies the
shot-noise contribution to the two-point correlation functions from the
contaminated galaxy sample.

For the analysis in this appendix, we shall assume that the galaxy distribution
is a Poisson sampling of the underlying galaxy density field. Note that in the
main text, when considering the relationship between the underlying, smooth
galaxy fields, we set the misidentification fractions $x_\lae$ and $x_\oii$ to
be constant. When dealing with the statistics of galaxies, however, we need to
take into account that the misidentification fractions are not constants
anymore.
When estimating the density contrast, the survey volume is often divided into a
grid of small cells.
If the grid is small enough, for example, then each cell will host
zero or one galaxy, and, in such an extreme case, the misidentification
fraction in each cell can be either $x=0$ (when the identification is correct),
or $x=1$ (when the identification is wrong).

Therefore, when calculating the contaminated power spectrum measured
from discrete points such as galaxies, we need to take into account the
distribution of misidentification fractions. We accommodate this by assuming
that the misidentification is a stochastic process governed by the probability
given by the mean misidentification fractions $\bar{x}_\lae$ and
$\bar{x}_\oii$.

In this section, we consider the statistics of two generic galaxy
populations that we refer to as 1 and 2, which can be, for example, LAEs and
OIIEs. Let's call the number of galaxy population 1 and 2 in a given cell $n_1$
and $n_2$, the probability of misidentification $p_1$ and $p_2$, and the true
number of misidentified galaxies $m_1$ and $m_2$. Then, in the observed sample,
we register
\ba
n_1^\obs &= n_1 - m_1 + m_2
\label{eq:n1obs} \\
n_2^\obs &= n_2 - m_2 + m_1\,.
\label{eq:n2obs}
\ea
galaxies as population 1 and 2.

While we only consider the case with constant mean misidentification fraction
$\bar{x}$ (thus, constant probability $p$) and the constant mean density $n$ in
the main text, here, we generalize the situation by considering their spatial
variation. That is usually the case for realistic galaxy surveys where the
survey conditions vary over different telescope pointings. We show that
including the spatial variation contributes to the survey window function.

\subsection{One-point statistics: distribution of misidentification fractions}
First, let us focus on a sufficiently small volume cells with a given mean
number of galaxies $\mu_n$. Under the assumption that the galaxies are Poisson
draws, the misidentified galaxies (with misidentification probability $p$) are
also Poisson draws with a modified mean number $\mu_m=p\mu_n$. We show that as
follows.

Consider a cell with $n$ galaxies. With the probability $p$ of
misidentification, the probability of misidentifying $m$ (that is, $n-m$
galaxies are correctly identified) is given by
\be
P(m|n) = 
\left(
\begin{array}{c}
n \\ m
\end{array}
\right) p^m (1-p)^{n-m}
=
\frac{n!}{m!(n-m)!} p^m (1-p)^{n-m}
\ee
when $n\ge m$, and $0$ otherwise.
We then calculate the probability of having $m$ misidentified galaxies by 
marginalizing the union probability $P(m\cap n)$ as follows:
\ba
P(m) 
=& 
\sum_{n=0}^\infty P(m\cap n)
=
\sum_{n=m}^\infty P(m|n)P(n)
=
\sum_{n=m}^\infty 
\frac{n!}{m!(n-m)!} p^m (1-p)^{n-m} e^{-\mu_n}\frac{\mu_n^n}{n!}
\vs
=&
\frac{(\mu_np)^m}{m!} 
e^{-\mu_n} 
\sum_{n=m}^\infty 
\frac{1}{(n-m)!} 
[\mu_n (1-p)]^{n-m} 
=
\frac{(\mu_np)^m}{m!} 
e^{-\mu_np}\,,
\ea
which completes the proof. It follows that the mean and the variance of the 
misidentified galaxies in the cell are 
\ba
\mu_m \equiv& \langle m \rangle
= p\mu_n
\\
\sigma_m^2 \equiv& \langle m^2 \rangle - \langle m \rangle^2
= p \mu_n
\label{eq:var_m1}
\ea
where $p$ is the misidentification probability in a given cell, which is the 
same as the mean misidentification fraction $\bar{x}$.

Using the Poisson probability distribution function, we also calculate the one-point covariance between the total number of galaxies $n$ and the misidentified galaxies $m$:
\ba
\langle nm \rangle - \left<n\right>\left<m\right>
&= \sum_{n=0}^\infty\sum_{m=0}^{n}
nm\,
P(n) \, P(m|n)
-
\mu_n\mu_m
= \sum_{n=0}^\infty
n\,
e^{-\mu_n}\,\frac{\mu_n^{n}}{n!}\,
\sum_{m=0}^{n}
m \,
\begin{pmatrix} n \\ m \end{pmatrix}\,
p^{m}\,
(1-p)^{n-m}
-
p\mu_n^2
\vs
&=
p\mu_n\,.
\label{eq:m1n1-avg}
\ea
Note that we use the following identity to calculate the second summation:
\ba
\label{eq:binomial}
a \, \partial_a \big[ (a+b)^n\big]
&= a n (a+b)^{n-1}
= \sum_{m=0}^n m \begin{pmatrix} n \\ m \end{pmatrix} a^m b^{n-m}\,.
\ea

\subsection{Two-point statistics: auto-correlation}
We calculate the two-point correlation function. Using \refeqs{n1obs}{n2obs}, we calculate the observed density contrast of the population $1$, $\delta_1^\obs(\vr)$, as
\ba
\label{eq:delta_obs}
\delta_1^\obs(\vr)
&= \frac{n_1^\obs(\vr) - \bar n_1^\obs(\vr)}{\bar n_1^\obs(\vr)}
= \frac{(n_1(\vr) - \bar n_1(\vr))
    - (m_1(\vr) - \bar m_1(\vr))
+ (m_2(\vr) - \bar m_2(\vr))}
{\bar n_1(\vr) - \bar m_1(\vr) + \bar m_2(\vr)}\,.
\ea
Note that we consider the cell small enough to define the local number density
as well as the local misidentification fraction. To calculate the observed
auto-correlation function $\xi_1^\obs(\vx) = \left\langle \delta_1^\obs(\vr)
\delta_1^\obs(\vr') \right\rangle$ we use the result for a Poisson-sampled
galaxy population in \citet{feldman/kaiser/peacock:1994},
\ba
\langle n(\vr) n(\vr') \rangle
&= \bar n(\vr) \bar n(\vr') [1 + \xi(\vr - \vr')]
+ \bar n(\vr)\delta^D(\vr - \vr')\,,
\ea
which must hold for both $n_1$ and $m_1$ as both are Poisson draws of
underlying smooth fields. That is,
\ba
\label{eq:n1n1}
\langle n_1(\vr) n_1(\vr') \rangle
&= \bar n_1(\vr) \bar n_1(\vr') [1 + \xi_1(\vr - \vr')]
+ \bar n_1(\vr)\delta^D(\vr - \vr')\,,
\\
\label{eq:m1m1}
\langle m_1(\vr) m_1(\vr') \rangle
&= \bar m_1(\vr) \bar m_1(\vr') [1 + \xi_1(\vr - \vr')]
+ \bar m_1(\vr)\delta^D(\vr - \vr')\,,
\\
\label{eq:m2m2}
\langle m_2(\vr) m_2(\vr') \rangle
&= \bar m_2(\vr) \bar m_2(\vr') [1 + \xi_2(\vr - \vr')]
+ \bar m_2(\vr)\delta^D(\vr - \vr')\,,
\ea
For the cross correlation function $\langle m_1(\vr) n_1(\vr')\rangle$, we use \refeq{m1n1-avg} to find
\ba
\label{eq:m1n1}
\langle m_1(\vr) n_1(\vr') \rangle
&= \bar m_1(\vr) \bar n_1(\vr') [1 + \xi_1(\vr - \vr')]
+ \bar m_1(\vr)\delta^D(\vr - \vr')\,.
\ea
Here, we implicitly assume that the classification does not yield extra bias;
thus, the misidentified galaxies have the same bias as the correctly identified
samples. Of course, this assumption can be violated when the misidentified
galaxies by themselves form a particular subclass with different bias
parameters. For example, we can think of the case where we misidentify
preferentially low-mass galaxies. Should it happen, we need to introduce the
new set of biased parameters for the misidentified samples. Note that we now
use the misidentification fraction $\bar{x}$ instead of the misidentification
probability $p$ that we use in the previous section.

The contribution from the cross-correlation between the two populations is
\ba
\langle m_2(\vr) n_1(\vr') \rangle
&= \bar m_2(\vr) \bar n_1(\vr') [1 + \xi_{12}(\vr - \vr')]\,,
\ea
and we assume that the cross correlation preserves parity,
$
\left<\delta_1(\vr)\delta_2(\vr')\right> 
=
\left<\delta_2(\vr)\delta_1(\vr')\right>,
$
or 
$\xi_{12}(\vr-\vr') = \xi_{21}(\vr-\vr')$, which leads to
\ba
\langle m_1(\vr) n_2(\vr') \rangle
&= \bar m_1(\vr) \bar n_2(\vr') [1 + \xi_{12}(\vr - \vr')]\,.
\ea
Note that the shot-noise term is absent because we assume that the two populations are statistically independent.

Using the two-point correlators we have calculated above, the observed
two-point correlation function
$\xi_1^\obs(\vx)=\left\langle\delta_1^\obs(\vr)\delta_1^\obs(\vr')\right\rangle$
as a function of the separation $\vx=\vr-\vr'$ becomes
\ba
\bar n_1^\obs(\vr) \bar n_1^\obs(\vr')
\xi_1^\obs(\vx)
&=
\left[
\bar n_1(\vr) 
-
\bar m_1(\vr) 
\right]
\left[
\bar n_1(\vr')
-
\bar m_1(\vr')
\right]
\xi_1(\vx)
+
\bar m_2(\vr) \bar m_2 (\vr')
\xi_2(\vx)
\vs
&+
\left\{
    \left[
\bar n_1(\vr) 
-
\bar m_1(\vr) 
\right]
\bar m_2(\vr')
+
\bar m_2(\vr) 
\left[
\bar n_1(\vr')
-
\bar m_1(\vr')
\right]
\right\}
\xi_{12}(\vx)
+
\bar n_1^{\rm obs}(\vr)\delta^D(\vx)
\ea
Now we define the local value of $\f(\vr)$ consistent with the one we define in \refeq{f} as
\be
\f(\vr)
= \frac{\bar{m}_2(\vr)}{\bar n_1^{\obs}(\vr)}
= \frac{\bar{m}_2(\vr)}{\bar n_1(\vr) - \bar m_1(\vr) + \bar m _2 (\vr)},
\ee
with which we simplify the expression for the observed two-point correlation function as
\be
\xi_1^\obs(\vx)
=
\left[1-\f(\vr)\right] \left[1-\f(\vr')\right] \xi_1(\vx)
+
\f(\vr)\f(\vr') \xi_2(\vx)
+
\left\{
\left[1-\f(\vr)\right]\f(\vr')
+
\f(\vr)\left[1-\f(\vr')\right]
\right\}
\xi_{12}(\vx)
+
\frac{1}{\bar{n}_1^\obs(\vr)}\delta^D(\vx)\,.
\ee
The spatially varying misidentification fraction, therefore, acts just like the survey window function effect \citep{feldman/kaiser/peacock:1994}.
Of course, by analogy we find the observed two-point correlation function of the second galaxy population as
\be
\xi_2^\obs(\vx)
=
\left[1-\g(\vr)\right] \left[1-\g(\vr')\right] \xi_2(\vx)
+
\g(\vr)\g(\vr') \xi_1(\vx)
+
\left\{
\left[1-\g(\vr)\right]\g(\vr')
+
\g(\vr)\left[1-\g(\vr')\right]
\right\}
\xi_{12}(\vx)
+
\frac{1}{\bar{n}_2^\obs(\vr)}\delta^D(\vx)\,.
\ee
For constant $\f$ and $\g$, the equations reduce to, respectively, \refeq{xiobs_all} and \refeq{xiobsoii_all}, except for the shot-noise contributions that we have not included in the main text. 
Note, however, that the shot-noise contribution is given by the total observed
number density of galaxies, not by taking \refeq{xiobs_all} and
\refeq{xiobsoii_all} replacing the individual two-point correlation function
with the respective shot noise contribution: that is, $\xi_1(\vx) \not\to
\xi_1(\vx) + 1/\bar{n}_1\delta^D(\vx)$.

As we have discussed in the main text, when considering two populations such as
LAEs and OIIEs that are far away, the direct correlation $\xi_{12}(\vx)$ must
be negligible compared to the their autocorrelations. We then find that, by
taking the Fourier transform, the observed power spectrum is given by the
convolution as 
\ba
\label{eq:pkLLcontam_general}
&\langle |\delta_1^\obs(\vk)|^2 \rangle
=
\int \frac{\dd^3\bfq}{(2\pi)^3}
\,\big|(2\pi)^3\delta^D(\vk - \bfq) - \f(\vk - \bfq)\big|^2 \, P_1(\bfq)
+
\int \frac{\dd^3\bfq}{(2\pi)^3}
\,\big|\f(\vk - \bfq)\big|^2 \, P_2(\bfq)
+
\int \dd^3\vr \, \frac{1}{\bar n_1^\obs(\vr)}\,.
\ea

\subsection{Two-point statistics: Cross-correlation}\label{app:cross-correlation}
Similarly, we calculate the cross-correlation function by 
\ba
\bar{n}_1^\obs(\vr)
\bar{n}_2^\obs(\vr')
\left<
\delta_1^\obs \delta_2^\obs(\vr')
\right>
= 
\left[1-\f(\vr)\right]\g(\vr')
\xi_1(\vx)
+
\f(\vr)\left[1-\g(\vr')\right]
\xi_2(\vx)
+
\left\{
\left[1-\f(\vr)\right]\left[1-\g(\vr')\right]
\f(\vr)\g(\vr')\right\}
\xi_{12}(\vx)\,.
\ea
which generalizes \refeq{xiobs_LO}.

\subsection{False detections}\label{app:falsedetections}
It is straightforward to model random, uncorrelated false detections. Here we
briefly show how they modify the observed power spectrum. If `0'
signifies false detections, `1' signifies LAEs, and `2' signifies OIIEs,
then the number of objects classified as false detections, LAEs, and OIIEs
are
\ba
n_0^\obs &= n_0 - m_{01} - m_{02} + m_{10} + m_{20} \,, \\
n_1^\obs &= n_1 - m_{10} - m_{12} + m_{01} + m_{21} \,, \\
n_2^\obs &= n_2 - m_{20} - m_{21} + m_{02} + m_{12} \,,
\ea
where $n_i$ are the true number of objects and $m_{ij}$ are the number of
objects of type $i$ misidentified as type $j$.
Introducing the six independent average interloper fractions
\ba
\f_{ij} &= \frac{\bar m_{ij}}{\bar n_j^\obs}\,,
\ea
we can write the observed power spectra as
\ba
\label{eq:P0obs}
P_0^\obs &= \f_{10}^2 P_1 + \f_{20}^2 P_2 \\
\label{eq:P1obs}
P_1^\obs &= (1-\f_{01}-\f_{21})^2 P_1 + \f_{21}^2 P_2 \\
P_2^\obs &= \f_{12}^2 P_1 + (1-\f_{02}-\f_{12})^2 P_2 \\
P_{01}^\obs &= (1-\f_{01}-\f_{21}) \f_{10} P_1 + \f_{20} \f_{21} P_2 \\
P_{02}^\obs &= \f_{10}\f_{12} P_1 + (1-\f_{02}-\f_{12}) \f_{20} P_2 \\
P_{12}^\obs &= (1-\f_{01}-\f_{21}) \f_{12} P_1 + (1-\f_{02}-\f_{12}) \f_{21} P_2\,,
\label{eq:P12obs}
\ea
 where we assume that the true cross-correlations all vanish, and
that the false detections do not cluster, i.e.\ $P_0=0$. To first order, the
last three equations will allow us to measure $\f_{10}$, $\f_{20}$, $\f_{12}$,
and $\f_{21}$. That is, it will be possible to measure the contribution of
LAEs and OIIEs in the false-detections sample ($\f_{10}$ and $\f_{20}$), but
the contribution of false detections in the two galaxy samples ($\f_{01}$ and
$\f_{02}$) will need to be assessed via other methods. For example,
in the context of HETDEX we can assume that $\f_{02}=0$, because false
detections will not show up in the continuum photometry, and, thus, will be
exclusively misclassified as LAEs.

Also apparent from \refeqs{P0obs}{P12obs} are the following two effects.
First, false detections reduce the observed power in the auto-correlations
by a factor $(1-\f_{01})^2$. This can be seen by writing in \refeq{P1obs}
$\f_{21}=(1-\f_{01})\f$, where $\f\equiv \bar m_{21}/\bar
n_1^{\obs,\mathrm{gal}}$ with $\bar n_1^{\obs,\mathrm{gal}}\equiv \bar
n_1^\obs-\bar m_{01}$ the number of galaxies (either LAE or OIIE) in
the LAE sample.
Second, the shot noise for the LAE sample will be $1/\bar
n_1^\obs=(1-\f_{01})/\bar n_1^{\obs,\mathrm{gal}}$. Thus, false detections
reduce the power of a survey by
\ba
\bar n_1 P_1  &\to (1-\f_{01}) \bar n_1  P_1 \,.
\ea

\section{Gaussian covariance of Power spectrum}
\label{app:pkerror}
We estimate the covariance of the galaxy auto- and cross-power spectra
by assuming that the galaxy density fields follow Gaussian statistics;
the connected higher-order correlators are determined by the multiplications 
among the disconnected two-point correlators that we calculate by Wick's 
theorem. This assumption works well in estimating the uncertainties 
(diagonal covariance) of nonlinear matter power spectrum in a suite of 
N-body simulations \citep{jeong/komatsu:2009}.

To get the general expression applicable for multiple galaxy populations, 
let us start from the cross power spectrum between the two populations,
labeled $x$ and $y$, for which we can estimate the cross power spectrum as
\ba
\label{eq:pkhat}
\hat P_{xy}(\vk)
&= \frac{1}{N_\vk V_s} \sum_{i=1}^{N_\vk} \frac12 
\left(
    \delta^*_{xi} \delta_{yi}
    + \delta_{xi} \delta_{yi}^*
\right)\,.
\ea
Here, $\delta_{xi}\equiv \delta_x(\vk_i)$ is the Fourier-space density 
contrast, $N_\vk$ is the total number of Fourier modes contributing the 
estimation. For example, when estimating the monopole, $P_{xy}(k)$,
$N_{\vk}$ equals to the number of discrete Fourier vectors satisfying
$|\vk_i|\sim k$, $N_k=4\pi k\Delta kV_s/(2\pi)^3$ for the Fourier space
radial bin size $\Delta k$; when estimating the two-dimensional power spectrum,
$P(k_\perp,k_\parallel)$, $N_{\vk}$ equals to the number of discrete 
Fourier vectors within the cylinders of total Fourier volume of 
$V_\vk = 4\pi k_\perp^2\Delta k_\perp \Delta k_\parallel$, including both 
positive and negative $k_\parallel$.
The configuration-space volume $V_s$ appears for the normalization.

The estimator of \refeq{pkhat} is unbiased because the statistical 
homogeneity demands that the ensemble average of each term 
in \refeq{pkhat} must be the cross power spectrum which is defined as
\be
\big\langle \delta^*_{xi} \delta_{yj} \big\rangle
= \delta^K_{i,j} V_s P_{xy}(\vk_i)\,,
\label{eq:defpxy}
\ee
where $\delta^K_{i,j}$ is a Kronecker delta. Furthermore, being the 
complex conjugate of each other, each contribution in \refeq{pkhat} is a 
real number; so is the cross power spectrum. We, therefore, take only the 
real part of the cross power spectrum $P_{xy}(\vk)$ that can be in general
a complex quantity \citep{bonvin/etal:2014}. Specifically, this is equivalent 
to assuming an even-parity cross power spectrum, $P_{xy}(\vk)=P_{xy}(-\vk)$.

In order to calculate the covariance matrix, we need other types of two-point
correlators that we can derive from \refeq{defpxy} using the reality of 
the galaxy density field: $\delta^*_i = \delta_{-i}$, with negative 
indices standing for the Fourier modes with negative wavevector 
($\vk_{-j}=-\vk_j$). They are 
\ba
\big\langle \delta_{xi} \delta_{yj}^* \big\rangle
=&\, \delta^K_{i,j} V_s P_{xy}(\vk_i)
\,,\vs
\big\langle \delta_{xi} \delta_{yj} \big\rangle
=
\big\langle \delta^*_{xi} \delta^*_{yj} \big\rangle
=&\, \delta^K_{i,-j} V_s P_{xy}(\vk_i)\,.
\ea
Using them, we finally calculate the covariance between two cross power 
spectra $\hat P_{xy}(\vk)$ and $\hat P_{zw}(\vk)$ with 
\refeq{pkhat} estimator as following. Note that we consider the same binning
scheme for $P_{xy}$ and $P_{zw}$ for which case the only non-zero covariance 
is
\ba
\big\langle \hat P_{xy}(\vk) \hat P_{zw}(\vk) \big\rangle
- \big\langle \hat P_{xy}(\vk) \big\rangle \big\langle \hat P_{zw}(\vk) \big\rangle
&= \frac{1}{N_\vk} \Big(
P_{xw}(\vk) P_{yz}(\vk)
+ P_{xz}(\vk) P_{yw}(\vk)
\Big)\,.
\label{eq:generalCij}
\ea
We shall apply the general expression in \refeq{generalCij} to the cases that 
we use in the main text of the paper.

\subsection{Auto power spectrum}
For the variance of the auto-power spectrum (with $x=y=z=w$),
\refeq{generalCij} reduces to 
\ba
\label{eq:deltaPkL}
\sigma_{P_{xx}(\vk)}^2
&= \frac{2}{N_\vk} \left(P_x(\vk) + \frac{1}{\bar n}\right)^2\,,
\ea
where we included shot noise as $P_{xx}(\vk) = P_x(\vk)+1/\bar{n}$, where 
$P_x(\vk)$ is the power spectrum of the underlying field.

\subsection{Cross power spectrum}
The cross-correlation (with $x=z$ and $y=w$) has the variance
\ba
\label{eq:pkLOerr}
\sigma_{P_{xy}(\vk)}^2
&= \frac{1}{N_\vk}
\left[
\(P_x(\vk) + \frac{1}{\bar n_x}\)
\(P_y(\vk) + \frac{1}{\bar n_y}\)
+ P_{xy}^2(\vk)
\right]\,.
\ea

\subsection{Variance of the multipole power spectra}\label{app:pkerror_multipole}
The multipole power spectrum is defined as the $k$-depending coefficient of the
multipole expansion:
\ba
P_{xy}(\vk) &= \sum_{\ell'=0}^\infty P_{\ell'}^{xy}(k) \, \mathcal{P}_{\ell'}(\mu)\,,
\ea
with the Legendre polynomials $\mathcal{P}_\ell(\mu)$.
Here, the assumption is that the power spectrum $P_{xy}(\vk)$ depends also 
on the wavenumber $k$ and the polar angle $\mu\equiv \nhat\cdot\khat$.
Using the orthogonality of the Legendre polynomials,
\ba
\int_{-1}^1 \dd\mu \, \mathcal{P}_\ell(\mu)\,\mathcal{P}_{\ell'}(\mu)
&= \frac{2}{2\ell+1}\,\delta^K_{\ell,\ell'}\,,
\ea
we find the expression for the multipole power spectrum as
\ba
P_\ell^{xy}(k)
&= \frac{2\ell+1}{2} \int_{-1}^1 \dd\mu\,\mathcal{P}_\ell(\mu)\,P_{xy}(\vk)\,.
\ea
The estimator for the multipole power spectrum is, using 
$\int_{-1}^1\dd\mu/2=1/N_k\sum_{i=1}^{N_k}$ that works for sufficiently 
large number of $N_k$,
\ba
\hat{P}_\ell^{xy}(k)
&= \frac{2\ell+1}{N_k V_s} \sum_{i=1}^{N_{k}}
\mathcal{P}_\ell(\mu)
\frac12 \(
\delta_{xi}^*\delta_{yi}
+
\delta_{xi}\delta_{yi}^*
\)\,.
\ea
Note that $\ell=0$ correspond to the monopole power spectrum. 
We calculate the covariance of this estimator using the same assumption that 
we have adopted earlier in this appendix and obtain that 
\ba
\left<
\hat{P}_\ell^{xy}(k)
\hat{P}_\ell^{zw}(k)
\right>
-
\left<
\hat{P}_\ell^{xy}(k)
\right>
\left<
\hat{P}_\ell^{zw}(k)
\right>
&=\frac{(2\ell+1)^2}{N_k}
\int_{-1}^1
\frac{\dd\mu}{2}
{\cal P}^2_\ell(\mu)
\Big(
P_{xw}(\vk) P_{yz}(\vk)
+ P_{xz}(\vk) P_{yw}(\vk)
\Big)\,.
\ea
This equation is consistent with the Eq.~(8) of \citet{taruya/etal:2011}.
Therefore, the variance of the multipole of the auto-power spectrum is 
\ba
\sigma_{P_\ell^{xx}(k)}^2
&=
\,\frac{2(2\ell+1)^2}{N_k}
\int_{-1}^1 \frac{\dd\mu}{2}\,\mathcal{P}_\ell^2(\mu)\,
\bigg[P_x(k,\mu) + \frac{1}{\bar n_x}\bigg]^2\,,
\ea
and that of the cross-power spectrum is
\ba
\sigma_{P_\ell^{xy}(k)}^2
&=
\frac{(2\ell+1)^2}{N_k}
\int_{-1}^1 \frac{\dd\mu}{2}\,\mathcal{P}_\ell^2(\mu)
\bigg[
\bigg(P_x(k,\mu) + \frac{1}{\bar n_x}\bigg)
\bigg(P_y(k,\mu) + \frac{1}{\bar n_y}\bigg)
+ P_{xy}^2(k,\mu)\bigg]\,.
\ea

\section{Systematic Bias}
\label{app:systematicbias}
In this appendix, we derive the systematic bias of a maximum likelihood
estimator relative to some reference maximum likelihood estimator.
Specifically, we consider two situations. First, when the measured power
spectrum differs from the model power spectrum by some $\Delta P(\vk)$, e.g.\
due to interlopers, and, second, when the $\Delta P(\vk)$ encapsulate
differences between realizations, e.g. to justfy our use of the
ensemble-averaged log-likelihood function.

In either case, our goal is to see how the estimated parameters $\hat\theta$
differ from those estimated by the reference $\hat\theta_\mathrm{ref}$. At
$\hat\theta$ the Jacobian $J=-(\ln\Like)_{,i}$ must vanish. Expanding in
$\Delta\hat\theta=\hat\theta-\hat\theta_\mathrm{ref}$, we get
\ba
\label{eq:jacobian-expansion}
0 =
J\big(\hat\theta\big)
&= J\big(\hat\theta_\mathrm{ref}\big)
+ F\big(\hat\theta_\mathrm{ref}\big) \Delta\hat\theta
+ \orderof\big(\Delta\hat\theta^2\big)\,,
\ea
where $F=-(\ln\Like)_{,ij}$ is the Fisher information matrix. Now we assume
that $\Delta P$ is small. That is, we assume that the Fisher information matrix
can be written as
\ba
F &= F_\mathrm{ref} + \orderof(\Delta P)\,,
\ea
where $F_\mathrm{ref}$ is the Fisher information matrix of the reference
log-likelihood function. \refeq{jacobian-expansion} can then be solved for
$\Delta\hat\theta$. Also, in all cases of interest to us, we have that the
Jacobian $J\big(\hat\theta_\mathrm{ref}\big)\propto\Delta P$ [see
\refeq{jacobian-interloper-bias} below]. Thus, the bias is
\ba
\label{eq:systematic_bias}
\Delta\hat\theta
&= - F_\mathrm{ref}^{-1}\big(\hat\theta_\mathrm{ref}\big) J\big(\hat\theta_\mathrm{ref}\big)
+ \orderof\big(\Delta P^2,\Delta\hat\theta^2\big)\,.
\ea

\subsection{Interloper Bias}
To predict the systematic bias due to interlopers using
\refeq{systematic_bias}, we need $J\big(\hat\theta_\mathrm{ref}\big)$. In this
case, the reference likelihood function is the unbiased interloper-free
likelihood, and the full likelihood is
\ba
\label{eq:nlnL-biased}
-\ln\mathfrak{L}(\theta)
&= \sum_\vk \frac{N_\vk}{2} \ln P(\vk;\theta)
+ \sum_\vk \frac{N_\vk}{2} \,
\frac{\hat P(\vk) + \Delta P(\vk)}{P(\vk;\theta)}
=
-\ln\mathfrak{L}_\mathrm{ref}(\theta)
+ \sum_\vk \frac{N_\vk}{2} \, \frac{\Delta P(\vk)}{P(\vk;\theta)}\,,
\ea
where the change in the observed power spectrum produced by interlopers is
\ba
\Delta P(\vk)
&= -\f\,(2-\f) P_\lae(\vk) + \f^2 P_\oii^\proj(\vk)
+ 2\f\,(1-\f)\,P_{\lae\times\oii}(\vk)
\,.
\ea
Thus, the Jacobian evaluated at $\hat\theta_\mathrm{ref}$ is
\ba
\label{eq:jacobian-interloper-bias}
J\big(\hat\theta_\mathrm{ref}\big)
&=
-\sum_\vk \frac{N_\vk}{2}
\, \frac{P_{,j}(\vk;\hat\theta_\mathrm{ref})}{P(\vk;\hat\theta_\mathrm{ref})}
\, \frac{\Delta P(\vk)}{P(\vk;\hat\theta_\mathrm{ref})}\,,
\ea
which allows us to calculate the systematic bias using
\refeq{systematic_bias}.

\subsection{Ensemble-Averaged Log-Likelihood Function}
The second application of \refeq{systematic_bias} is to justify using the
ensemble-averaged log-likelihood function $\<-\ln\Like\>$ to assess the bias of
our estimator. In this case, the reference estimator is the maximum of the
ensemble-averaged log-likelihood function (which may be biased to begin with),
and $\Delta P(\vk)$ encapsulates the differences
between realizations. That is,
\ba
\Delta P(\vk) \equiv P(\vk)-\<P(\vk)\>\,.
\ea
The likelihood function [\refeq{nlnL}] can be written similarly to
\refeq{nlnL-biased} and, thus, $J\propto\Delta P$. However, this time we are
interested in the ensemble average $\big\langle\Delta\hat\theta\big\rangle$,
which will tell us whether the ensemble of Monte Carlo realizations will give
the same bias as a single MCMC run on the ensemble average. Since, $\<\Delta
P\>=0$, \refeq{systematic_bias} becomes
\ba
\label{eq:avgdeltathetahat}
\big\langle\Delta\hat\theta\big\rangle
&= 0 + \orderof\big(\Delta P^2,\Delta\hat\theta^2\big)\,,
\ea
which confirms that the two methods agree to first order. We have also verified
this result by generating 100 realizations of the power spectra and
comparing with a single MCMC run on the ensemble-averaged log-likelihood. We
performed this both with and without misidentification. The biases and
covariance matrices agree to within \SI{\sim10}{\percent} in both cases.
This is consistent with the expected sampling variances for 100 realizations
and negligible second-order terms in \refeq{avgdeltathetahat}.

\end{document}